\documentclass{aa}

\usepackage{graphicx}
\usepackage{picture}
\usepackage{color}
\definecolor{orange}{rgb}{0.8,0.4,0.0}
\usepackage[colorlinks=true,
citecolor=blue,
draft=false]{hyperref}
\usepackage[normalem]{ulem}
\usepackage{xspace}
\usepackage{tikz}
\usepackage{txfonts}
\usepackage{helvet}
\newcommand{\kmpers}{\ensuremath{\mathrm{km \, s^{-1}}}\xspace} 

\newcommand{\cmthree}{cm$^{-3}$}

\newcommand{\vlsr}{$\upsilon_{\rm LSR}$}              

\newcommand{\about}{$\sim$}                       
\newcommand{\expo}[1]{$10^{#1}$}
\newcommand{\texpo}[1]{$\,\times\,10^{#1}$}

\newcommand{\nhtva}{$n$($\mathrm{H_2}$)}           

\newcommand{\htva}{H$_2$}

\newcommand{\tolvco}{$^{12}$CO\xspace}
\newcommand{\trettenco}{$^{13}$CO\xspace}
\newcommand{\catteno}{C$^{18}$O\xspace}

\newcommand{\asec}{$^{\prime \prime}$}
\newcommand{\adeg}{$^{\circ}$}

\newcommand{\msun}{\ensuremath{M_{\odot}}\xspace}

\usepackage{xcolor}

\authorrunning{Per Bjerkeli, et al.}
\titlerunning{Episodic infall towards a compact disk in B335?}

\begin{document} 

  \title{Episodic infall towards a compact disk in B335?
}

   \author{Per Bjerkeli \inst{1}
          \and Jon P. Ramsey  \inst{2}
          \and Daniel Harsono \inst{3}
          \and Adele Plunkett \inst{4}
          \and Zhi-Yun Li \inst{2}
          \and \\
          Matthijs H. D., van der Wiel \inst{5}
          \and Hannah Calcutt \inst{6}
          \and Jes K. J\o rgensen \inst{7}
          \and Lars E. Kristensen \inst{7}}

   \institute{Chalmers University of Technology,
              Department of Space, Earth and Environment, SE-412 96 Gothenburg, Sweden\\
              \email{per.bjerkeli@chalmers.se}
             \and Department of Astronomy, University of Virginia, Charlottesville, VA 22904, USA
              \and
              Institute of Astronomy, Department of Physics, National Tsing Hua University, Hsinchu, Taiwan, R.~O.~C.
              \and
              National Radio Astronomy Observatory, 520 Edgemont Rd, Charlottesville, VA 22903, USA
              \and
              Independent researcher, Zwolle, the Netherlands
\and
Institute of Astronomy, Faculty of Physics, Astronomy and Informatics, Nicolaus Copernicus University in
Toru{\'n}, ul. Grudzi\k{a}dzka 5, 87--100 Toru{\'n}, Poland
              \and
              Niels Bohr Institute, University of Copenhagen, \O ster Voldgade 5--7, DK-1350 Copenhagen K, Denmark. 
             }

   \date{Submitted October 12, 2022; Accepted June 16, 2023}


  \abstract
  {Previous observations of the isolated Class~0 source B335 have presented evidence of ongoing infall in various molecular lines, e.g., HCO$^+$, HCN, CO. There have been no confirmed observations of a rotationally supported disk on scales greater than \about~12~au.}
  {The presence of an outflow in B335 suggests that also a disk should be present or in formation. To constrain the earliest stages of protostellar evolution and disk formation, we aim to map the region where gas falls inwards and observationally constrain its kinematics. Furthermore, we aim to put strong limits on the size and orientation of any disk-like structure in B335.}
  {We use high angular resolution \trettenco\ data from the Atacama Large Millimeter/submillimeter Array (ALMA), and combine it with shorter-baseline archival data to produce a high-fidelity image of the infall in B335. We also revisit the imaging of high-angular resolution Band 6 continuum data to study the dust distribution in the immediate vicinity of B335. }
  {Continuum emission shows an elliptical structure (10 by 7 au) with a position angle 5 degrees east of north, consistent with the expectation for a forming disk in B335. A map of the infall velocity (as estimated from the \trettenco\ emission), shows evidence of asymmetric infall, predominantly from the north and south. Close to the protostar, infall velocities appear to exceed free-fall velocities. 3D radiative transfer models, where the infall velocity is allowed to vary within the infall region, can explain the observed kinematics.}
  {The data suggests that a disk has started to form in B335 and that gas is falling towards that disk. However, kinematically-resolved line data towards the disk itself is needed to confirm the presence of a rotationally supported disk around this young protostar. The measured high infall velocities are not easily reconcilable with a magnetic braking scenario and suggest that there is a pressure gradient that allows the infall velocity to vary in the region. }

\maketitle  
\section{Introduction}
\label{sec:intro}
When dark clouds collapse under their gravity \citep{Shu:1987fk}, the conservation of angular momentum  leads to the formation of a protoplanetary disk in orbit around a forming protostar. Material with little angular momentum migrates inward and onto the protostar, while material with excess angular momentum contributes to the disk's build-up. Via the launching of magnetically-powered winds, further angular momentum can be removed from the disk  \citep{Blandford:1982fj,Pudritz:1983fv,Shu:1994kx} and allow disk material to migrate inwards towards the protostar \citep{Gravity-Collaboration:2020tu}.

Infall of material from envelope to disk scales is a prerequisite to star formation and is the primary signature used to determine if a core is collapsing to form a protostar. Nevertheless, detecting such motions turns out to be surprisingly difficult and, indeed, it was during a search for infall signatures that outflows were first discovered \citep{Snell:1979fk,Snell:1980lr}. With improvements in telescopes, however, infall is now frequently detected towards the youngest sources \citep[e.g.][]{Kristensen:2012kx}. 

Infall motion is typically studied through spectrally resolved gas observations and the detection of a special type of self-absorbed, optically thick spectral line profile. In a collapsing core, the temperature of the gas increases towards the centre of the core, and the blue-shifted part of the spectrum will be enhanced with respect to the red-shifted part \citep{Leung:1977gs}. 
However, such infall profiles have also been observed in regions where infall is known not to be present, for example, towards molecular outflows and shocks \citep{Bjerkeli:2012fk,Bjerkeli:2011qy} where red-shifted and blue-shifted emission can mimic the characteristic shape of infall profiles. Furthermore, the absence of infall profiles does not rule out infall. In fact, previous surveys revealed that most young stellar objects do not show evidence of infalling material despite the high accretion rates inferred from outflows and/or infrared emission \citep[e.g.][]{Carney:2016kf, Heiderman:2015fm}. Therefore, one should be careful when interpreting infall profiles, and high angular resolution observations are generally needed to distinguish between different kinematical components. Such high angular and spectral resolution is available with the Atacama Large Millimeter/submillimeter Array (ALMA). 

In the general picture of star formation, we expect infall to be most prominent during the earliest stages of star formation, in other words, during the prestellar core and Class~0 \citep{Andre:1990fk} phases, when there is still ample material available in the envelope. A representative example is the infall profile in the water line observed with {\it Herschel} towards the prestellar core L1544 \citep{Caselli:2012rt}. The ground-state \textit{ortho-}water line (at 557~GHz) turns out to be an excellent tracer of infall, and such profiles are more common towards Class 0 sources than Class I sources \citep{Kristensen:2012kx,Mottram:2013qy}. Observing conditions when studying infall water line profiles are, however, not ideal due to the limited spatial resolution available with satellites, the opaqueness of the Earth's atmosphere when observing from the ground, and the known high abundance of water in outflows. Therefore, to characterise the inward motions on the smallest scales, and to be able to separate emission from different components (for example disk, envelope, outflow), high angular resolution observations are needed.

Such observations were recently presented towards one of the youngest known protostellar regions, B335, which is an isolated, dense, and almost spherical Bok globule. In a study by \citet{Evans:2015qp}, it was shown from the line profiles of HCO$^+$ and HCN that matter is falling inwards and line profiles were consistent with previously published models \citep{Evans:2005qy} where the radius of the infall region is 4000 au. In a follow-up study, combination of data from \textit{Spitzer}, \textit{Herschel}, \textit{ALMA} and the literature, showed that B335 underwent an increase in luminosity over the last few years \citep{Evans:2023aa}. This suggests that accretion is time-variable in B335, which could be a result of an episodic supply of infalling material, and is in line with another study of dense gas tracers (e.g.\ C$^{17}$O) on 100 -- 860 au scales, that revealed a complex velocity pattern indicative of asymmetric infall of material towards the central region \citep{Cabedo:2021hw}. Furthermore, based on continuum measurements, \citet{Evans:2015qp} showed that any disk in the centre is of very low mass,$~$\expo{-3}~\msun. This is consistent with the results of \citet{Yen:2015vf}, specifically that if a rotationally-supported disk is present in B335, it must be very small. It should be noted, however, that such a low value should be viewed with some skepticism. The latest modeling by \citet{Evans:2023aa} emphasize that part of the disk mass could be hidden behind optically thick dust. When it comes to the size of the disk in B335, a study carried out by \citet{Bjerkeli:2019ip} at 0.035\asec\ resolution revealed a dust continuum peak no larger than 12 au (0.07\asec) in diameter, where rotation is hinted at but not confirmed. On much larger scales, extended continuum emission is detected over an extent of \about 1500 au. Spectral line profiles of \trettenco\ revealed clear red-shifted absorption features in the vicinity of the continuum peak.

Detailed studies of infall in B335 are also important because it has been proposed that B335 may be an example of magnetic braking during disk formation in action \citep[e.g.][]{Yen:2018rt,Maury:2018qf,Cabedo:2023aa}. In this scenario, when a sufficiently strong magnetic field is present and aligned with the rotational axis of the core, disk formation can be suppressed through efficient removal of the angular momentum of the gas \citep[e.g.][]{Galli:1993oq, Galli:1993qe,Allen:2003mf, Li:2014jm,Wurster:2018re}. If the magnetic field becomes strong enough, it can remove nearly all of the angular momentum in the infall region \citep{Galli:2006jr}. Whether such magnetic braking occurs in B335 is not yet clear, but the relatively small disk \citep{Yen:2015vf, Bjerkeli:2019ip}, together with a known magnetically-powered outflow, plus apparent magnetically regulated collapse \citep{Maury:2018qf} may suggest that this is the case. 

In this paper, we combine the \trettenco\,(2--1) long-baseline ALMA observations presented in \citet{Bjerkeli:2019ip}, with previous observations taken with ALMA in a more compact configuration in order to examine the velocity profile on small scales in B335, and where infall velocities are highest. We also revisit the imaging of the combined continuum data presented in \citet{Bjerkeli:2019ip} to further illuminate the small-scale structure of the region. Even though infall has previously been detected in B335, it has not previously been resolved on very small spatial scales. A kinematically resolved infall velocity map can shed light on how matter is falling towards the protostar on au scales and timescales of only a few years. Indeed, accretion streamers towards disks have recently been suggested for multiple targets \citep[e.g.][]{Alves:2020vo, Segura-Cox:2020wc, Pineda:2020xm, Valdivia-Mena:2022aa}, where in some cases inflowing material has been shown to impact the disk \citep[e.g.][]{Garufi:2022aa}. Meanwhile, recent observations of B335 have shown that outflow emission show variability on very short timescales \citep[e.g.][]{Bjerkeli:2019ip}. Variable ejection is typically tied to variability close to the protostar \citep{Audard:2014xy}, but it is important to investigate whether such variability manifests itself also in the infall region on slightly larger scales. 

It is worth noting that there has been some uncertainty about the distance to B335 over the years; recently, \citet{Watson:2020wp} demonstrated that B335 was connected to a reflection nebula associated with the star HD184982 at a distance of 164.5$\pm$1.2 pc, based on Gaia measurements. In this paper, we adopt this (new) distance to B335.

The structure of the paper is as follows: The observations are described in Sec.\ 2. In Sec.\ 3, we discuss the geometry and the kinematics of the infalling material first from a purely observational point of view and secondly through comparisons with 3D radiative transfer models. Our summary and conclusions are presented in Sec.\ 4. 

\section{Observations and results}
\label{sec:observations}
Our long baseline observations were carried out between October 21--29 2017 as part of the ALMA Cycle 5 program 2017.1.00288.S. The observational details of that observing program, as well as details regarding calibration and imaging \citep[carried out in CASA;][]{McMullin:2007nr}, are described thoroughly in \citet{Bjerkeli:2019ip}.
We summarize them here for completeness. The angular resolution of the observations was 0.032 by 0.023\asec and the largest recoverable scale is of the order 0.3\asec (\about 50 au), much smaller than the \about 4000 au extent of the infall region as inferred in \citet{Evans:2015qp}. Our goal is, therefore, to combine our long-baseline observations with shorter baseline observations from 2013 (Project ID: 2013.1.00879.S), with observational details given in \citet{Yen:2015vf}. We chose to include these data to also recover emission on scales that are a factor of ten larger. 
Unlike the case of \tolvco, there is, however, no \trettenco\ 2--1 data available on intermediate scales. Hence, it is worth noting that sensitivity to emission on \about0.3\asec\ scales will be slightly suppressed in the present work. In addition, we revisited the imaging of the long-baseline continuum data that was presented in \cite{Bjerkeli:2019ip}. Specifically, we lowered the robust parameter during cleaning to 0.0 and reduced the cell size from 0.01\arcsec\ to 0.001\arcsec\ to recover small scale emission and improve image quality in an attempt to recover any disk-like structure. All molecular line spectral cubes were imaged after continuum subtraction in the $(u,v)$ domain.  The analysis of all the cleaned data products presented in this study was carried out in Matlab and images were convolved to a common angular resolution of 0.035 arcseconds to allow for direct comparisons. This resolution matches the resolution of the cubes presented in \citet{Bjerkeli:2019ip}.

\begin{figure*}[ht]
   \flushleft
   \begin{tabular}{c c}
    \includegraphics[width=0.49\hsize]{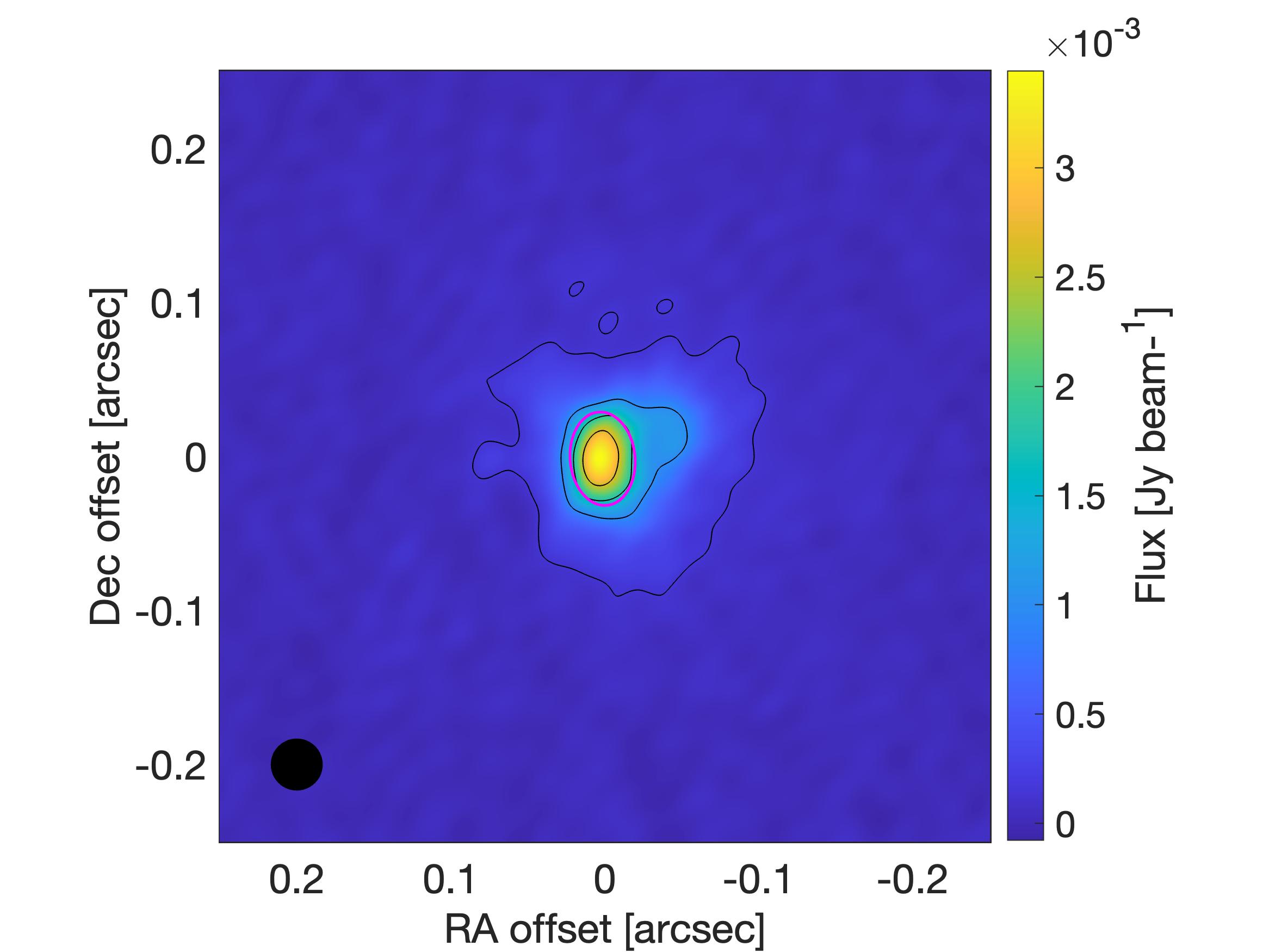} & \includegraphics[width=0.49\hsize]{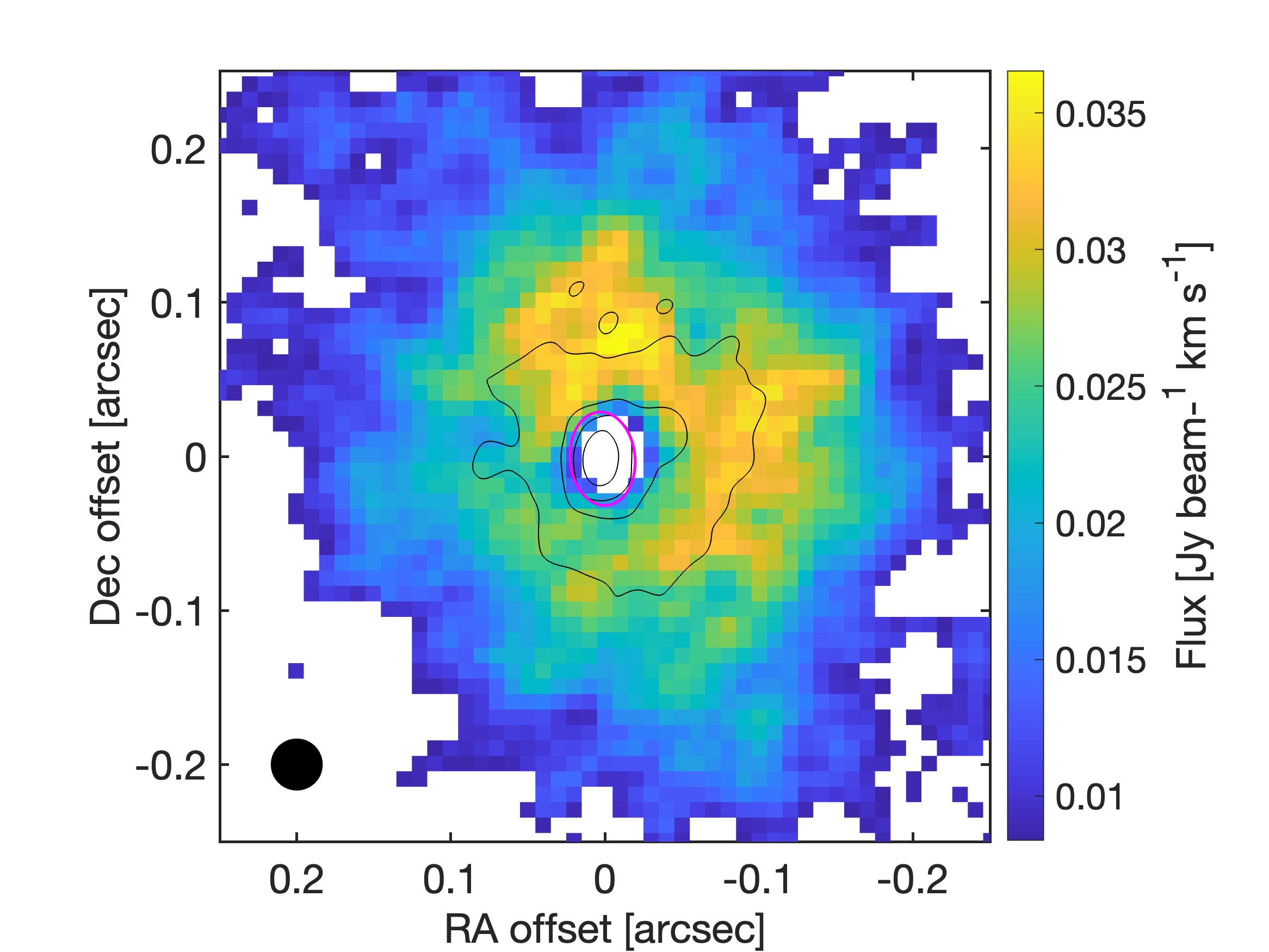}
    \end{tabular}
    \caption{Continuum and \trettenco\ emission around B335. \textit{Left:} Re-imaged continuum with line contours from 5~$\sigma$ in steps of 30~$\sigma$. Purple ellipse show a 2D Gaussian fit to the upper 40\% of the continuum emission. All data was convolved to a common resolution of 0.035\asec with beamsize shown in lower left of each panel. \textit{Right:} Moment 0 map of the \trettenco\ emission, integrated from -5 to +5 \kmpers with respect to the systemic velocity of 8.3~\kmpers, and  overlaid with the continuum in black contours. Only regions with \trettenco\ emission above 3$\sigma$ are included.}
  \label{fig:continuumcofigure}
\end{figure*}

Inspection of our improved map of the long-baseline dust continuum emission in B335 reveals a single-peaked structure elongated in the north-south direction and peaking towards the position of B335. Moreover, we observe complicated extended structures that are better revealed in our improved map. Specifically, we detect extended emission down to the 5$\sigma$ level in a region extending out to a distance of \about 0.1\asec\ (17~au) from the protostar (Fig.~\ref{fig:continuumcofigure}).

\begin{figure}[ht]
   \flushleft
   \hspace{-0.00cm}
   \includegraphics[width=1.0\hsize]{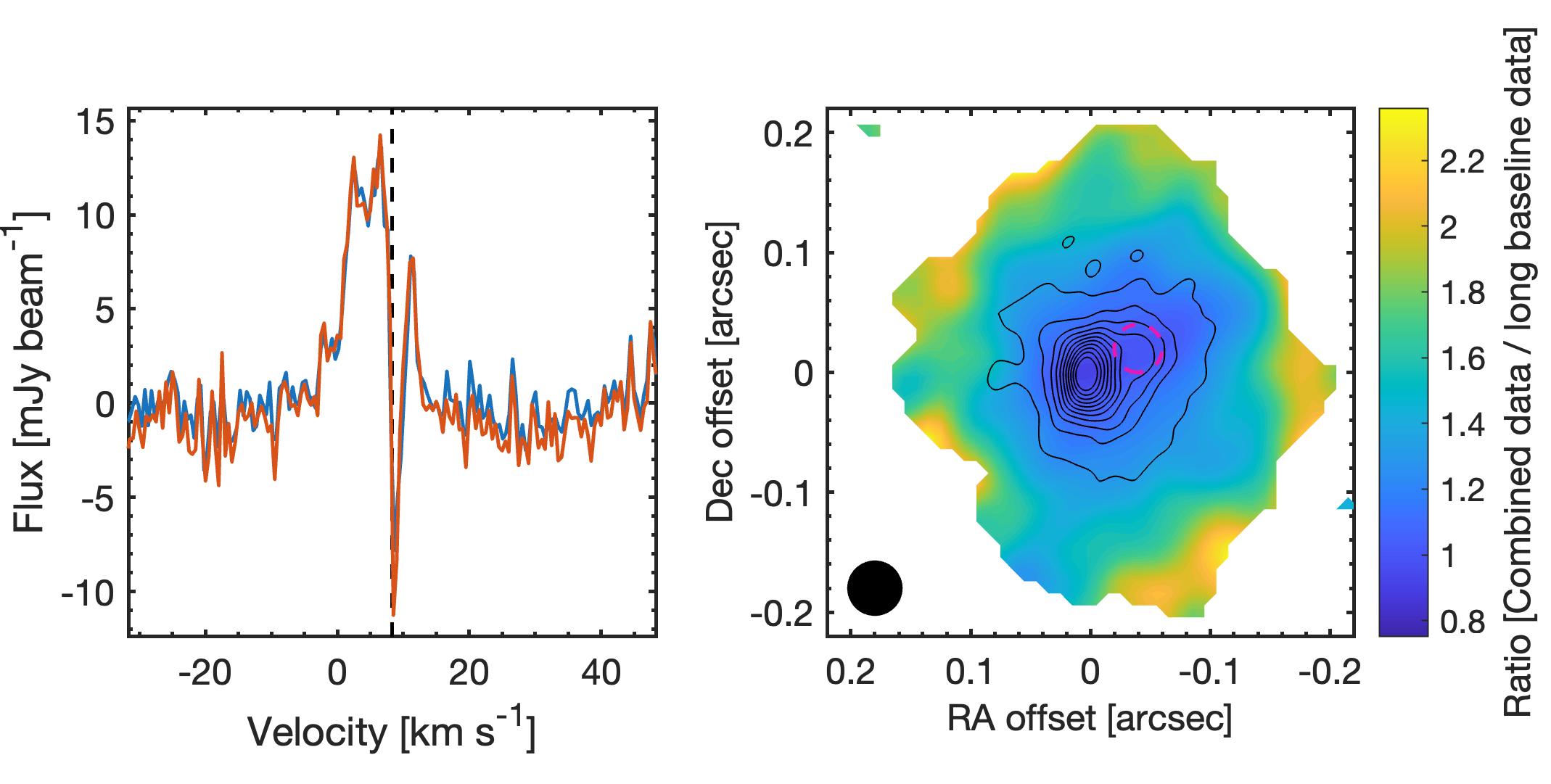} \\
   \includegraphics[width=1.0\hsize]{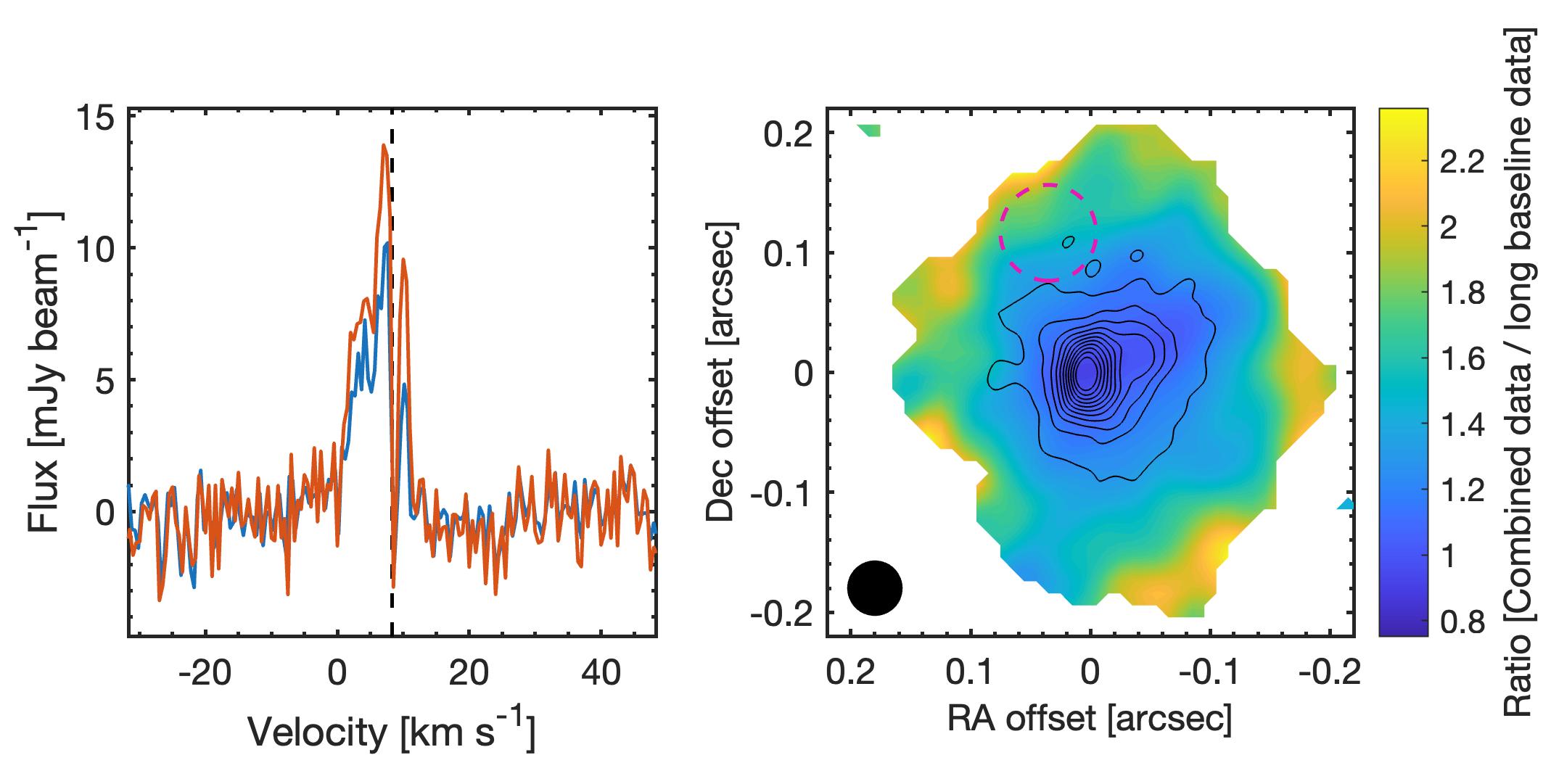} \\
   \includegraphics[width=1.0\hsize]{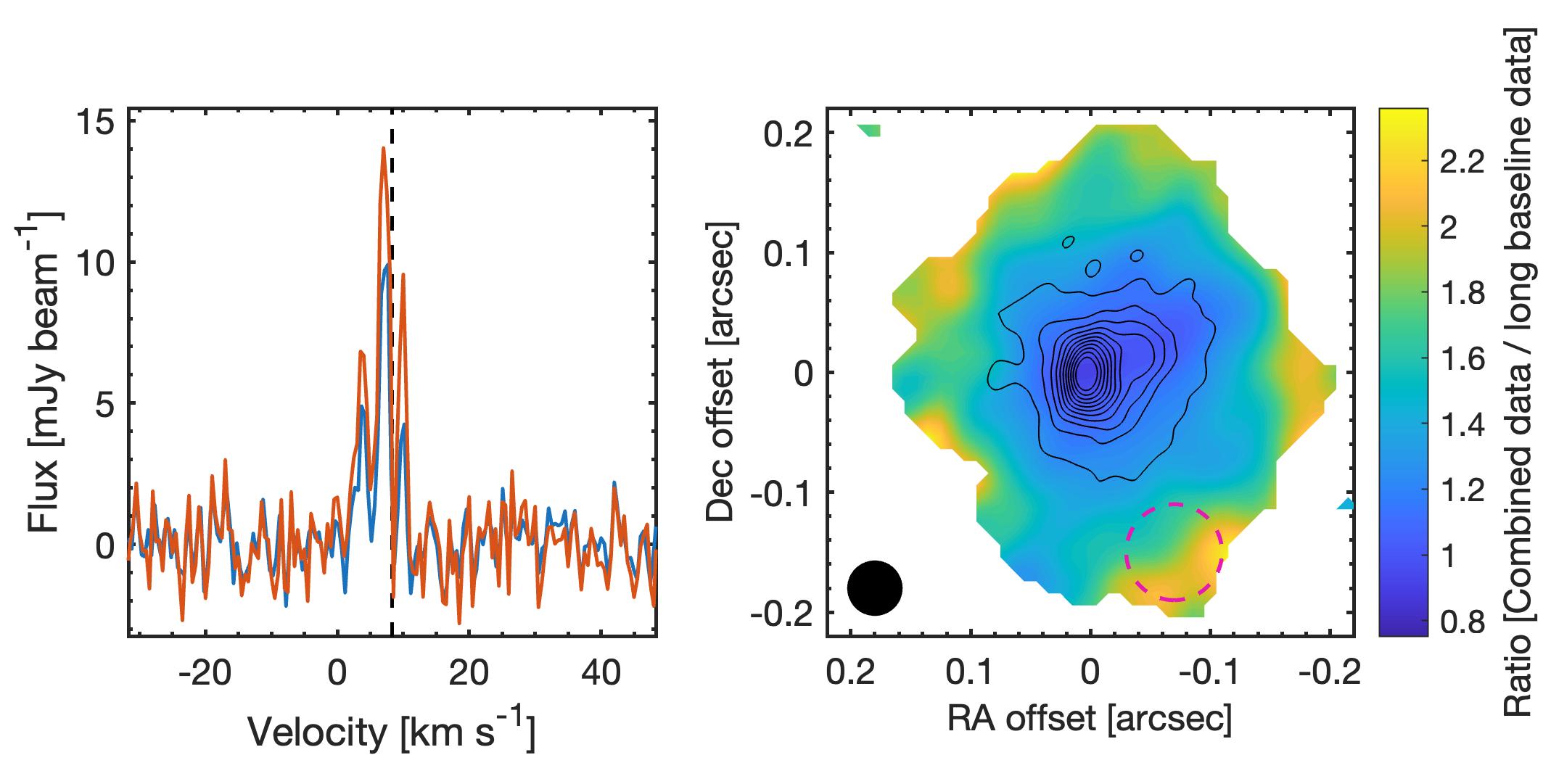} \\
      \caption{Comparisons of long-baseline and combined data line profiles \textit{Left:} Long-baseline (blue) and combined data (red) line profiles for three different positions (averaged over dashed purple circles in right panels). \vlsr~=~8.3~\kmpers, is indicated with black dashed lines in the left panels. \textit{Right:} The ratio of the \trettenco\ line flux between combined and long baseline data as a function of position, where both combined and long baseline emission exceeds 3$\sigma$. The line ratio maps suggest that information on larger scales is filtered out by the interferometer. Note that the regions used for averaging in the lower two panels are slightly bigger to improve the signal-to-noise ratio. Black contours show the continuum starting at 5 $\sigma$ in steps of 10$\sigma$. }
  \label{fig:lineprofiles}
\end{figure}

The combined \trettenco\ emission (Fig.~\ref{fig:continuumcofigure}), exhibits infall-type profiles  (Fig.~\ref{fig:lineprofiles}) characterized by an absorption feature at the systemic velocity of 8.3~\kmpers\ \citep{Evans:2005qy}. These infall profiles persist throughout the entire region surrounding the protostar \citep[see also Fig.~8 in][]{Bjerkeli:2019ip}. Notably, the infall profiles are also present in the northern and southern regions, where no \tolvco\ outflow emission is present. In addition, no significant \trettenco\ contribution from the outflow can be detected on the investigated scales. Neither does the \trettenco\ data show any signs of rotation on any scales. As such, we conclude that most (if not all) of the material traced by \trettenco\ is falling towards the central region. We consequently conclude the infall shape of the line profiles is an optical depth effect. Furthermore, careful inspection of line profiles throughout the region (Fig.~\ref{fig:lineprofiles}) and comparison between integrated long-baseline and combined data emission show that line profiles differ most at larger distances from the protostar (while the ratio approaches one towards the protostellar position). This suggests that a fraction of the emission originates from scales that are not picked up in our long baseline observations. The effect seems to be more prominent for the low-velocity gas, at greater distances from the protostar, but may still indicate that infall takes place on larger scales than what is probed here. This finding is in line with earlier, lower resolution studies of infall towards B335 \citep[e.g.][]{Evans:2015qp} where the infall region was estimated to be 4000~au in size.

\section{Discussion}
\label{sec:discuss}
\subsection{1.3 mm continuum and \trettenco\ emission}
\label{sub:emission}
In Fig.~\ref{fig:continuumcofigure}, the continuum emission is presented together with a 2D Gaussian fit to the centrally peaked part of the continuum (where the emission is above 1.4 mJy/beam). The best fit to the continuum yields a FWHM of 0.059\asec\ by 0.041\asec\ (9.7 by 6.7 au), and a position angle of 5 degrees east of north (purple ellipse in Fig.~\ref{fig:continuumcofigure}). This size and position angle are both consistent with what one would expect for a disk-like structure in B335 \citep{Yen:2015vf,Bjerkeli:2019ip}, if present. The elongation of the centrally peaked continuum emission is perpendicular to the direction of the outflow to within the uncertainty of the outflow position angle as inferred from Herbig-Haro knots \citep[$\sim 80$ \adeg $\pm 10$ \adeg E of N, see e.g.][]{Galfalk:2007lr}. Although it has been shown previously that the inner structure on scales of 10 -- 15 au is consistent with rotation \citep{Yen:2015vf,Bjerkeli:2019ip}, an optically-thin tracer of the kinematics on even smaller scales would be required to confirm the presence of a rotationally supported disk around B335. Our observations do include the \catteno\ 2--1 line, but the signal-to-noise ratio of the data is not sufficient to reveal the kinematics of such a disk-like structure.

Also presented in Fig.~\ref{fig:continuumcofigure} is the integrated \trettenco\ emission map (from -5~\kmpers~--~+5~\kmpers) above 3$\sigma$ (colours) for the region surrounding the continuum peak. This region is extended compared to the continuum emitting region, but nevertheless shows self-absorbed line profiles, suggesting that a not insignificant contribution to the line profiles comes from the surrounding cloud material in the region in the form of absorption.

\subsection{Geometry and kinematics of infalling material}
\label{subub:infallgeometry}
In a scenario where infall is spherically symmetric and increasing towards the centre, the emission will be detected across a range of velocities and a traditional intensity-weighted velocity map (e.g.,\ Fig.~\ref{fig:momentmap}), a channel map (e.g.,\ Fig.~\ref{fig:channelmap}) or a position-velocity diagram \citep[see Fig.~8 in][]{Bjerkeli:2019ip} may not provide a clear picture of the velocity field within each beam. However, the width of a line profile at each position of the infalling gas will provide a lower limit to the true infall velocity in that position. In the case where this position is directly towards the protostar (the dominant source of gravity on the small scales examined here) one would expect the width of the line profile to almost correspond to the maximum infall velocity (at least if the signal-to-noise ratio is sufficiently high). Meanwhile, in the case where the position is offset from the protostar, one would expect the extent of the line profile to be narrower than the maximum infall velocity. The reason for this is that since material can move at an angle with respect to the line of sight, the line profile will not capture the true maximum infall velocities (since they are at an angle with respect to the line-of-sight direction). In the spherically symmetrical case, this can be accounted for, but if infall is asymmetric, it becomes more difficult. 

To further investigate the geometry and kinematics of the infalling material in B335, we fit a Gaussian profile to the blue-shifted part of the line profile in each position of the \trettenco\ map. Since -- in an infall scenario -- the blue-shifted emission would suffer less from absorption than the red-shifted emission, this provides a better method to constrain the velocity field in the region. After careful comparison between observed spectra and Gaussian fits, we decide to take the velocity where the Gaussian fit meets the observed noise level at each position (Fig.~\ref{fig:infallvelocity})  
as a measure of the maximum observed motion in each position of the map -- in other words, to estimate a lower limit on the true infall velocity. From visual inspection of selected Gaussian fits, we conclude that a coefficient of determination (``R squared´´) higher than 0.25 is desirable in order to be able to accurately determine (i.e., to within approximately 1~\kmpers) the extent of the line profile. Hence and in the following, we only include the positions where this requirement is fulfilled. This method of producing a velocity map is useful to assess the morphology of the infalling region as described below. While it does not account for the skewness of the line-profiles, we note that the results remain consistent with our method even if we fit the blue-shifted line profiles at each position using a log-normal distribution (Fig.~\ref{fig:infallvelocity}), which give slightly higher velocity estimates. 


In the case of spherical infall -- and as mentioned before -- velocity measurements from Gaussian fitting to the emission line profiles would for most of the positions be lower than the true infall velocity in each given position due to projection effects. 
In the region around B335, we can furthermore assume that thermal broadening does not contribute significantly (less than 1~\kmpers) to the width of line profiles due to the relatively low temperatures \citep[200~K, see:][]{Bjerkeli:2019ip}. Similarly there is no reason to believe that there should be a significant contribution from turbulence, since there is no source that can contribute in the region outside the outflow. 
As such, and since we do not know the morphology of the infalling region, velocities estimated from Gaussian fits
\begin{figure*}[ht]
   \flushleft
   \begin{tabular}{c c}
   \hspace{-1.5cm}
   \includegraphics[width=0.55\hsize]{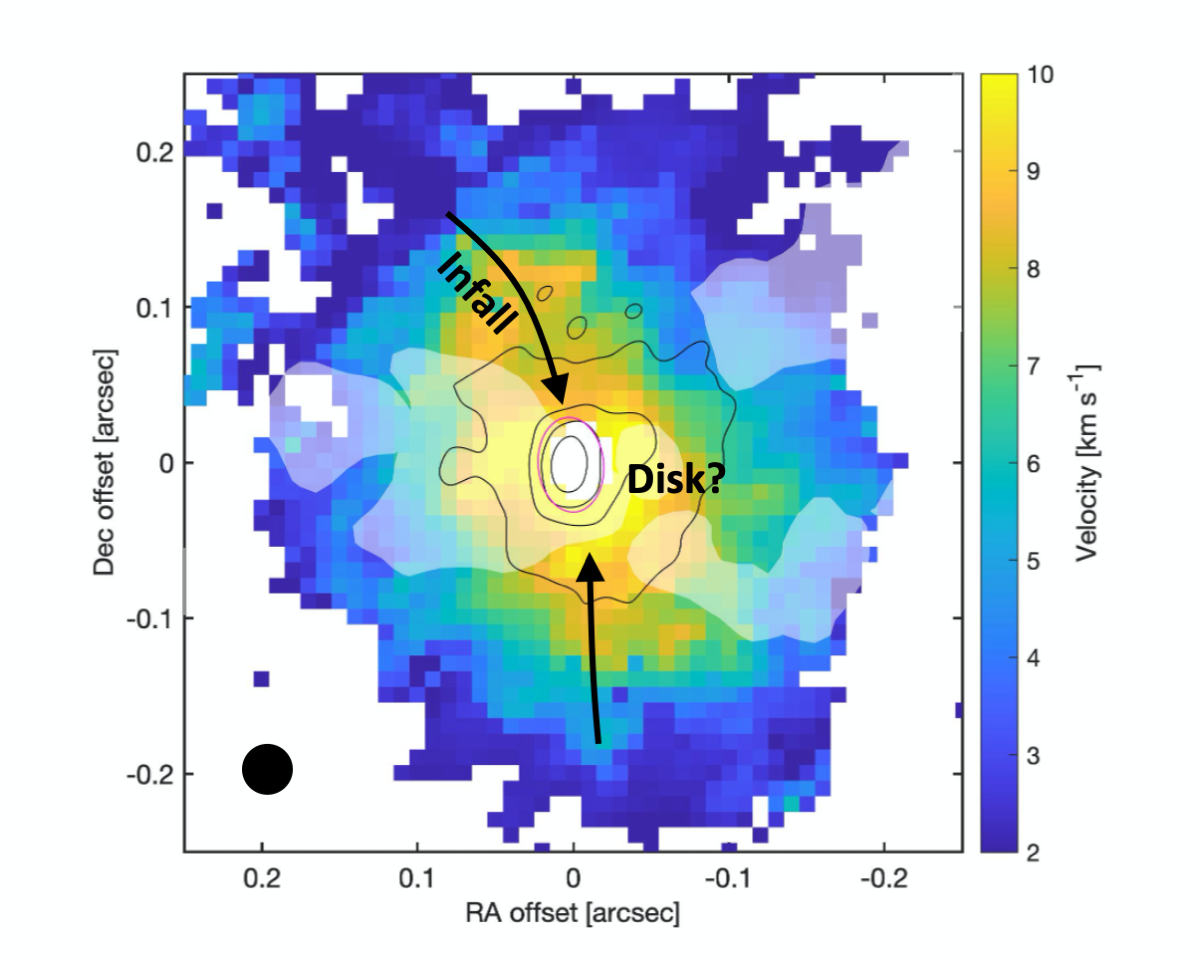} &
   \hspace{-0.9cm}\vspace{+0.4cm}
  \includegraphics[width=0.58\hsize]{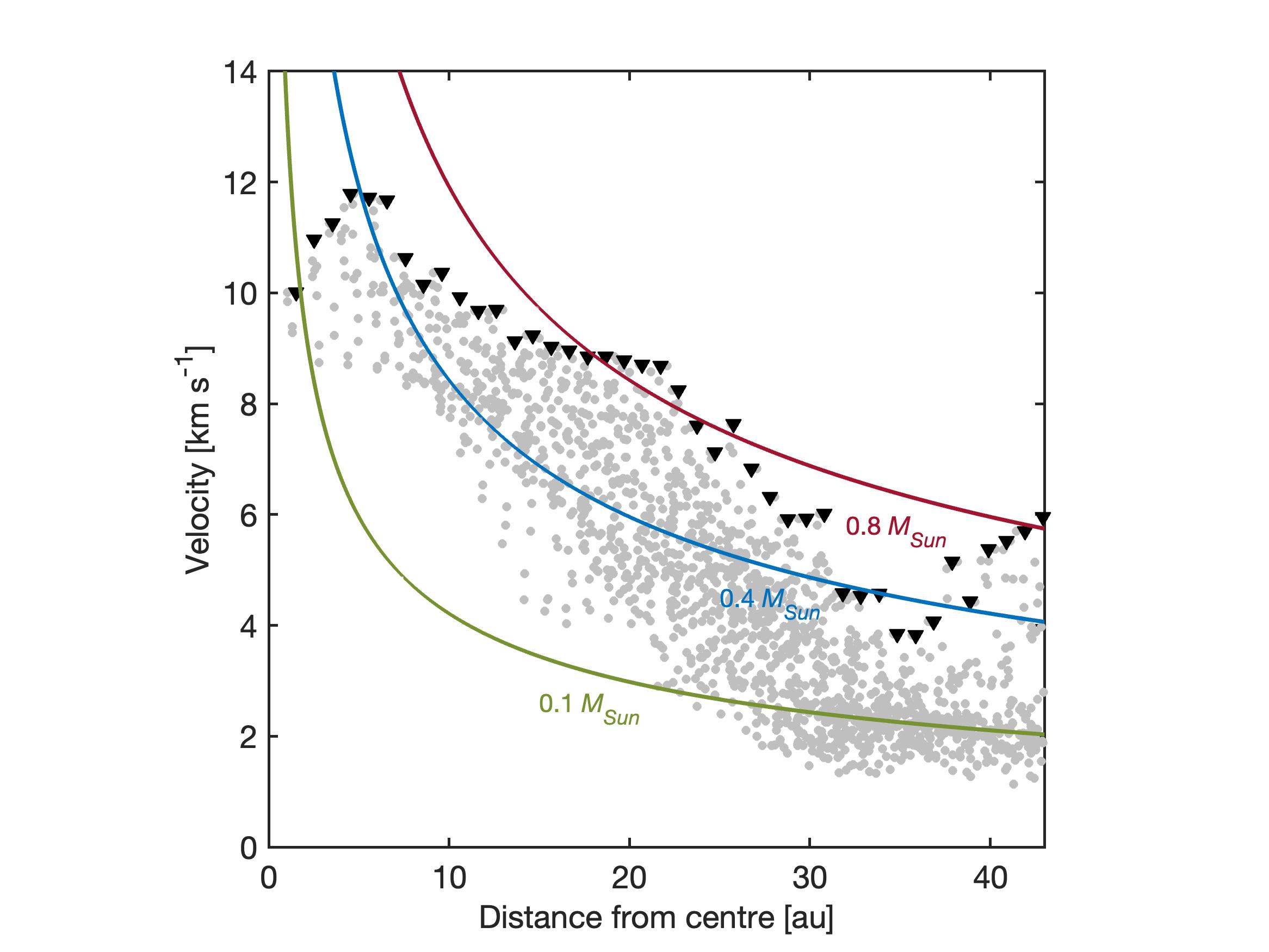} \\
  \end{tabular}
      \caption{Infall velocity field. \textit{Left:} Lower limit on the infall velocity (colour) as estimated from the FWHM of Gaussian fits to the blue-shifted components of line profiles at each position. Translucent white regions denote where \tolvco\ emission is detected \citep{Bjerkeli:2019ip}, and therefore where contamination from the outflow could contribute. Black contours show continuum starting at 5$\sigma$ in steps of 30$\sigma$, and the red ellipse show a 2D Gaussian fit to the upper 40\% of the continuum emission. For the velocity map, only positions where the coefficient of determination is higher than 0.25 are included.      
      \textit{Right:} Lower limit on the infall velocity as a function of distance from the continuum peak (grey dots). Maximum measured velocities at each radii are shown with black triangles. Note that this plot takes into account every pixel in the map, and therefore some of the gray dots correspond to regions coincident with the outflow. The observed lower limit on the infall velocities are compared with the spherically-symmetric free-fall scenario for three different protostellar masses (green, blue and red solid lines). 
      }
  \label{fig:infallfigure}
\end{figure*}
(Fig.~\ref{fig:infallfigure}) to the blue-shifted line profiles in the spatially resolved map should be considered \emph{lower limits} to the true velocity at each given position. In fact, higher velocities are reasonable to expect given the asymmetric nature of the emission.

From Fig.~\ref{fig:infallfigure}, it can be seen that \mbox{\trettenco\ 2--1} does not trace the outflow emanating from B335. In fact, high velocities in \trettenco\ are found predominantly to the north and south of B335, perpendicular to the direction of the outflow. This is expected in the case of infall, since this region is not actively participating in the outflow. We stress that the B335 region is spatially resolved in these ALMA observations and, in the following, we focus our analysis on the regions to the north and the south of the protostar, where the outflow is not present. In particular (Sec.~\ref{sec:radiativetransfer}), we focus on a position located 20 au to the north of the protostar and more than 10 au away from the outflow cavity wall.

\subsection{High velocity components and episodic flows on small scales?}
\label{sub:smallscalecomponents}
The lower limits on the infall velocities presented in Fig.~\ref{fig:infallfigure} can be compared to the expected infall velocities at corresponding radii. 
It is reasonable to assume that most of the mass in the B335 system is in the protostar and not the (disk-)like structure surrounding it. The protostellar mass was estimated by \citet{Yen:2015vf} and \citet{Evans:2015qp} to be 0.05~\msun and 0.15~\msun, respectively. Taking the new distance estimate of 164.5~pc into account (versus the prior distance, 100 pc), this translates to a central mass in the 0.1 -- 0.4~\msun\ range. In the case of free-fall, this implies maximum infall velocities of:
\begin{equation}
    \upsilon_{inf}~\mathbf{(r)} = \sqrt{\frac{2G(M_{\rm{env}}+M_{*})}{r}} \simeq\ {(3-6)}\,\left(\frac{r}{20~ \rm{au}}\right)^{-0.5}\,\kmpers,
    \label{eq:maxinfall}
\end{equation}
where $r$ is the distance from the protostellar position, $G$ is the gravitational constant, $M_{\rm{star}}$ is the mass of the central protostar, and $M_{\rm{env}}$ is the mass of the envelope inside $r$, which is negligible on the small scales discussed here \citep[less than 0.1\% of the total envelope mass assuming the density profile of][Appendix C]{Kristensen:2012kx}. At 20~au separation from the protostar, this yields velocities below the \about6~\kmpers\ level. Based on the observations, the lower limits that we estimate for the infall velocities (Fig.~\ref{fig:infallfigure}), are up to 50\% higher than 6~\kmpers, in particular towards positions located to the north and the south of the protostar. These measurements are not easily reconcilable with a magnetic braking scenario, where, on the contrary, velocities are expected to be lower than the free-fall velocities \citep[see e.g.,][Figs. 4, 7 \& 12]{Lam:2019aa}. 

We note that our infall map in Fig.~\ref{fig:infallfigure} suggests velocity variations on the level of a few \kmpers. However, given the uncertainties described above, it cannot be confirmed from the velocity map alone that fluctuations are due to episodic infall rather than noise or random motions. We also note that estimates of the infall velocity in the overlap region between the outflow and the envelope are particularly uncertain. What is clear from the map, however, is that velocities to the north and to the south are systematically higher than what is expected in the free-fall scenario, assuming the mass estimate is correct. Furthermore, velocities are not decreasing as \textit{r}$^{-0.5}$ as expected in a free-fall scenario. For the case in which the protostellar mass is higher than expected (e.g., 0.8~\msun), we derive velocities that are too low at small radii close to the protostar (see Fig.~\ref{fig:infallfigure}). 

We conclude that it is unlikely that the observed high infall velocities can be explained by spherical infall. Additionally, we note that anisotropic accretion along the cavity walls is observed on slightly larger scales \citet{Cabedo:2021hw} and that B335 recently underwent a burst in luminosity \citep{Evans:2023aa} accompanied by an ejection event \citep{Bjerkeli:2019ip}. Thus, it would not be entirely surprising if asymmetric and time-variable accretion also propagated to smaller scales.




\subsection{Radiative transfer modeling}
\label{sec:radiativetransfer}
To constrain the nature of the observed \trettenco\ infall velocities, we used the 3D line radiative transfer code LIME \citep[][v1.4.3]{Brinch:2010rm}. LIME does not put any constraints on the geometry and complexity of the models and different components can be formulated analytically. The density, temperature, abundance and velocity structures are specified and LIME generates an unstructured Delaunay grid from the input model. Here, 50 000 grid points are used and the sampling probability is weighted by the density structure so that the grid is finer towards the centre of the model. The \trettenco\ collisional rate coefficients data file from the LAMBDA\footnote{http://home.strw.leidenuniv.nl/~$\tilde{ }$~moldata/} database, which is derived from \citet{Yang:2010vn}, was used. After convergence is reached, the model is ray-traced with velocity resolution similar to the observations. In all models, the micro-turbulent velocity is set to 200 m s$^{-1}$ and the spatial resolution of the output is set to 0.01\asec.

\begin{figure}[h]
   \centering
    \includegraphics[width=0.9\hsize]{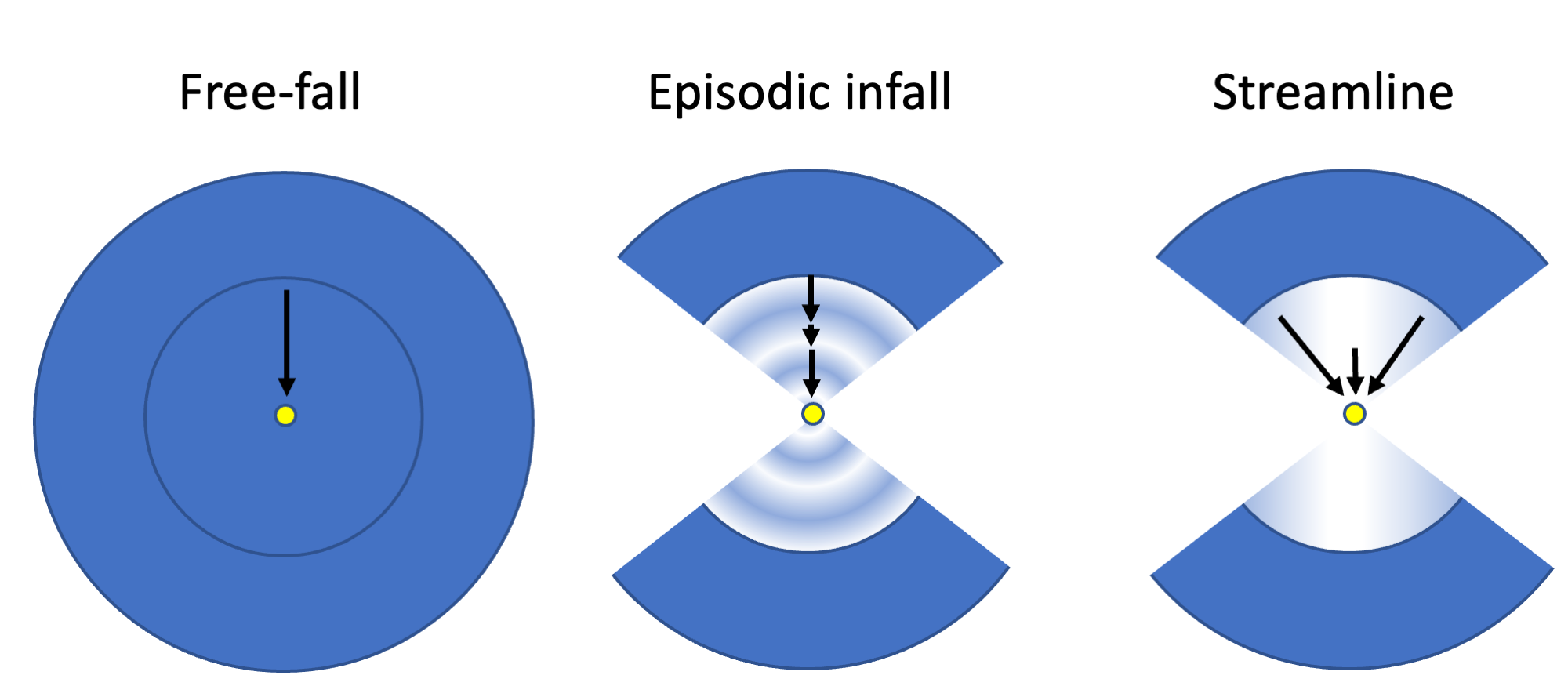} 
    \caption{Illustration depicting the three distinct models under consideration. The yellow dot marks the protostar, the arrows denote the direction and variation in velocity, while the shading denotes deviations from spherical free-fall.}
  \label{fig:modelcartoon}
\end{figure}
We consider three different types of models (see Fig.~\ref{fig:modelcartoon}) to investigate various scenarios that can explain the observed velocities. They are: i) spherical free-fall towards a central protostar, ii) episodic infall in the regions where outflow motions are not present, iii) free-fall towards a central protostar, but where the velocity is higher along the cavity walls, to mimic the possible occurrence of streamers in B335. Two cases for each model are considered, one where the mass of the central protostar is 0.1\msun\ and one where the mass is 0.4\msun,  making for a total of six cases. A protostellar mass of 0.4\msun is considerably higher than what was previously estimated in \citet{Yen:2015vf}, but allows us to investigate to what degree the protostellar mass affects the infall velocity morphology on slightly larger scales. Note that no disk is included in any of the models. Leaving out this component will affect the computed line profiles close to the protostar, in particular since its rotation will not be taken into account. It will also affect the line profile shapes particularly towards red-shifted velocities since the presence of a disk increases the optical depth towards certain positions. This choice will not, however, affect the infall velocities and line profiles on larger scales ($>$ 10 au) where we focus our investigation, since any disk, if present, is less $<$ 10 au.

\begin{figure}[ht]
   \centering
   \begin{tabular}{c c}
    \hspace{-0.4cm}\includegraphics[width=0.55\hsize,trim={160 0 60 100},clip]{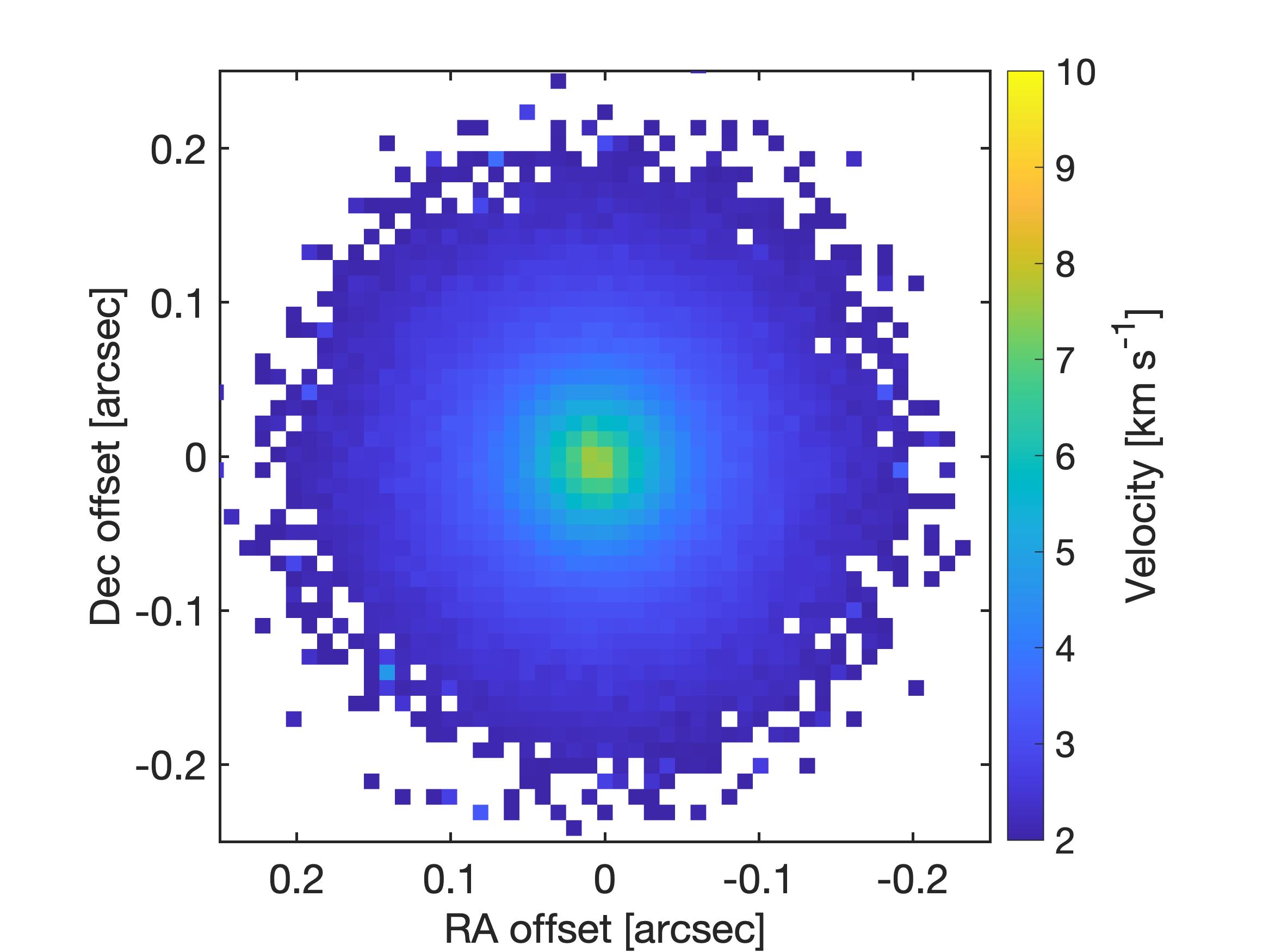}
    \put(-80,20){\makebox(0,0){\colorbox{white}{\fontsize{5}{10}\selectfont\textcolor{black}{\sffamily Free-fall, \textit{M} = \textrm{\sffamily 0.1~\textit{M}$_{\sun}$}   }}}}  
    & \hspace{-0.6cm}\includegraphics[width=0.55\hsize,trim={160 0 60 100},clip]{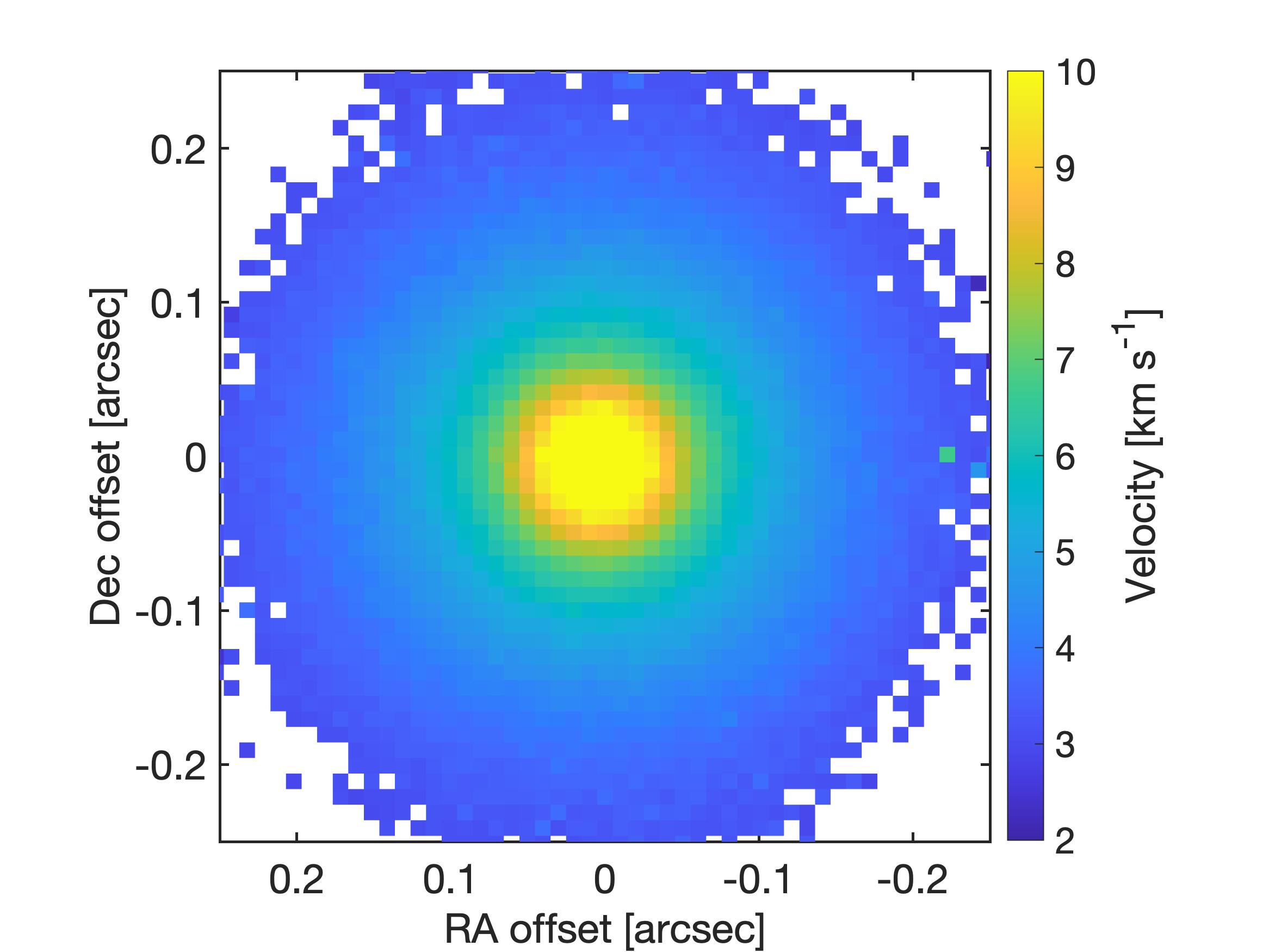}\put(-80,20){\makebox(0,0){\colorbox{white}{\fontsize{5}{10}\selectfont\textcolor{black}{\sffamily Free-fall, \textit{M} = \textrm{\sffamily 0.4}~\textit{M}$_{\sun}$}   }}} \\
    \hspace{-0.4cm}\hspace{0cm}\includegraphics[width=0.55\hsize,trim={160 0 60 100},clip]{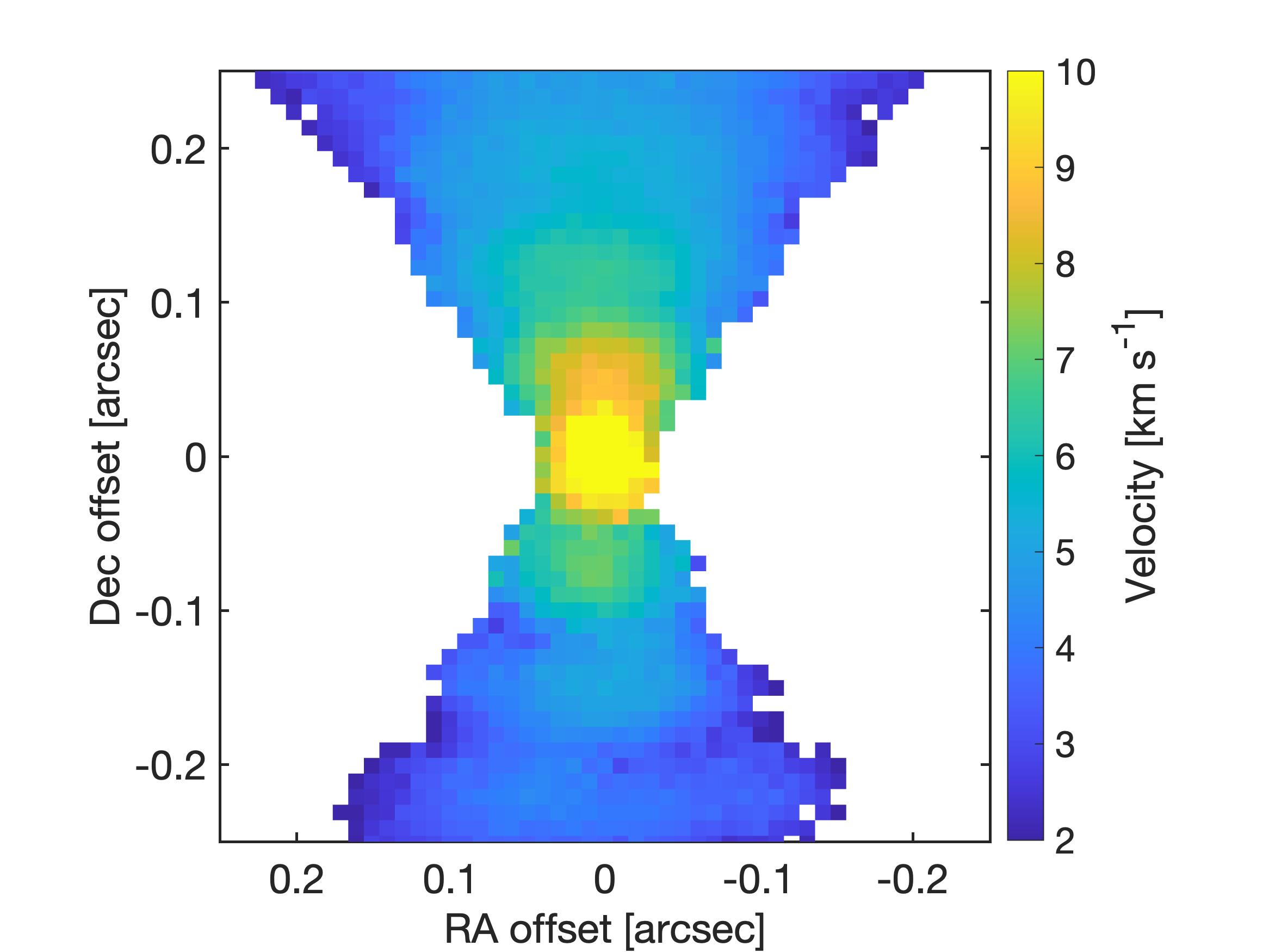}\put(-80,20){\makebox(0,0){\colorbox{white}{\fontsize{5}{10}\selectfont\textcolor{black}{\sffamily Episodic infall, \textit{M} = \textrm{\sffamily 0.1}~\textit{M}$_{\sun}$}   }}} & \hspace{-0.6cm}\includegraphics[width=0.55\hsize,trim={160 0 60 100},clip]{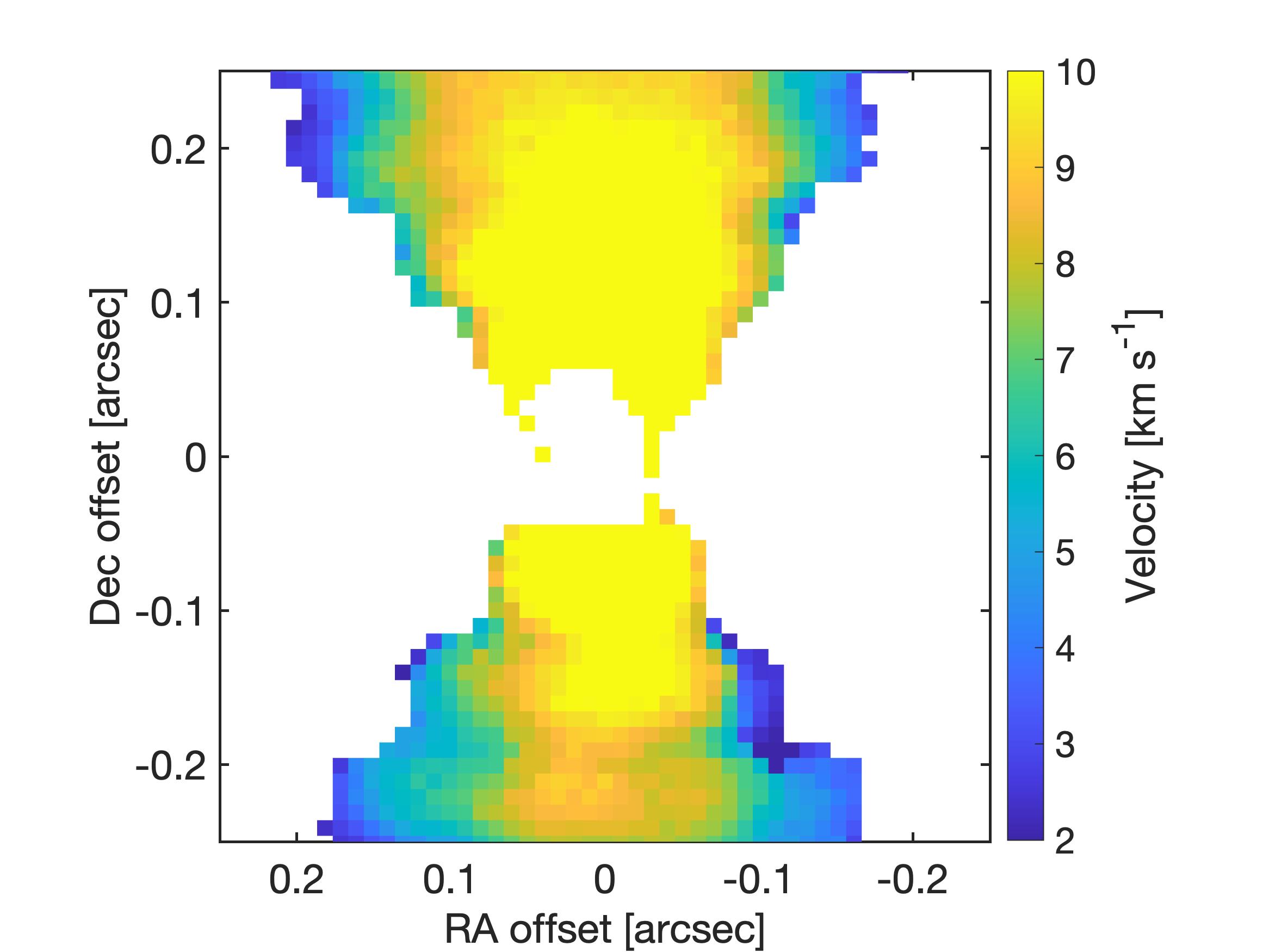}\put(-80,20){\makebox(0,0){\colorbox{white}{\fontsize{5}{10}\selectfont\textcolor{black}{\sffamily Episodic infall, \textit{M} = \textrm{\sffamily 0.4}~\textit{M}$_{\sun}$}   }}} \\
    \hspace{-0.4cm}\hspace{0cm}\includegraphics[width=0.55\hsize,trim={160 0 60 100},clip]{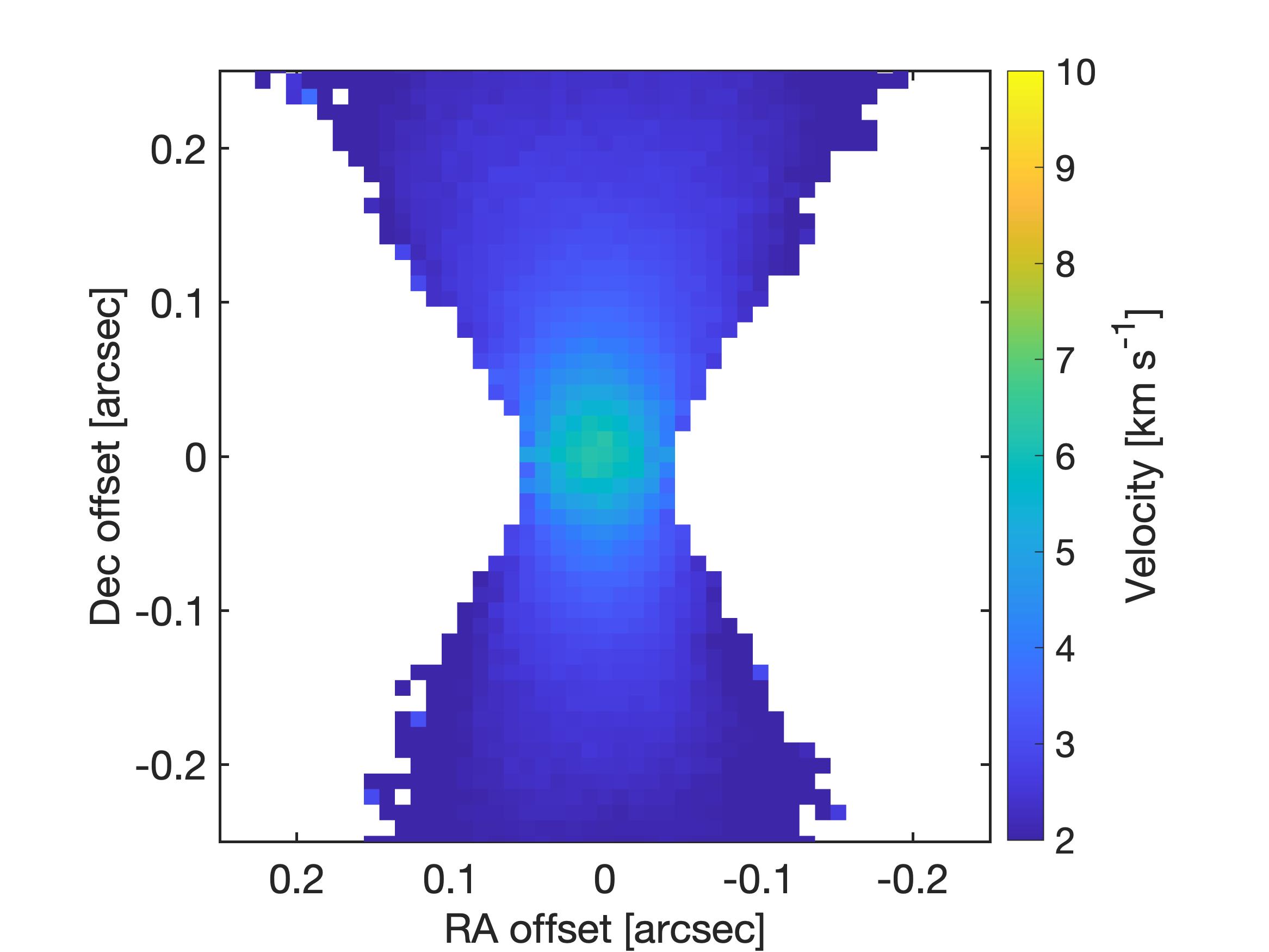}\put(-80,20){\makebox(0,0){\colorbox{white}{\fontsize{5}{10}\selectfont\textcolor{black}{\sffamily Streamline, \textit{M} = \textrm{\sffamily 0.1}~\textit{M}$_{\sun}$}   }}} & \hspace{-0.55cm}\includegraphics[width=0.55\hsize,trim={160 0 60 100},clip]{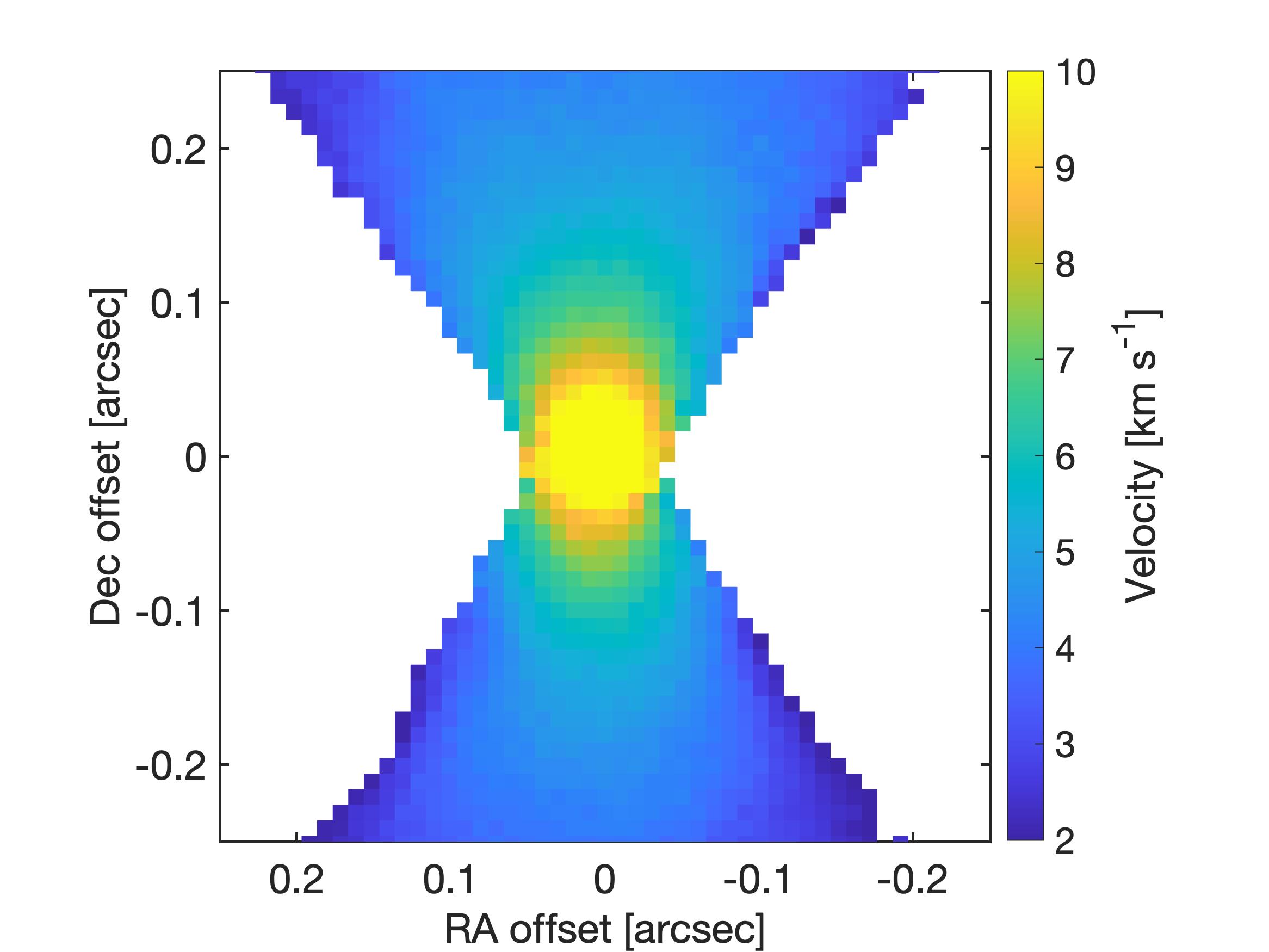}\put(-80,20){\makebox(0,0){\colorbox{white}{\fontsize{5}{10}\selectfont\textcolor{black}{\sffamily Streamline, \textit{M} = \textrm{\sffamily 0.4}~\textit{M}$_{\sun}$}   }}} \\
    \end{tabular}
   \caption{Infall velocity field LIME models for B335 for two different protostellar masses (0.1\msun\ in the left column and 0.4\msun\ in the right column). The models were post-processed with \texttt{simalma} and all plots have been convolved to a common resolution of 0.035\asec. Only positions where the coefficient of determination is higher than 0.25 are included and due to our relatively simple description for the envelope, our method to extract the infall velocity field does not perform sufficiently well in the regions where infall is not present (hence the white regions).}
   \label{fig:limemodels}
\end{figure}

\begin{figure}[ht!]
   \centering
   \begin{tabular}{c c}
   \hspace{-0.3cm}\includegraphics[width=0.6\hsize,trim={160 0 60 100},clip]{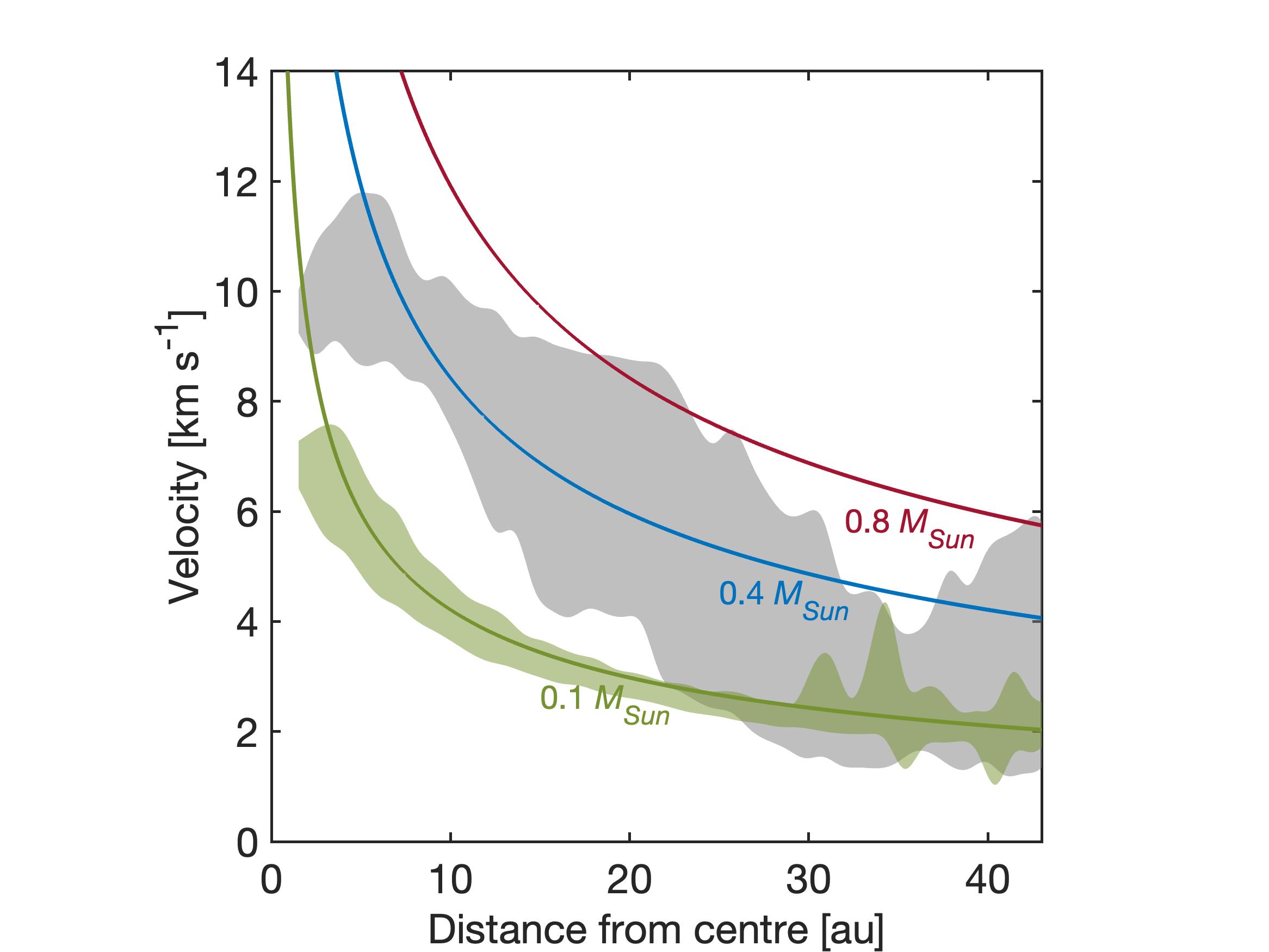}\put(-80,20){\makebox(0,0){\fontsize{5}{10}\selectfont\textcolor{black}{\sffamily Free-fall, \textit{M} = \textrm{\sffamily 0.1}~{\textit{M}$_{\odot}$}\xspace}}} & \hspace{-1.3cm}\includegraphics[width=0.6\hsize,trim={160 0 60 100},clip]{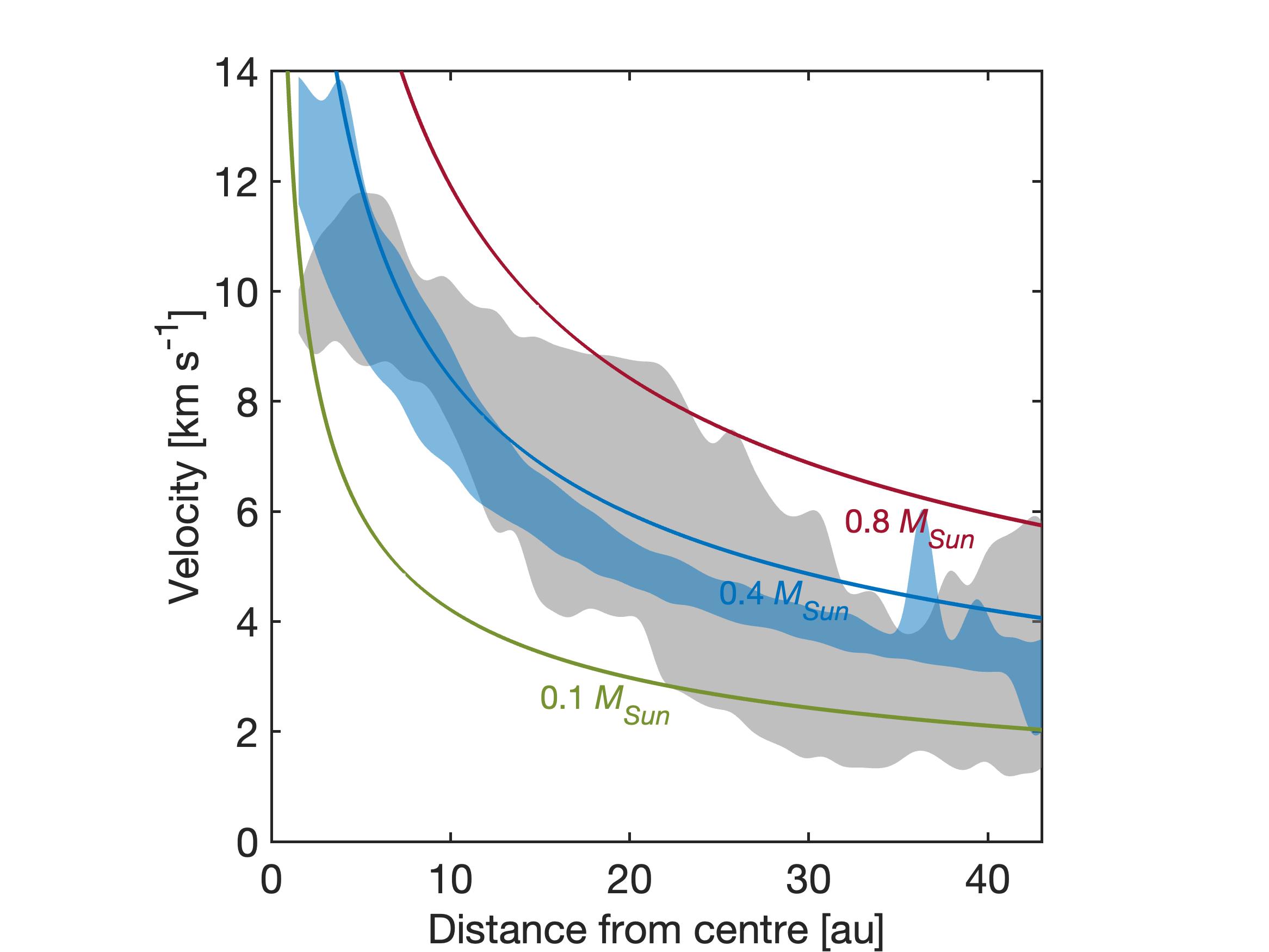}\put(-80,20){\makebox(0,0){\fontsize{5}{10}\selectfont\textcolor{black}{\sffamily Free-fall, \textit{M} = \textrm{\sffamily 0.4}~{\textit{M}$_{\odot}$}\xspace}}} \\
    \hspace{-0.3cm}\includegraphics[width=0.6\hsize,trim={160 0 60 100},clip]{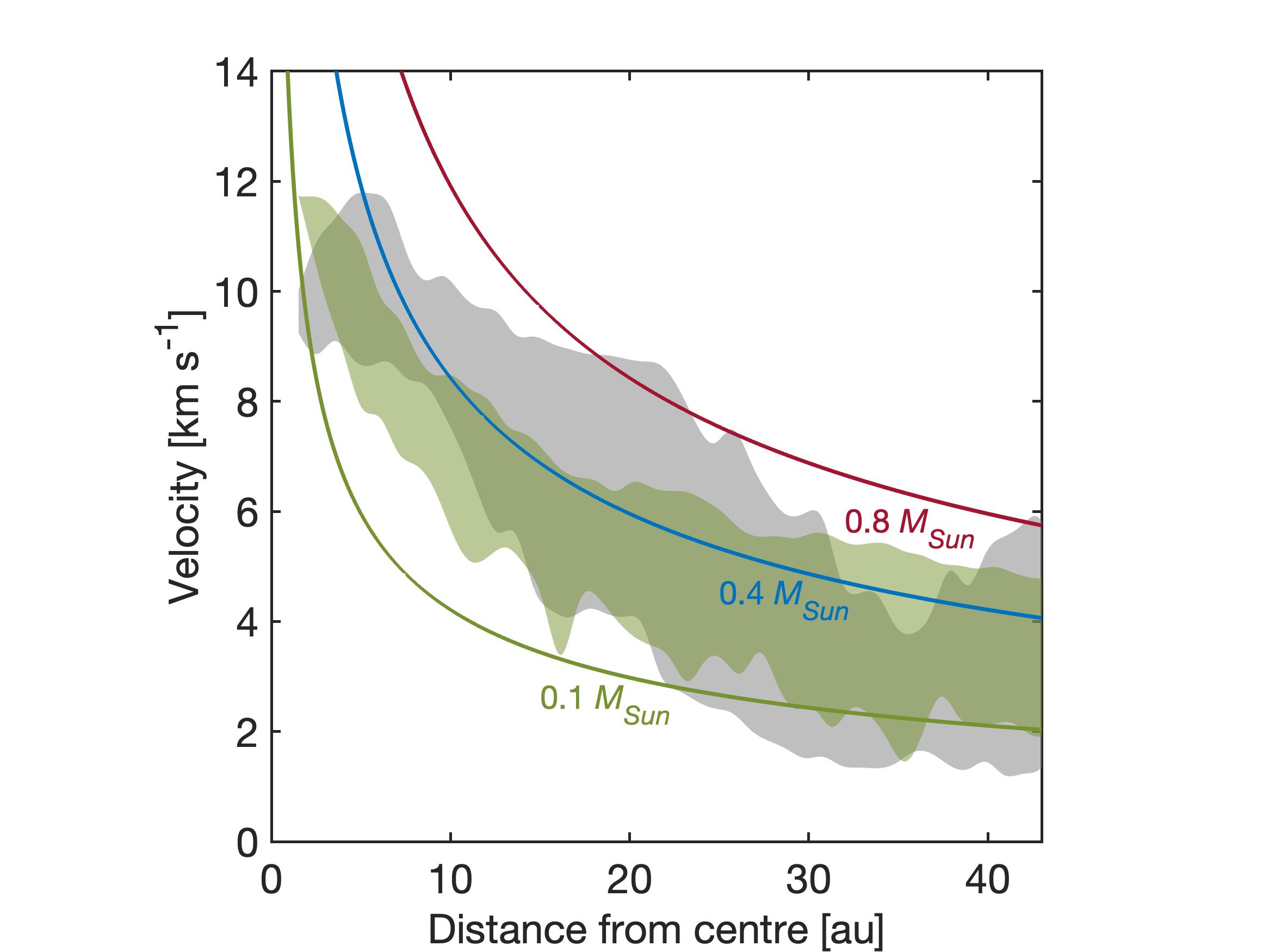}\put(-80,20){\makebox(0,0){\fontsize{5}{10}\selectfont\textcolor{black}{\sffamily Episodic infall, \textit{M} = \textrm{\sffamily 0.1}~{\textit{M}$_{\odot}$}\xspace}}} & \hspace{-1.3cm}\includegraphics[width=0.6\hsize,trim={160 0 60 100},clip]{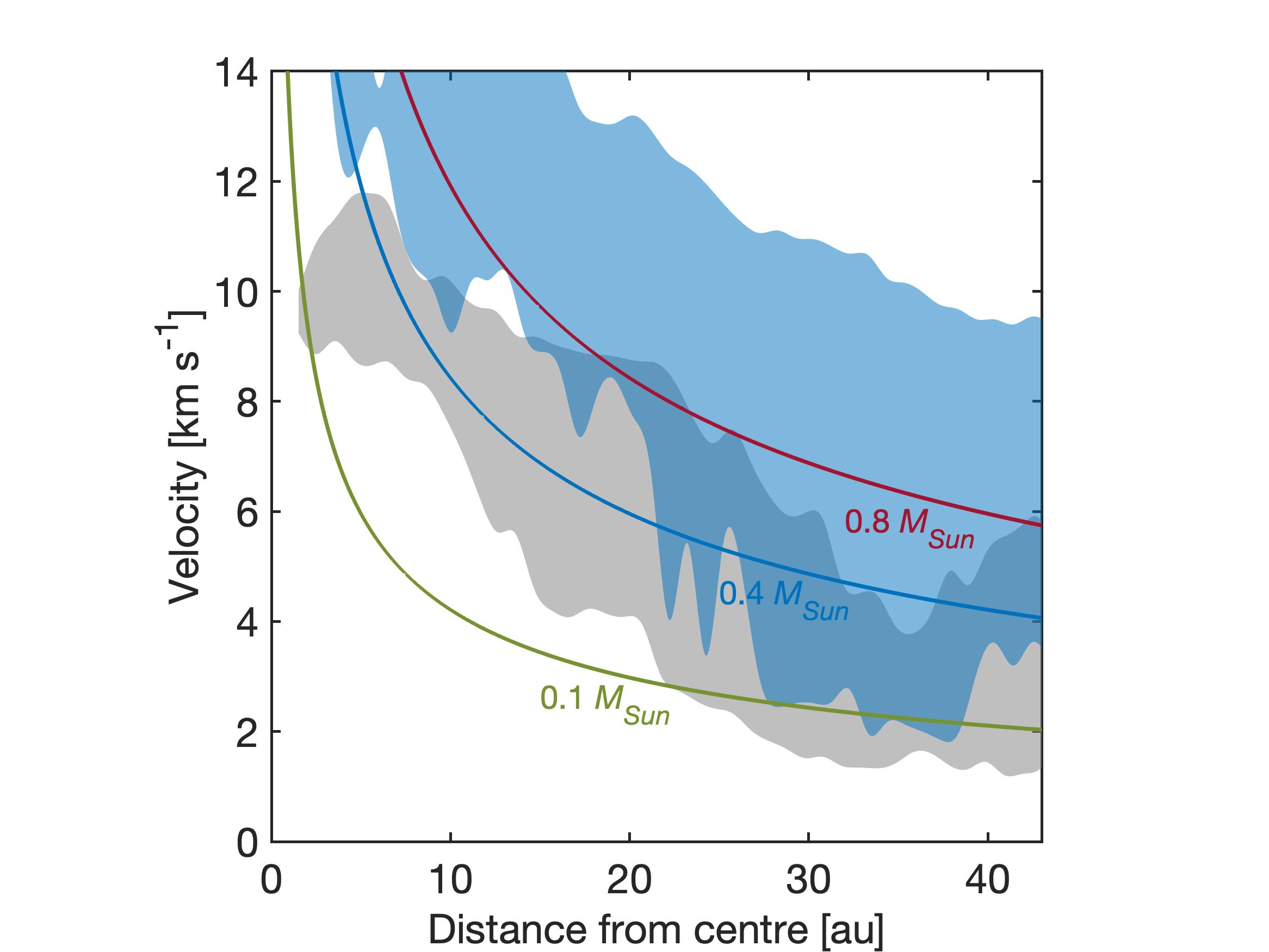}\put(-80,20){\makebox(0,0){\fontsize{5}{10}\selectfont\textcolor{black}{\sffamily Episodic infall, \textit{M} = \textrm{\sffamily 0.4}~{\textit{M}$_{\odot}$}\xspace}}} \\
    \hspace{-0.3cm}\includegraphics[width=0.6\hsize,trim={160 0 60 100},clip]{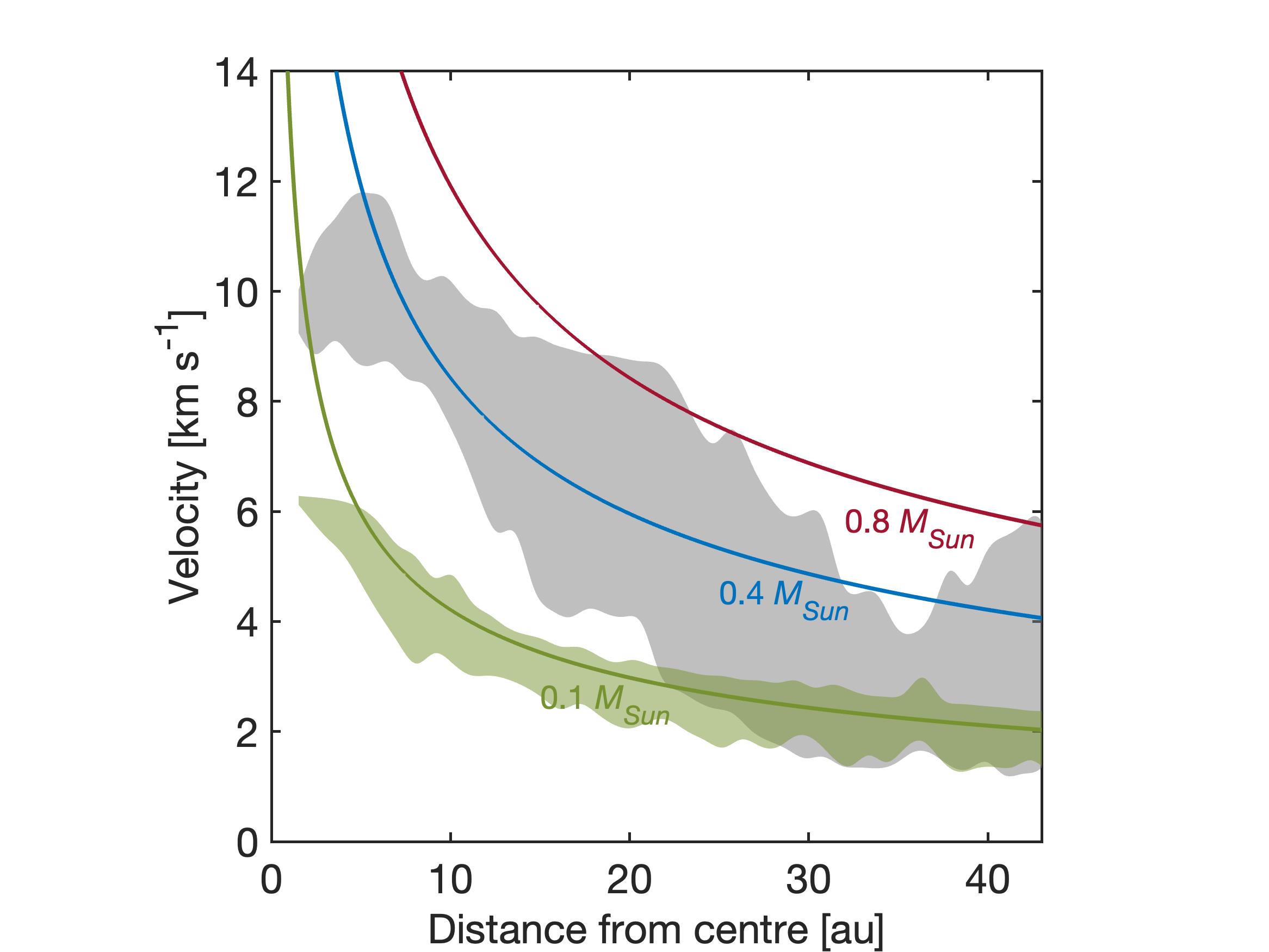}\put(-80,20){\makebox(0,0){\fontsize{5}{10}\selectfont\textcolor{black}{\sffamily Streamline, \textit{M} = \textrm{\sffamily 0.1}~{\textit{M}$_{\odot}$}\xspace}}} & \hspace{-1.3cm}\includegraphics[width=0.6\hsize,trim={160 0 60 100},clip]{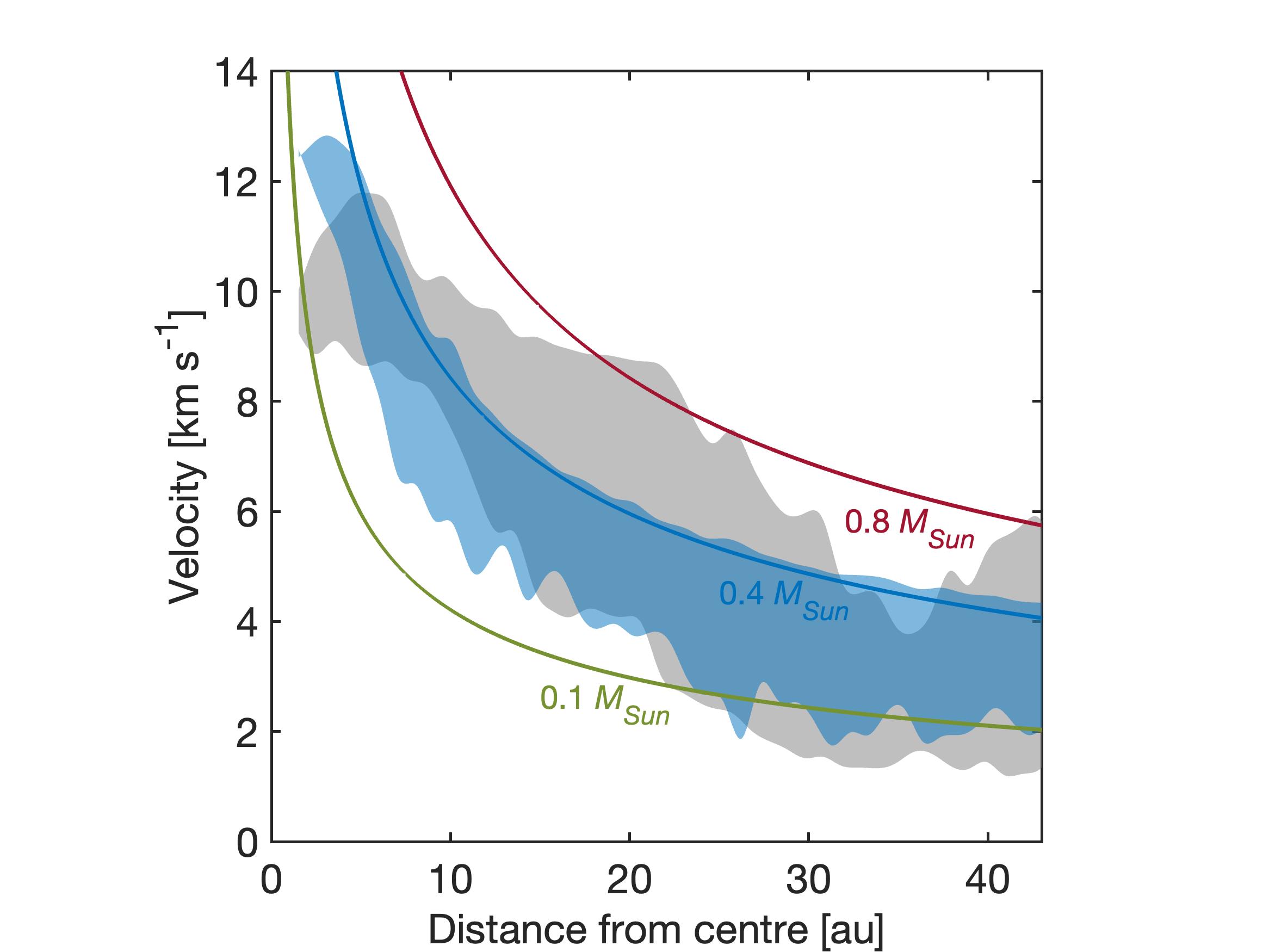}\put(-80,20){\makebox(0,0){\fontsize{5}{10}\selectfont\textcolor{black}{\sffamily Streamline, \textit{M} = \textrm{\sffamily 0.4}~{\textit{M}$_{\odot}$}\xspace}}} \\
    \end{tabular}
   \caption{Lower-limit on the infall velocity as a function of distance. Gray regions show the measured velocities from observations (same as dots in Fig.~\ref{fig:infallfigure}) while green (0.1~\msun) and blue (0.4~\msun) regions show the measured velocities from the models. The theoretical spherically-symmetric free-fall curves (Fig.~\ref{fig:infallfigure}) are included for comparison. Note that the spread on measured velocities increases with distance due to lower signal-to-noise ratio. In addition, Gaussian fitting is not possible for the few spectra very close to the protostar (hence no velocity estimates in that region). Furthermore, the number of points used to create the colored colored regions vary with distance from the centre (c.f., Fig.~\ref{fig:infallfigure}). For example, the ``bumps'' at large distances in the models are caused by a relatively small number of line-width measurements.}
   \label{fig:infallfigure2}
\end{figure}

When defining the physical conditions in the models, we choose all parameters to be as close as possible to what has previously been published. For the first model, the density follows a power-law profile \citep{Kristensen:2012kx} out to 6600 au distance from the protostar:
\begin{equation}
    n_{\rm{H_2}} = 10^9 \left( \frac{(164.5/250)\, r}{9.8~\rm{au}} \right)^{-1.4} \rm{cm}^{-3}.
    \label{eq:densityspherical}
\end{equation}
We set the abundance of \trettenco\ to a constant value of 1.7\texpo{-6} with respect to \htva\ \citep{Frerking:1987mu}. A surrounding envelope with a density of \expo{9}~\cmthree\ is included and extends from 6600 to 10 000 au. This density is unrealistically high, but chosen so that the envelope produces absorption in the line profiles near the systemic velocity that corresponds reasonably well with the observations. The velocity in the envelope ($\mathbf{r}>$6600~au) is set to zero. Within the infall region, the gas is assumed to be in free-fall (see eq.~\ref{eq:maxinfall}) and the temperature follows a power-law:
\begin{equation}
    T = 30 \left( \frac{(164.5/100)\,r}{100\,\rm{au}}  \right)^{-0.4} \rm{K}.
    \label{eq:temperature}
\end{equation}
This profile is taken from \citet{Yen:2015vf}, and not in perfect agreement with the observed temperatures in the inner region as derived from the CH$_3$OH emission \citep{Bjerkeli:2019ip}. This is, however, of less importance here since we focus on the line profiles on larger scales.

To account for the response of the interferometer, we employed \texttt{simalma} in CASA to post-process our model emission maps using the ALMA Cycle 9 configuration and array set up corresponding to the observations in this study. After applying the same process as we did for the actual observations to extract the lower limit on the infall velocity (see Fig.~\ref{fig:infallvelocity}), we obtain the infall velocity maps presented in Fig.\ \ref{fig:limemodels}.

In the second (``episodic'') model, the infall velocity is allowed to vary in the radial direction. In addition, we assume no infall takes place in the region where the outflow is present (in the east-west directions), and the temperature and density in this region is set to be constant ($T$ = 100~K, \nhtva~=~\expo{4}~\cmthree, respectively). As a result, the infall region is now effectively two cones with 45\adeg~opening angles to the north and south (middle column of Fig.\ \ref{fig:limemodels}). The velocity is varied relative to the free-fall velocity using a cosine function in radius, where the amplitude is allowed to change by 50\% and where the period is 20 au. The mean velocity in the region is, in addition, increased by 40\% relative to the free-fall velocity to account for the fact that infall only occurs in a fraction of the total region (i.e.\ to make up for the region where the outflow is present). The standard spherical envelope model from \citet{Kristensen:2012kx}) does not account for this modified geometry, and it is unknown what effect our modifications will have on the large-scale envelope. We do find, however, that we need to vary the density in the envelope by a factor of up to five in order to match the observed line profiles. 
The temperature follows the same power-law profile as in the first model, but we acknowledge that it may also vary. It is worth noting that the variability in this particular model corresponds to a time-scale of approximately 20 years. This timescale is chosen to match the observed infall velocities, but is also in line with the studies of  \citet{Bjerkeli:2019ip} and \citet{Evans:2023aa}, where the episodic ejection event that took place in 2015 can be connected to a luminosity burst that peaked in 2019. The resulting infall velocities are presented in the middle row of Fig.~\ref{fig:limemodels}.

In the third model, we include the outflow cavity as above, but instead vary the velocities and densities in the azimuthal direction (i.e.\ angle with respect to the north-south axis). The absolute value of the velocity is varied with a cosine where it is increased by a factor of two along the cavity walls and goes to zero along north and south directions. Densities and velocities are also adjusted to keep the mass and mean velocities constant, while the temperature continues to follow the power-law profile from eq.\ \ref{eq:temperature}. The resulting infall velocites are presented in the bottom row of Fig.\ \ref{fig:limemodels}.

It should be noted that we have found
it difficult to fully account for the fluxes observed close to \vlsr, while simultaneously fitting the line flux in the wings (see Fig.~\ref{fig:LIMElineprofiles}). For the cases presented here, this amounts to model fluxes near \vlsr\ that are up to 50\% lower than the observations. The flux near \vlsr\ is particularly low in the episodic infall scenario, while fluxes roughly correspond to observations in the streamline scenario. The reason for this is not entirely understood (but also keeping in mind that some of the observed emission could be filtered out by the interferometer), but could be due to low-velocity emission that is not captured in our simplistic description of the envelope, and/or due to optical depth effects. That said, we want to emphasize that reproducing the exact line profiles is not the main purpose of this study. Instead, we are aiming to understand which scenarios can match the observed line profile widths. In fact, while the line profile width is determined by the maximum infall velocity at each position, a complex interplay between several different parameters (e.g., density, temperature, molecular abundance) determines the line profile shapes.

To test how well the derived lower limits on the infall velocity (Fig.~\ref{fig:limemodels}) correspond with real line widths, we also extract limits from the models before employing \texttt{simalma}, by taking the maximum extent of the line profiles at each position. This recovers the results of the simulated models, demonstrating that our measurement technique is robust on the scales studied here. How well the derived velocities correspond with the actual infall velocities can also be asserted since we know the true infall velocities in the models. In Fig.~\ref{fig:infallfigure2}, we present how the derived infall velocities from the different models compare with observations and theoretical free-fall scenarios. Although we estimate that infall velocities exceed the theoretical values in approximately 10\% of positions in the 0.1~\msun\ free-fall scenario and 1\% of positions in the 0.4~\msun\ free-fall scenario, we attribute these discrepancies to a few instances of bad fitting. It is important to note that these discrepancies do not significantly affect the overall conclusions and for most of the positions in the map, estimated infall velocities are below the theoretical value. This is consistent with the assumption that the width of the line provide a lower limit to the true infall velocity in each position.


While our models are simplistic, they allow us to test whether a higher protostellar mass can account for the infall velocities observed at larger distances from the protostar in the free-fall scenario. However, both low and high protostellar mass scenarios in the spherical free-fall models produce velocities that are not consistent with observations. In particular, a low protostellar mass produces too low velocities at all radii while a high protostellar mass produce too high velocities in the central region. We also test whether episodic infall and/or enhanced infall along the cavity walls can account for the observed high velocities. Infall velocity maps for the six different cases (and three different models) are presented in Fig.~\ref{fig:limemodels}. Here, the images have been convolved to a common resolution of 0.035\asec.

From the different cases, a few important conclusions can be drawn. It is not possible to reproduce the high infall velocities observed on large scales with a pure free-fall scenario. Even in the case where the protostellar mass is high (0.4\msun), the velocities do not match the observed (lower limit) 8~\kmpers\ infall velocities on 20~au scales. To obtain such high velocities, a central protostellar mass of $M$~=~0.8\msun\ would be required in the free-fall scenario. That would, however, significantly increase the infall velocity close to the protostar to \about 20~\kmpers, which is inconsistent with the observations. In fact, no matter how the protostellar mass is varied in the free-fall scenario, the velocity profile with distance from the protostar cannot be reproduced. Hence, we conclude that a pure free-fall scenario does not provide a good match to the data at hand.

The streamline model provides a better fit to the observations, but only in the case where the protostellar mass is 0.4\msun (Figs.~\ref{fig:limemodels} \& \ref{fig:infallfigure2}). The scenario where velocities and densities are allowed to vary episodically in the radial direction also provides a better fit to the observations, but in the case where the protostellar mass is 0.1\msun~(Figs.~\ref{fig:limemodels} \& \ref{fig:infallfigure2}). For both the aforementioned models, high-velocity gas is present on scales that are larger than what is seen in the observations, and varying parameters suggests that the temperature profile is a defining parameter in this instance. A steeper slope for the temperature would provide a better match to the observed velocity field, but would on the other hand reduce the intensity of the \trettenco line and not match the temperature in the central region derived from the methanol emission. We note that matching the high velocities observed at $\sim$20 au distance is particularly challenging for any of our models. The fit may be improved with a much more complex, fine-tuned velocity drop-off with distance, but such a model is outside the purpose of this study.

While no exact match to the observed line profiles is obtained in either of these two scenarios (Fig.~\ref{fig:LIMElineprofiles}), it is, however, clear that models where the infall velocities is allowed to deviate from free-fall velocities, can both better reproduce the relatively high infall velocities on larger scales while simultaneously keeping velocities sufficiently low close to the protostar (Fig.~\ref{fig:limemodels}). At the same time, they show a spread in velocity at each radii, that is in better agreement with observations (Fig.~\ref{fig:infallfigure2}). Our modelling therefore suggests that a characteristic variation in the infall velocity on small scales is present in B335. We cannot, however, from these observations alone put constraints on the cause of such variations. High temperatures and Kelvin-Helmholtz instabilities in the vicinity of the cavity wall could perhaps explain high infall velocities at an angle towards the disk. Similarly, if matter is falling in towards the disk in "clumps" from the north and south, it is plausible that density or pressure fluctuations will alter the infall velocity. To better understand infall in B335, and in particular to confirm that the infall fluctuates in time, deep follow-up observations with ALMA in its longest baselines configurations and at different epochs will be required. Finally, if the current data indeed indicate that infall is occurring episodically and with a variation similar to the one we included in our modelling, it would be intriguing to monitor B335 over time for additional outbursts.

\section{Summary and conclusions}
\label{sec:conclude}
In this paper, we have presented observations of \trettenco (2--1) and dust continuum emission in B335. Emission line profiles were analysed and compared with 3D radiative transfer models. Our main findings are:
\begin{itemize}
    \item The peak of the 1.3 mm continuum emission towards B335 shows an elongated structure, 10 by 7 au in size, with a position angle 5\adeg\ east of north. This is consistent with a scenario where a disk has started to form.
    \item \trettenco\ emission towards B335 shows line profiles where the blue-shifted emission component is enhanced with respect to the red-shifted component. We interpret the observations of \trettenco\ as being material falling inwards towards the protostar from the north and south. Velocities are not following a $r^{0.5}$ profile as would be expected in the case of spherically symmetric infall.
    \item The relatively high velocities are not easily reconcilable with a scenario where disk growth is actively prohibited via magnetic braking.
    \item Since there is no reason to believe that the infall velocity should exceed the free-fall velocity at any given location, this suggests asymmetric, perhaps time-variable infall. A scenario where the central mass is higher than 0.4\msun\ could, in principle, account for slightly higher velocity emission on 20~au scales but at the same time would not account for the velocity decrease with distance and the relatively low velocities at small distances from the protostar. From radiative transfer modeling using toy models (free-fall, episodic infall, infall along a streamline) we favour a scenario where the infall velocity is varying either in the radial direction or with angle from the north-south axis.
\end{itemize}

\begin{acknowledgements}
The authors would like to thank the anonymous referee for a thorough review that helped improve the quality of the paper. In addition, the authors would like to thank Neal Evans, Yao-Lun Yang, John Tobin, Johanna Matero and Hannah Shoemaker for interesting discussions. This paper makes use of the following ALMA projects data: 2013.1.00879.S and 2017.1.00288.S. ALMA is a partnership of ESO (representing its member states, NSF (USA) and NINS (Japan), together with NRC (Canada), MOST and ASIAA (Taiwan), and KASI (Republic of Korea), in cooperation with the Republic of Chile. The Joint ALMA Observatory is operated by ESO, AUI/NRAO and NAOJ. 
We acknowledge support from the Nordic ALMA Regional Centre (ARC) node based
at Onsala Space Observatory. The Nordic ARC node is funded through Swedish Research Council grant No 2017-00648. PB acknowledges the support of the Swedish Research Council (VR) through contract 2017-04924. JPR and ZYL are supported in part by NSF grant AST-1910106 and NASA grant 80NSSC20K0533. JPR would like to further acknowledge the support of the Virginia Initiative on Cosmic Origins (VICO). HC's research group is supported by an OPUS research grant (2021/41/B/ST9/03958) from the Narodowe Centrum Nauki. The research of JKJ is supported the Independent Research Fund Denmark (grant No. DFF0135-00123B).The research of LEK is supported by a research grant (19127) from VILLUM FONDEN. DH is supported by Center for Informatics and Computation in Astronomy (CICA) grant and grant number 110J0353I9 from the Ministry of Education of Taiwan. DH also acknowledges support from the National Science and Technology Council of Taiwan through grant number 111B3005191. The astrophysics HPC facility at the University of Copenhagen, supported by a research grant (VKR023406) from VILLUM FONDEN, was used for carrying out the radiative transfer modeling.
\end{acknowledgements}
\newpage
\bibliographystyle{aa}
\bibliography{papers.bib}

\begin{thebibliography}{44}
\expandafter\ifx\csname natexlab\endcsname\relax\def\natexlab#1{#1}\fi

\bibitem[{{Allen} {et~al.}(2003){Allen}, {Li}, \& {Shu}}]{Allen:2003mf}
{Allen}, A., {Li}, Z.-Y., \& {Shu}, F.~H. 2003, \apj, 599, 363

\bibitem[{{Alves} {et~al.}(2020){Alves}, {Cleeves}, {Girart}, {Zhu}, {Franco},
  {Zurlo}, \& {Caselli}}]{Alves:2020vo}
{Alves}, F.~O., {Cleeves}, L.~I., {Girart}, J.~M., {et~al.} 2020, \apjl, 904,
  L6

\bibitem[{{Andr{\'e}} {et~al.}(1990){Andr{\'e}}, {Martin-Pintado}, {Despois},
  \& {Montmerle}}]{Andre:1990fk}
{Andr{\'e}}, P., {Martin-Pintado}, J., {Despois}, D., \& {Montmerle}, T. 1990,
  \aap, 236, 180

\bibitem[{{Audard} {et~al.}(2014){Audard}, {{\'A}brah{\'a}m}, {Dunham},
  {Green}, {Grosso}, {Hamaguchi}, {Kastner}, {K{\'o}sp{\'a}l}, {Lodato},
  {Romanova}, {Skinner}, {Vorobyov}, \& {Zhu}}]{Audard:2014xy}
{Audard}, M., {{\'A}brah{\'a}m}, P., {Dunham}, M.~M., {et~al.} 2014, Protostars
  and Planets VI, 387

\bibitem[{{Bjerkeli} {et~al.}(2012){Bjerkeli}, {Liseau}, {Larsson}, {Rydbeck},
  {Nisini}, {Tafalla}, {Antoniucci}, {Benedettini}, {Bergman}, {Cabrit},
  {Giannini}, {Melnick}, {Neufeld}, {Santangelo}, \& {van
  Dishoeck}}]{Bjerkeli:2012fk}
{Bjerkeli}, P., {Liseau}, R., {Larsson}, B., {et~al.} 2012, \aap, 546, A29

\bibitem[{{Bjerkeli} {et~al.}(2011){Bjerkeli}, {Liseau}, {Nisini}, {Tafalla},
  {Benedettini}, {Bergman}, {Dionatos}, {Giannini}, {Herczeg}, {Justtanont},
  {Larsson}, {McOey}, {Olberg}, \& {Olofsson}}]{Bjerkeli:2011qy}
{Bjerkeli}, P., {Liseau}, R., {Nisini}, B., {et~al.} 2011, \aap, 533, A80+

\bibitem[{{Bjerkeli} {et~al.}(2019){Bjerkeli}, {Ramsey}, {Harsono}, {Calcutt},
  {Kristensen}, {van der Wiel}, {J{\o}rgensen}, {Muller}, \&
  {Persson}}]{Bjerkeli:2019ip}
{Bjerkeli}, P., {Ramsey}, J.~P., {Harsono}, D., {et~al.} 2019, \aap, 631, A64

\bibitem[{{Blandford} \& {Payne}(1982)}]{Blandford:1982fj}
{Blandford}, R.~D. \& {Payne}, D.~G. 1982, \mnras, 199, 883

\bibitem[{{Brinch} \& {Hogerheijde}(2010)}]{Brinch:2010rm}
{Brinch}, C. \& {Hogerheijde}, M.~R. 2010, \aap, 523, A25

\bibitem[{{Cabedo} {et~al.}(2021){Cabedo}, {Maury}, {Girart}, \&
  {Padovani}}]{Cabedo:2021hw}
{Cabedo}, V., {Maury}, A., {Girart}, J.~M., \& {Padovani}, M. 2021, \aap, 653,
  A166

\bibitem[{{Cabedo} {et~al.}(2023){Cabedo}, {Maury}, {Girart}, {Padovani},
  {Hennebelle}, {Houde}, \& {Zhang}}]{Cabedo:2023aa}
{Cabedo}, V., {Maury}, A., {Girart}, J.~M., {et~al.} 2023, \aap, 669, A90

\bibitem[{{Carney} {et~al.}(2016){Carney}, {Y{\i}ld{\i}z}, {Mottram}, {van
  Dishoeck}, {Ramchandani}, \& {J{\o}rgensen}}]{Carney:2016kf}
{Carney}, M.~T., {Y{\i}ld{\i}z}, U.~A., {Mottram}, J.~C., {et~al.} 2016, \aap,
  586, A44

\bibitem[{{Caselli} {et~al.}(2012){Caselli}, {Keto}, {Bergin}, {Tafalla},
  {Aikawa}, {Douglas}, {Pagani}, {Y{\'{\i}}ld{\'{\i}}z}, {van der Tak},
  {Walmsley}, {Codella}, {Nisini}, {Kristensen}, \& {van
  Dishoeck}}]{Caselli:2012rt}
{Caselli}, P., {Keto}, E., {Bergin}, E.~A., {et~al.} 2012, \apjl, 759, L37

\bibitem[{{Evans} {et~al.}(2023){Evans}, {Yang}, {Green}, {Zhao}, {Di
  Francesco}, {Lee}, {J{\o}rgensen}, {Choi}, {Myers}, \&
  {Mardones}}]{Evans:2023aa}
{Evans}, Neal~J., I., {Yang}, Y.-L., {Green}, J.~D., {et~al.} 2023, \apj, 943,
  90

\bibitem[{{Evans} {et~al.}(2015){Evans}, {Di Francesco}, {Lee}, {J{\o}rgensen},
  {Choi}, {Myers}, \& {Mardones}}]{Evans:2015qp}
{Evans}, II, N.~J., {Di Francesco}, J., {Lee}, J.-E., {et~al.} 2015, \apj, 814,
  22

\bibitem[{{Evans} {et~al.}(2005){Evans}, {Lee}, {Rawlings}, \&
  {Choi}}]{Evans:2005qy}
{Evans}, II, N.~J., {Lee}, J.-E., {Rawlings}, J.~M.~C., \& {Choi}, M. 2005,
  \apj, 626, 919

\bibitem[{{Frerking} {et~al.}(1987){Frerking}, {Langer}, \&
  {Wilson}}]{Frerking:1987mu}
{Frerking}, M.~A., {Langer}, W.~D., \& {Wilson}, R.~W. 1987, \apj, 313, 320

\bibitem[{{G{\aa}lfalk} \& {Olofsson}(2007)}]{Galfalk:2007lr}
{G{\aa}lfalk}, M. \& {Olofsson}, G. 2007, \aap, 475, 281

\bibitem[{{Galli} {et~al.}(2006){Galli}, {Lizano}, {Shu}, \&
  {Allen}}]{Galli:2006jr}
{Galli}, D., {Lizano}, S., {Shu}, F.~H., \& {Allen}, A. 2006, \apj, 647, 374

\bibitem[{{Galli} \& {Shu}(1993{\natexlab{a}})}]{Galli:1993oq}
{Galli}, D. \& {Shu}, F.~H. 1993{\natexlab{a}}, \apj, 417, 220

\bibitem[{{Galli} \& {Shu}(1993{\natexlab{b}})}]{Galli:1993qe}
{Galli}, D. \& {Shu}, F.~H. 1993{\natexlab{b}}, \apj, 417, 243

\bibitem[{{Garufi} {et~al.}(2022){Garufi}, {Podio}, {Codella}, {Segura-Cox},
  {Vander Donckt}, {Mercimek}, {Bacciotti}, {Fedele}, {Kasper}, {Pineda},
  {Humphreys}, \& {Testi}}]{Garufi:2022aa}
{Garufi}, A., {Podio}, L., {Codella}, C., {et~al.} 2022, \aap, 658, A104

\bibitem[{{Gravity Collaboration} {et~al.}(2020){Gravity Collaboration},
  {Garcia Lopez}, {Natta}, {Caratti o Garatti}, {Ray}, {Fedriani},
  {Koutoulaki}, {Klarmann}, {Perraut}, {Sanchez-Bermudez}, {Benisty},
  {Dougados}, {Labadie}, {Brandner}, {Garcia}, {Henning}, {Caselli}, {Duvert},
  {de Zeeuw}, {Grellmann}, {Abuter}, {Amorim}, {Baub{\"o}ck}, {Berger},
  {Bonnet}, {Buron}, {Cl{\'e}net}, {Coud{\'e} Du Foresto}, {de Wit}, {Eckart},
  {Eisenhauer}, {Filho}, {Gao}, {Garcia Dabo}, {Gendron}, {Genzel},
  {Gillessen}, {Habibi}, {Haubois}, {Haussmann}, {Hippler}, {Hubert},
  {Horrobin}, {Jimenez Rosales}, {Jocou}, {Kervella}, {Kolb}, {Lacour}, {Le
  Bouquin}, {L{\'e}na}, {Ott}, {Paumard}, {Perrin}, {Pfuhl}, {Ramirez}, {Rau},
  {Rousset}, {Scheithauer}, {Shangguan}, {Stadler}, {Straub}, {Straubmeier},
  {Sturm}, {van Dishoeck}, {Vincent}, {von Fellenberg}, {Widmann}, {Wieprecht},
  {Wiest}, {Wiezorrek}, {Woillez}, {Yazici}, \&
  {Zins}}]{Gravity-Collaboration:2020tu}
{Gravity Collaboration}, {Garcia Lopez}, R., {Natta}, A., {et~al.} 2020, \nat,
  584, 547

\bibitem[{{Heiderman} \& {Evans}(2015)}]{Heiderman:2015fm}
{Heiderman}, A. \& {Evans}, Neal~J., I. 2015, \apj, 806, 231

\bibitem[{{Kristensen} {et~al.}(2012){Kristensen}, {van Dishoeck}, {Bergin},
  {Visser}, {Y{\i}ld{\i}z}, {San Jose-Garcia}, {J{\o}rgensen}, {Herczeg},
  {Johnstone}, {Wampfler}, {Benz}, {Bruderer}, {Cabrit}, {Caselli}, {Doty},
  {Harsono}, {Herpin}, {Hogerheijde}, {Karska}, {van Kempen}, {Liseau},
  {Nisini}, {Tafalla}, {van der Tak}, \& {Wyrowski}}]{Kristensen:2012kx}
{Kristensen}, L.~E., {van Dishoeck}, E.~F., {Bergin}, E.~A., {et~al.} 2012,
  \aap, 542, A8

\bibitem[{{Lam} {et~al.}(2019){Lam}, {Li}, {Chen}, {Tomida}, \&
  {Zhao}}]{Lam:2019aa}
{Lam}, K.~H., {Li}, Z.-Y., {Chen}, C.-Y., {Tomida}, K., \& {Zhao}, B. 2019,
  \mnras, 489, 5326

\bibitem[{{Leung} \& {Brown}(1977)}]{Leung:1977gs}
{Leung}, C.~M. \& {Brown}, R.~L. 1977, \apjl, 214, L73

\bibitem[{{Li} {et~al.}(2014){Li}, {Banerjee}, {Pudritz}, {J{\o}rgensen},
  {Shang}, {Krasnopolsky}, \& {Maury}}]{Li:2014jm}
{Li}, Z.-Y., {Banerjee}, R., {Pudritz}, R.~E., {et~al.} 2014, Protostars and
  Planets VI, 173

\bibitem[{{Maury} {et~al.}(2018){Maury}, {Girart}, {Zhang}, {Hennebelle},
  {Keto}, {Rao}, {Lai}, {Ohashi}, \& {Galametz}}]{Maury:2018qf}
{Maury}, A.~J., {Girart}, J.~M., {Zhang}, Q., {et~al.} 2018, \mnras, 477, 2760

\bibitem[{{McMullin} {et~al.}(2007){McMullin}, {Waters}, {Schiebel}, {Young},
  \& {Golap}}]{McMullin:2007nr}
{McMullin}, J.~P., {Waters}, B., {Schiebel}, D., {Young}, W., \& {Golap}, K.
  2007, in Astronomical Society of the Pacific Conference Series, Vol. 376,
  Astronomical Data Analysis Software and Systems XVI, ed. R.~A. {Shaw},
  F.~{Hill}, \& D.~J. {Bell}, 127

\bibitem[{{Mottram} {et~al.}(2013){Mottram}, {van Dishoeck}, {Schmalzl},
  {Kristensen}, {Visser}, {Hogerheijde}, \& {Bruderer}}]{Mottram:2013qy}
{Mottram}, J.~C., {van Dishoeck}, E.~F., {Schmalzl}, M., {et~al.} 2013, \aap,
  558, A126

\bibitem[{{Pineda} {et~al.}(2020){Pineda}, {Segura-Cox}, {Caselli},
  {Cunningham}, {Zhao}, {Schmiedeke}, {Maureira}, \& {Neri}}]{Pineda:2020xm}
{Pineda}, J.~E., {Segura-Cox}, D., {Caselli}, P., {et~al.} 2020, Nature
  Astronomy, 4, 1158

\bibitem[{{Pudritz} \& {Norman}(1983)}]{Pudritz:1983fv}
{Pudritz}, R.~E. \& {Norman}, C.~A. 1983, \apj, 274, 677

\bibitem[{{Segura-Cox} {et~al.}(2020){Segura-Cox}, {Schmiedeke}, {Pineda},
  {Stephens}, {Fern{\'a}ndez-L{\'o}pez}, {Looney}, {Caselli}, {Li}, {Mundy},
  {Kwon}, \& {Harris}}]{Segura-Cox:2020wc}
{Segura-Cox}, D.~M., {Schmiedeke}, A., {Pineda}, J.~E., {et~al.} 2020, \nat,
  586, 228

\bibitem[{{Shu} {et~al.}(1994){Shu}, {Najita}, {Ostriker}, {Wilkin}, {Ruden},
  \& {Lizano}}]{Shu:1994kx}
{Shu}, F., {Najita}, J., {Ostriker}, E., {et~al.} 1994, \apj, 429, 781

\bibitem[{{Shu} {et~al.}(1987){Shu}, {Adams}, \& {Lizano}}]{Shu:1987fk}
{Shu}, F.~H., {Adams}, F.~C., \& {Lizano}, S. 1987, \araa, 25, 23

\bibitem[{{Snell} {et~al.}(1979){Snell}, {Loren}, \& {Plambeck}}]{Snell:1979fk}
{Snell}, R.~L., {Loren}, R.~B., \& {Plambeck}, R.~L. 1979, in Bulletin of the
  American Astronomical Society, Vol.~11, Bulletin of the American Astronomical
  Society, 713--+

\bibitem[{{Snell} {et~al.}(1980){Snell}, {Loren}, \& {Plambeck}}]{Snell:1980lr}
{Snell}, R.~L., {Loren}, R.~B., \& {Plambeck}, R.~L. 1980, \apjl, 239, L17

\bibitem[{{Valdivia-Mena} {et~al.}(2022){Valdivia-Mena}, {Pineda},
  {Segura-Cox}, {Caselli}, {Neri}, {L{\'o}pez-Sepulcre}, {Cunningham},
  {Bouscasse}, {Semenov}, {Henning}, {Pi{\'e}tu}, {Chapillon}, {Dutrey},
  {Fuente}, {Guilloteau}, {Hsieh}, {Jim{\'e}nez-Serra}, {Marino}, {Maureira},
  {Smirnov-Pinchukov}, {Tafalla}, \& {Zhao}}]{Valdivia-Mena:2022aa}
{Valdivia-Mena}, M.~T., {Pineda}, J.~E., {Segura-Cox}, D.~M., {et~al.} 2022,
  \aap, 667, A12

\bibitem[{{Watson}(2020)}]{Watson:2020wp}
{Watson}, D.~M. 2020, Research Notes of the American Astronomical Society, 4,
  88

\bibitem[{{Wurster} \& {Li}(2018)}]{Wurster:2018re}
{Wurster}, J. \& {Li}, Z.-Y. 2018, Frontiers in Astronomy and Space Sciences,
  5, 39

\bibitem[{{Yang} {et~al.}(2010){Yang}, {Stancil}, {Balakrishnan}, \&
  {Forrey}}]{Yang:2010vn}
{Yang}, B., {Stancil}, P.~C., {Balakrishnan}, N., \& {Forrey}, R.~C. 2010,
  \apj, 718, 1062

\bibitem[{{Yen} {et~al.}(2015){Yen}, {Takakuwa}, {Koch}, {Aso}, {Koyamatsu},
  {Krasnopolsky}, \& {Ohashi}}]{Yen:2015vf}
{Yen}, H.-W., {Takakuwa}, S., {Koch}, P.~M., {et~al.} 2015, \apj, 812, 129

\bibitem[{{Yen} {et~al.}(2018){Yen}, {Zhao}, {Koch}, {Krasnopolsky}, {Li},
  {Ohashi}, \& {Takakuwa}}]{Yen:2018rt}
{Yen}, H.-W., {Zhao}, B., {Koch}, P.~M., {et~al.} 2018, \aap, 615, A58

\end{thebibliography}
\clearpage
\begin{appendix}
\section{Supplementary material}
\label{sec:appendixa}
\renewcommand{\thesubsection}{\Alph{section}}
\begin{figure*}[htb]
   \centering
     \includegraphics[width=0.49\hsize]{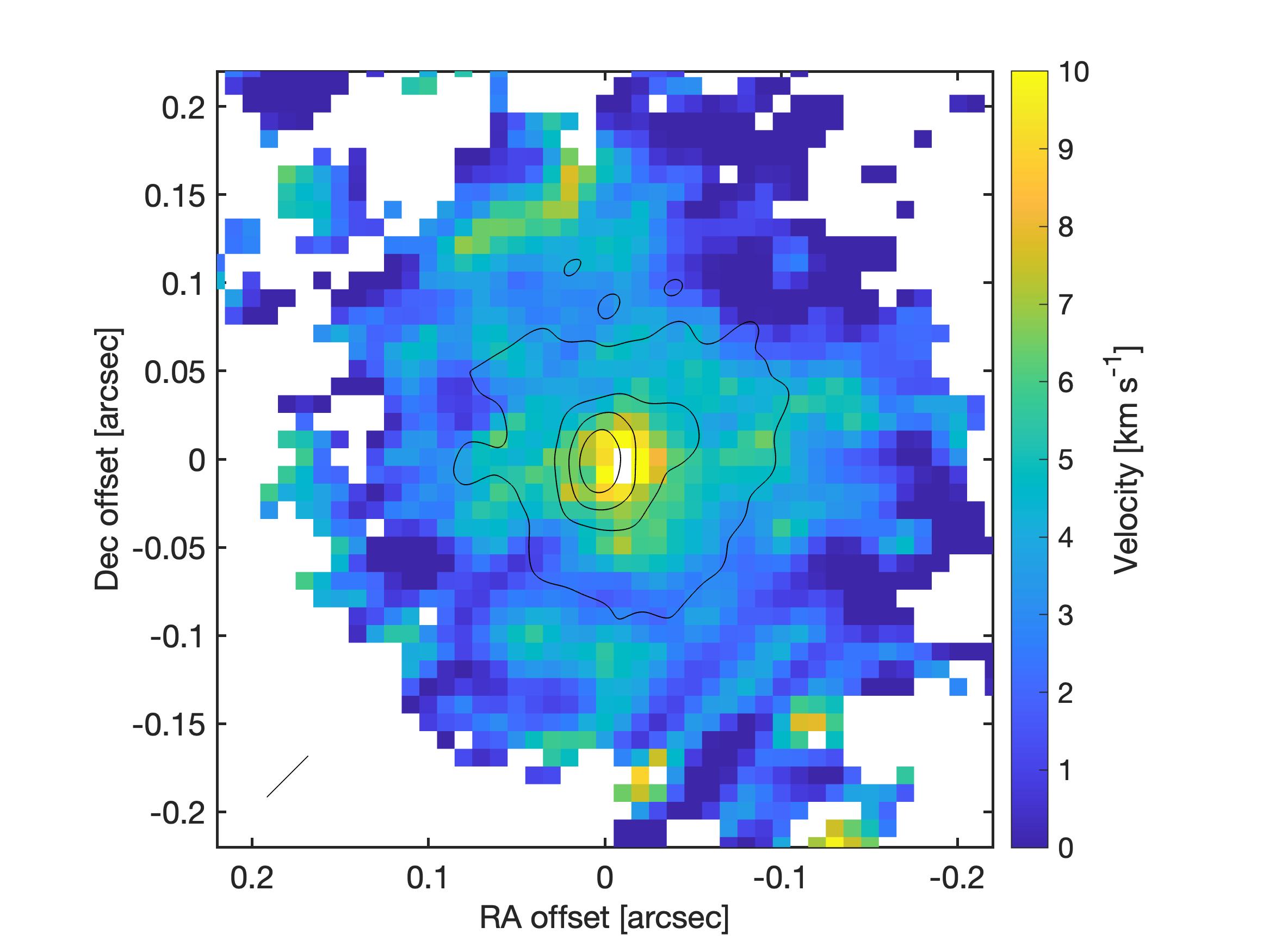} 
    \caption{Observed \trettenco\ intensity weighted velocity map (moment 1). Note that velocities have been corrected w.r.t \vlsr\ and the sign changed to allow comparison with Fig.~\ref{fig:infallfigure}. The first moment is calculated in a velocity range from --15 to +15~\kmpers\ with respect to \vlsr, and pixels where emission is below 3$\sigma$ ($\sigma~=~0.012~\rm{Jy~beam}^{-1}~\rm{\kmpers}$) are masked out. }
    \label{fig:momentmap}
\end{figure*}
\begin{figure*}[htb]
   \centering
    \begin{tabular}{c c}
     \includegraphics[width=0.48\hsize]{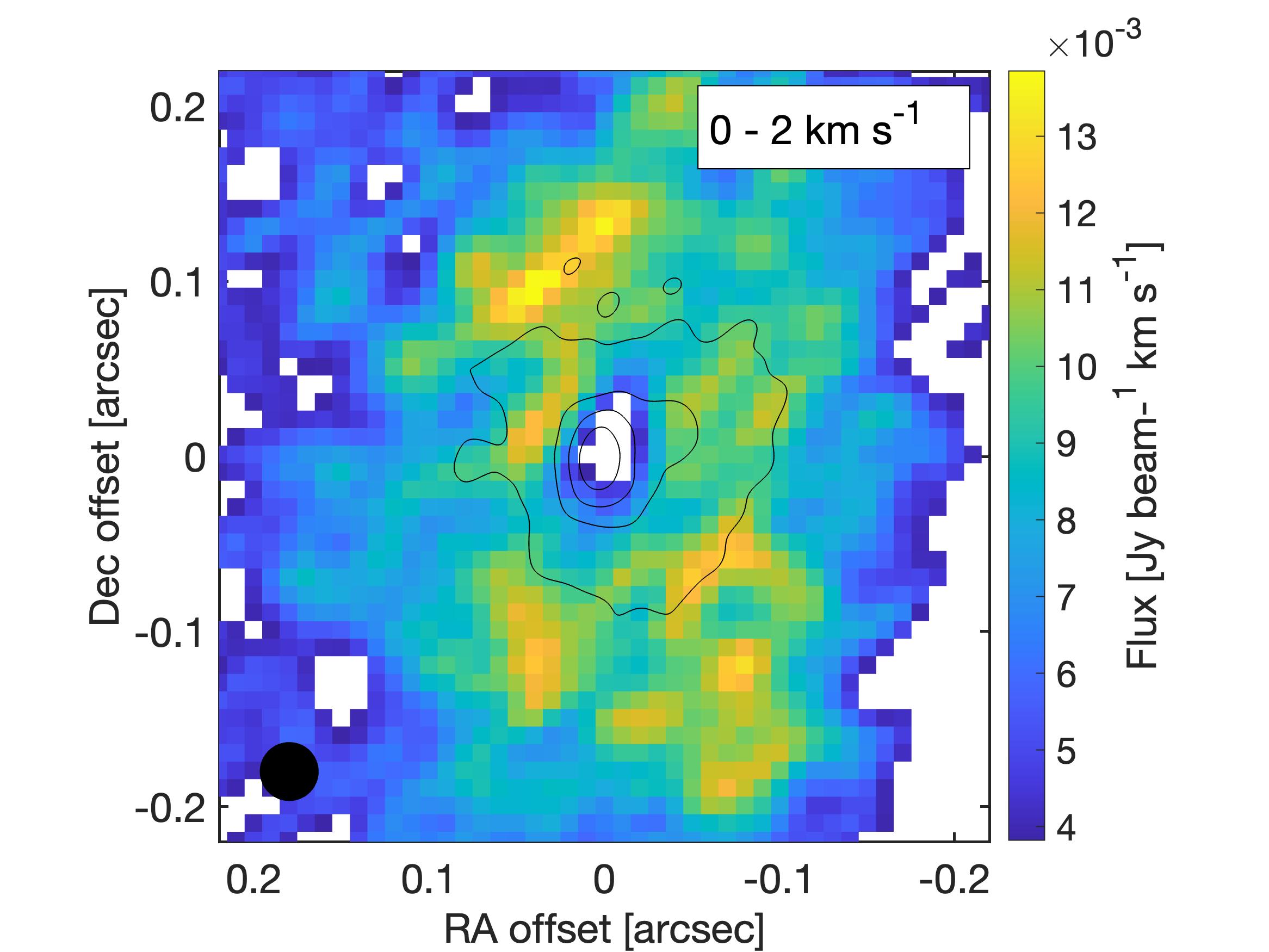} & \includegraphics[width=0.48\hsize]{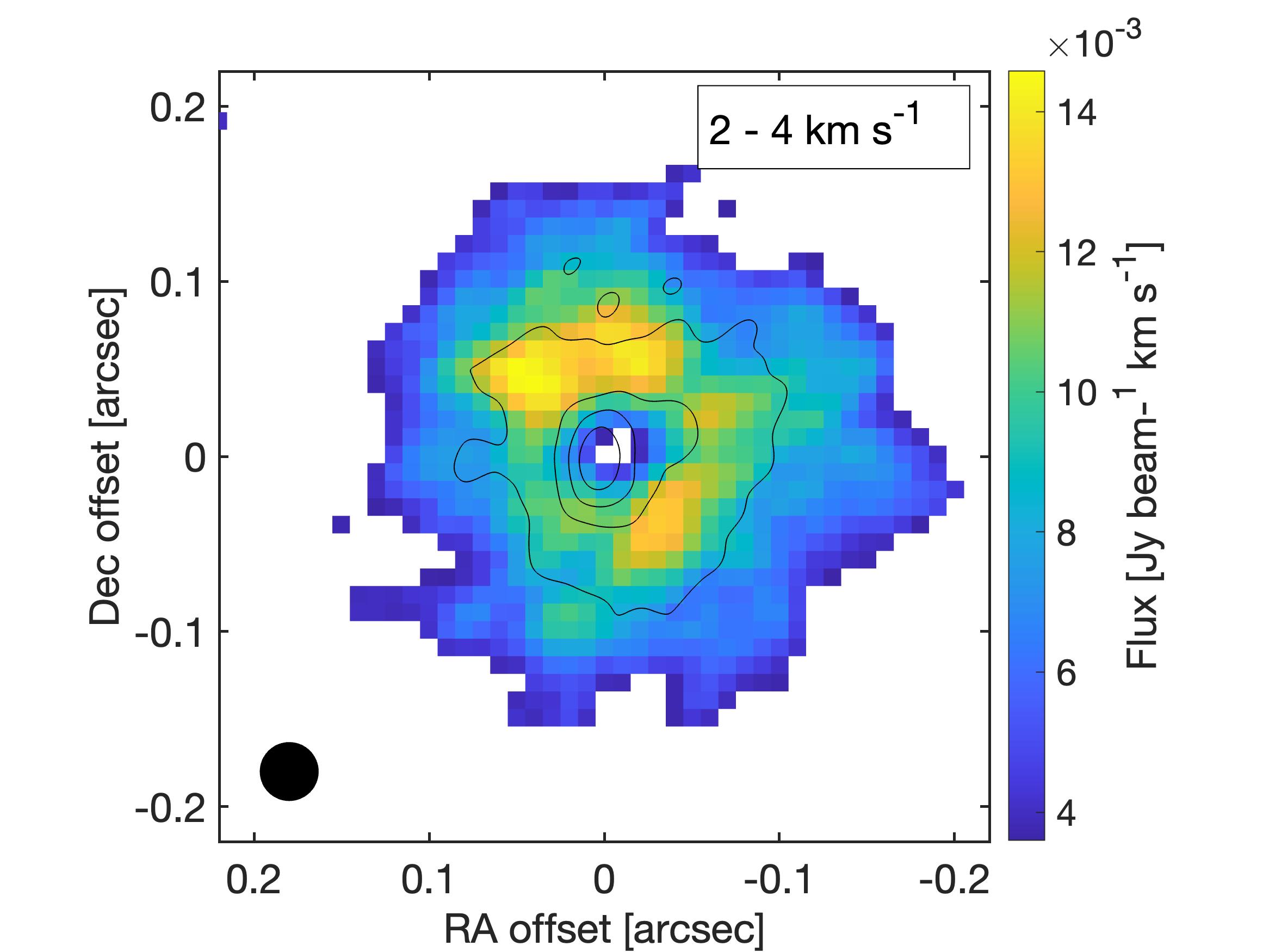} \\ 
     \includegraphics[width=0.48\hsize]{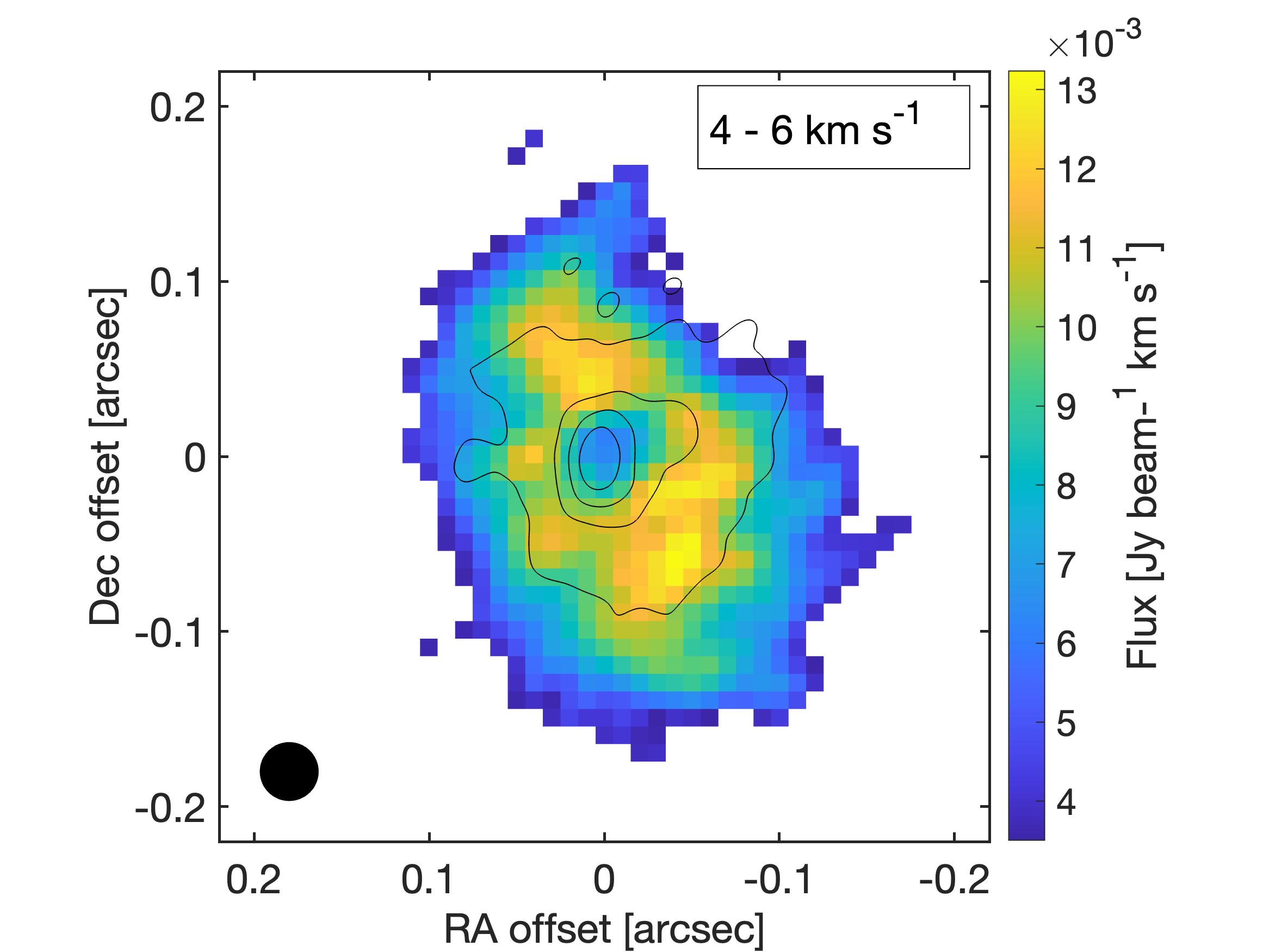} & \includegraphics[width=0.48\hsize]{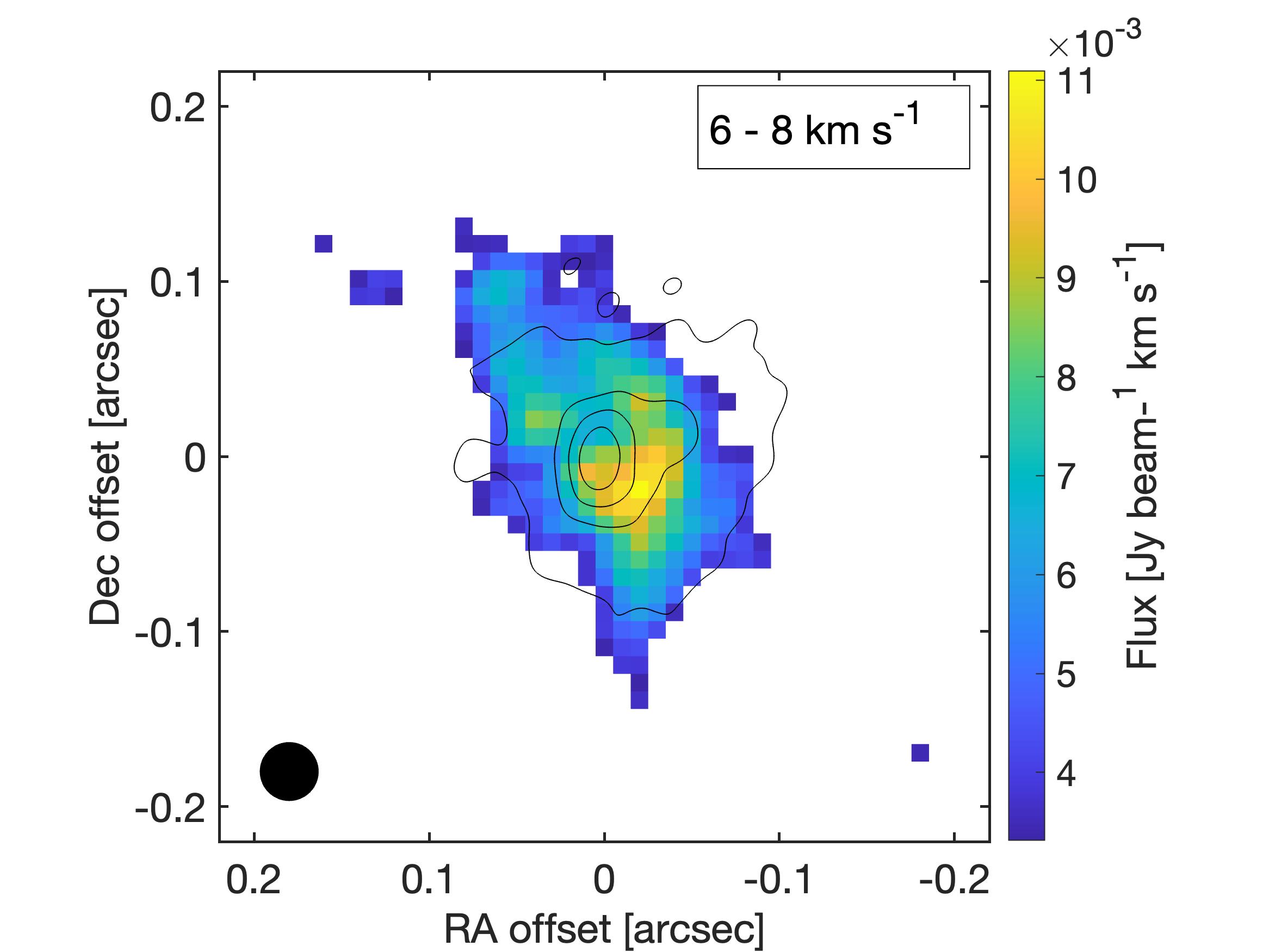} \\ 
    \end{tabular}
    \caption{Channel maps of the observed \trettenco\ emission in four different velocity bins. Only regions with emission above 3$\sigma$ are included, and hence it should be noted that the range of the color bars slightly varies between panels, with the lower end indicating the 3$\sigma$ level. Furthermore, velocities have been corrected w.r.t \vlsr\ and the sign changed to allow comparison with Fig.~\ref{fig:infallfigure}. }
    
    \label{fig:channelmap}
\end{figure*}
\begin{figure*}[htb]
   \centering
   \begin{tabular}{c c}
     \includegraphics[width=0.49\hsize]{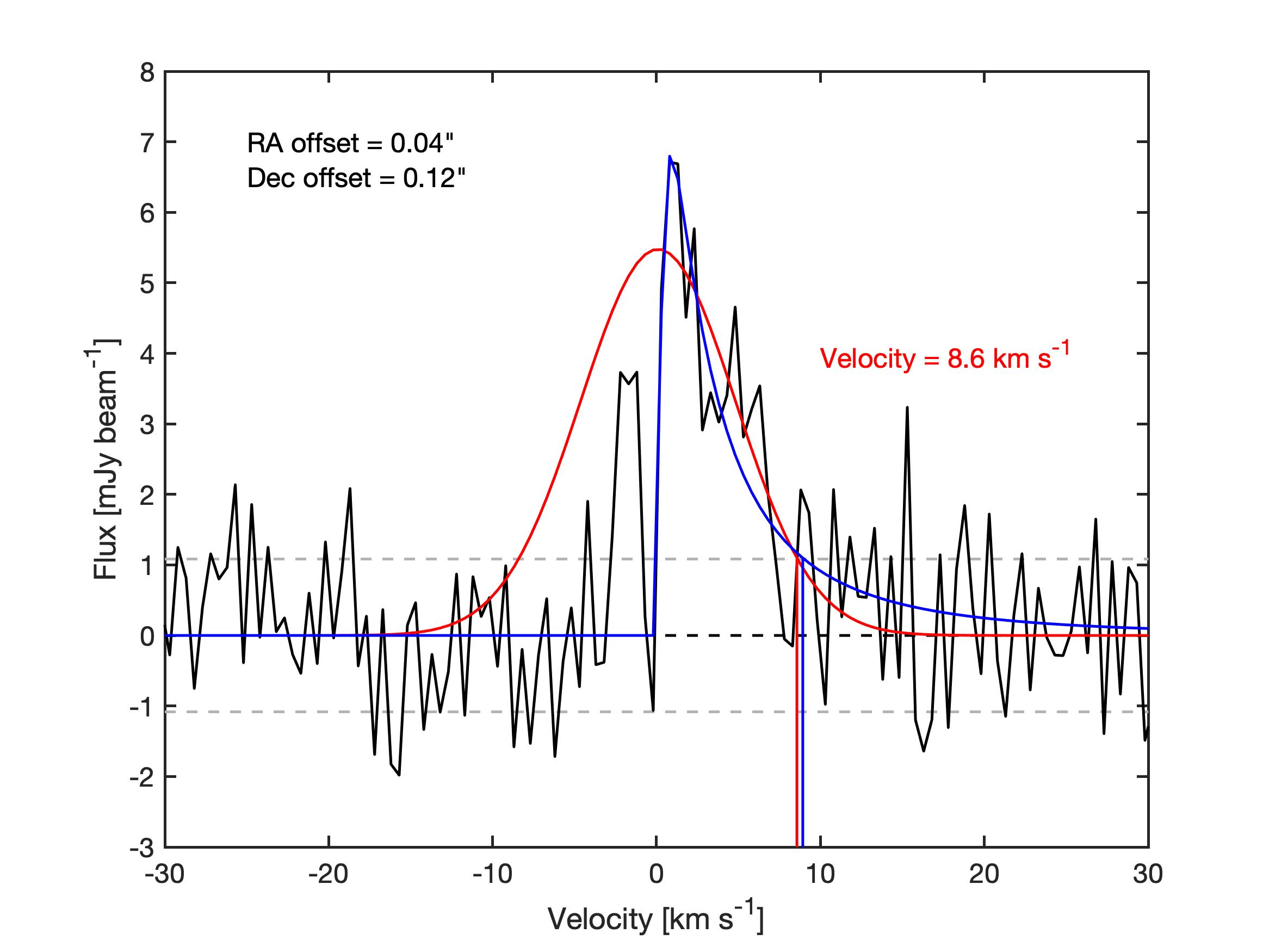} & \includegraphics[width=0.49\hsize]{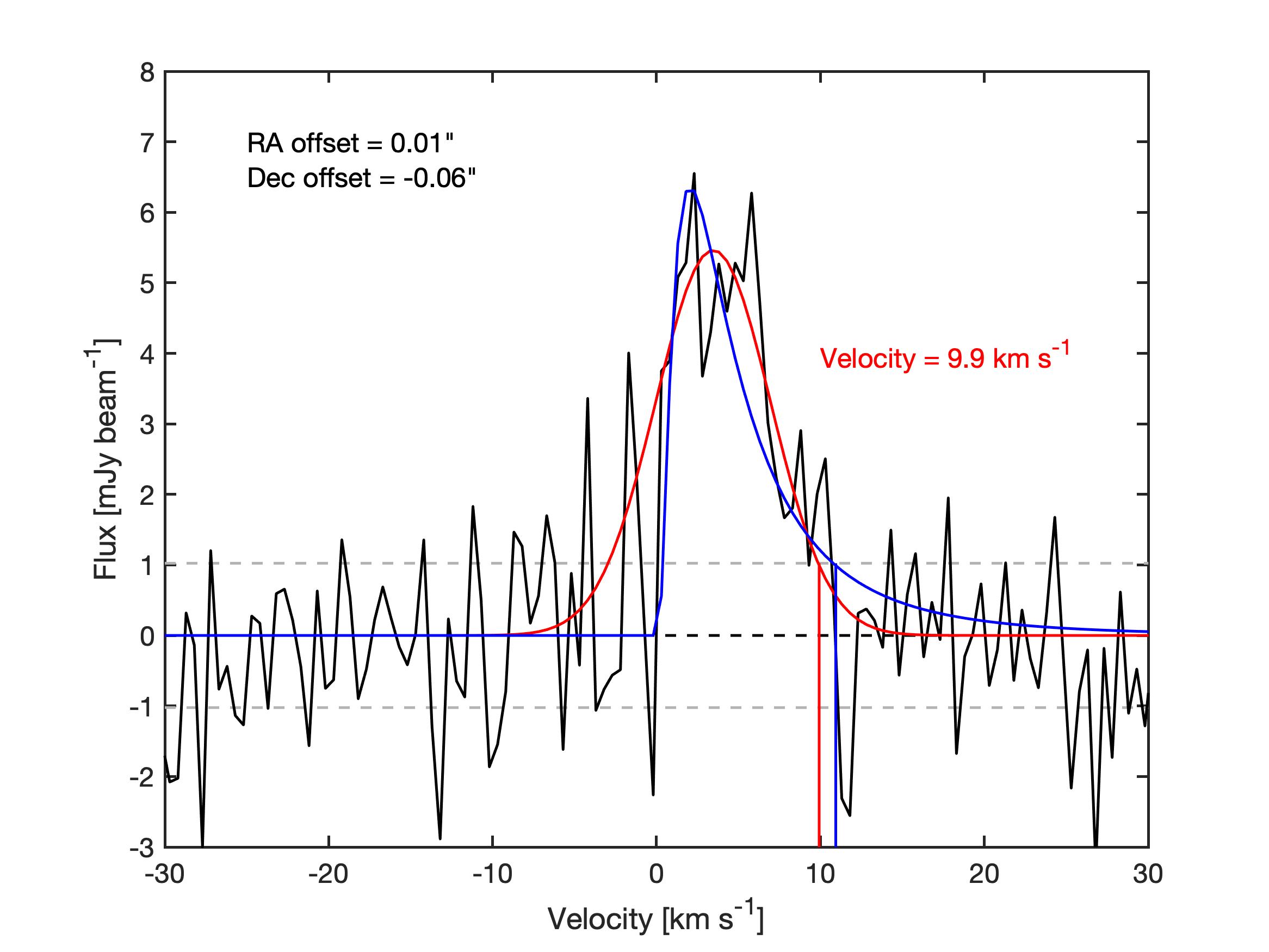} \\
     \includegraphics[width=0.49\hsize]{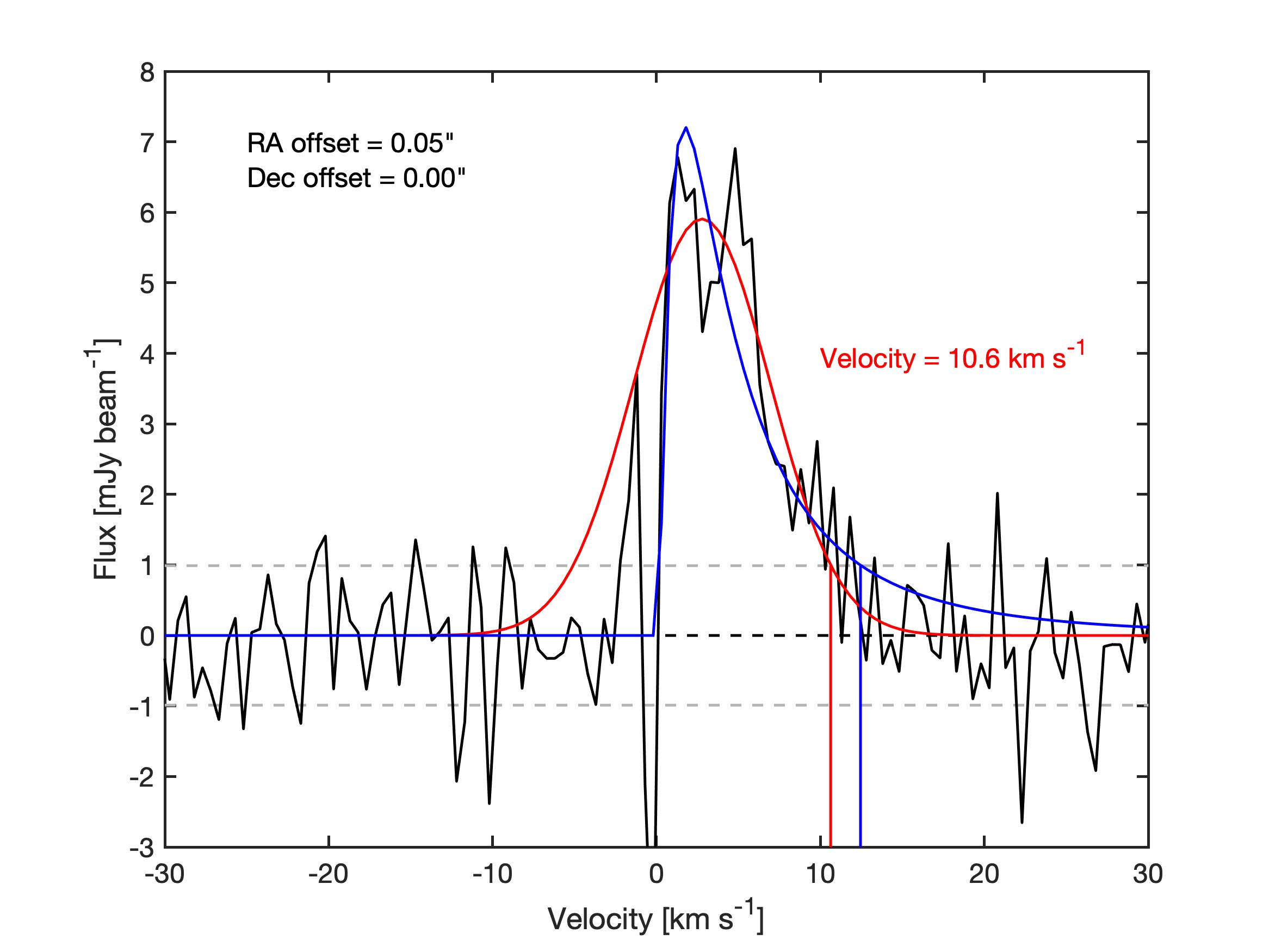} & \includegraphics[width=0.49\hsize]{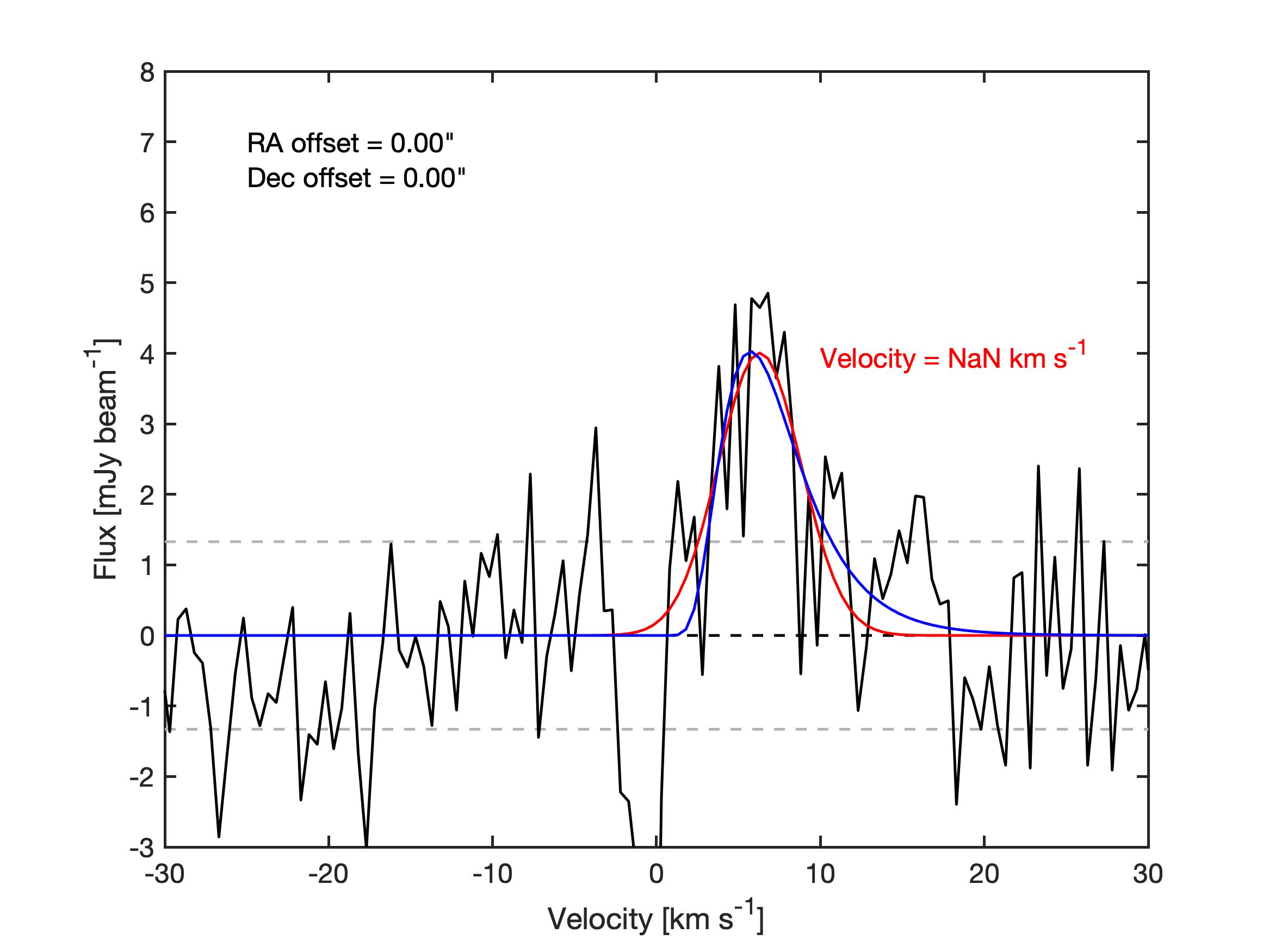} \\
    \end{tabular}
    \caption{Gaussian (red) and log-normal (blue) fits to the emission (black) in four different example positions at native spatial resolution. Note that the sign of the velocity scale has been changed and corrected w.r.t.\ \vlsr\ in order to capture a lower limit on the infall velocity for each position. The log-normal fits generally provide slightly higher velocity estimates (vertical blue lines) than the Gaussian fits (vertical red lines). The $\sigma$ level for each spectrum is indicated with a light grey dashed line and is calculated in the line emission free regions. Estimated velocities from the Gaussian fits for each of the spectra are indicated with red text. Note that the fit towards the central position is considered to be too poor for further analysis.}
    \label{fig:infallvelocity}
\end{figure*}

\begin{figure*}[ht]
   \centering
   \begin{tabular}{c c}
    a & b \\
    \includegraphics[width=0.45\hsize]{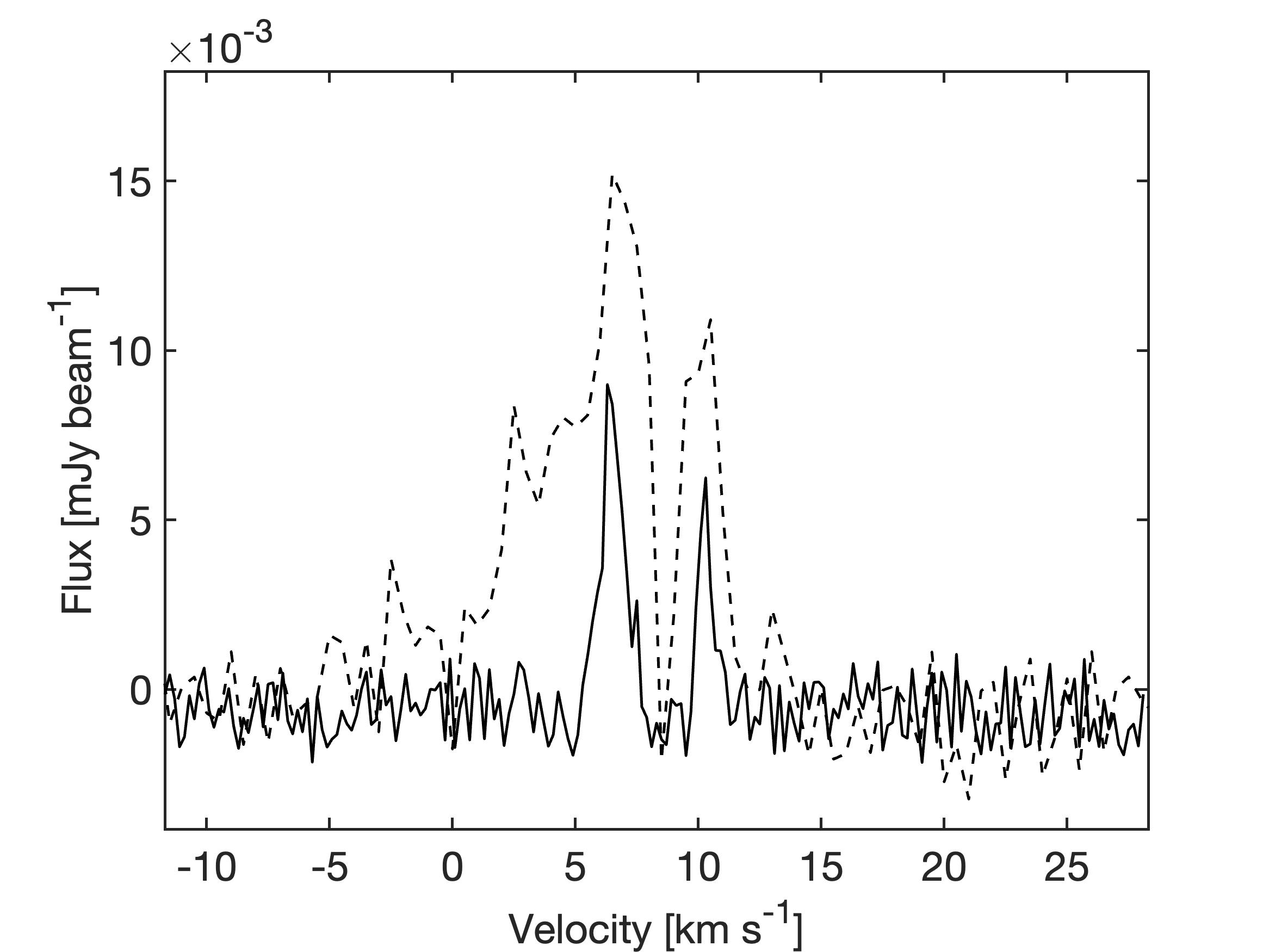}\put(-70,150){\makebox(0,0){Free-fall, \textit{M} = \textrm{0.1}~\msun}} & \includegraphics[width=0.45\hsize]{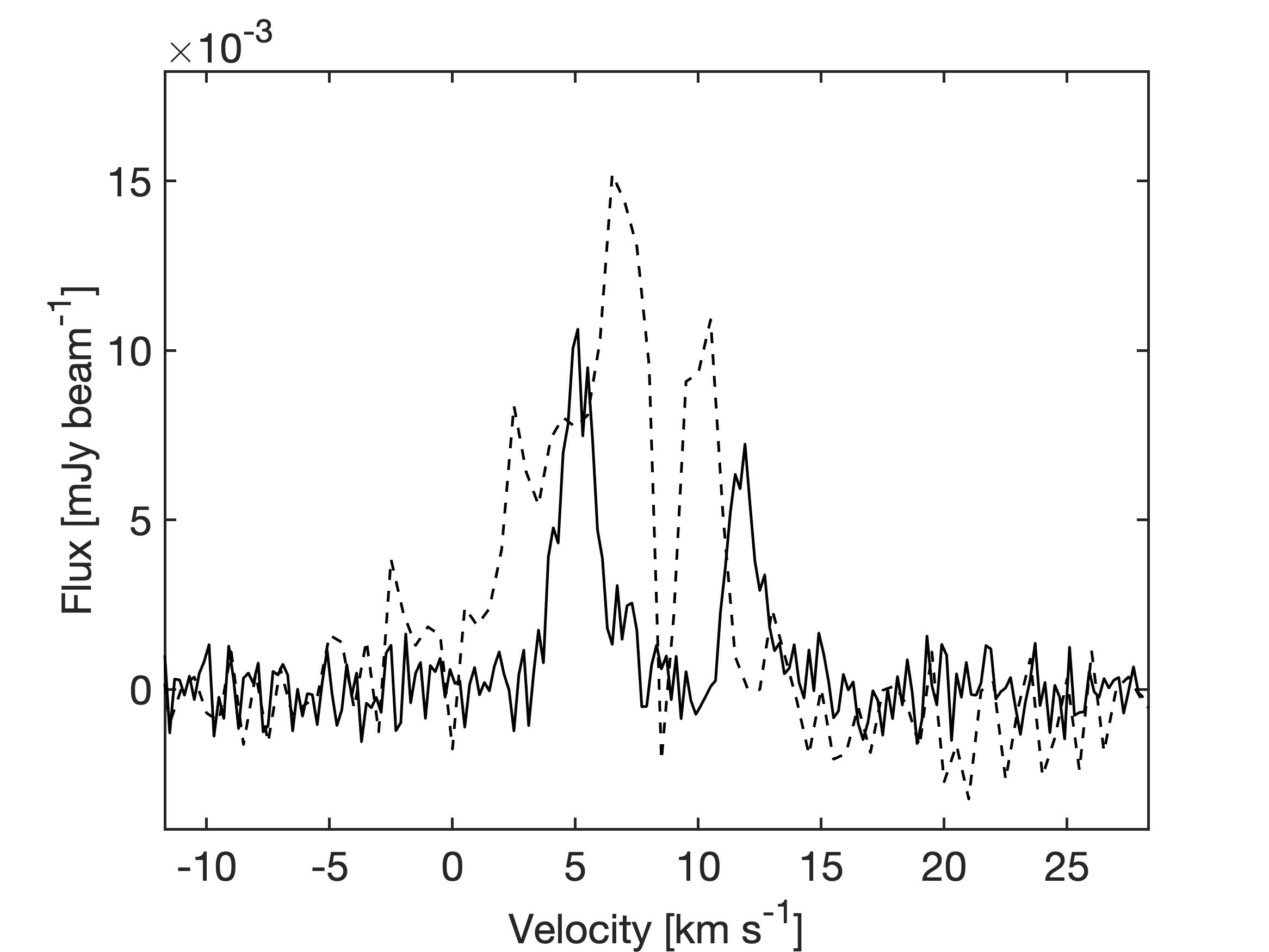}\put(-70,150){\makebox(0,0){Free-fall, \textit{M} = \textrm{0.4}~\msun}} \\
    c & d \\
    \includegraphics[width=0.45\hsize]{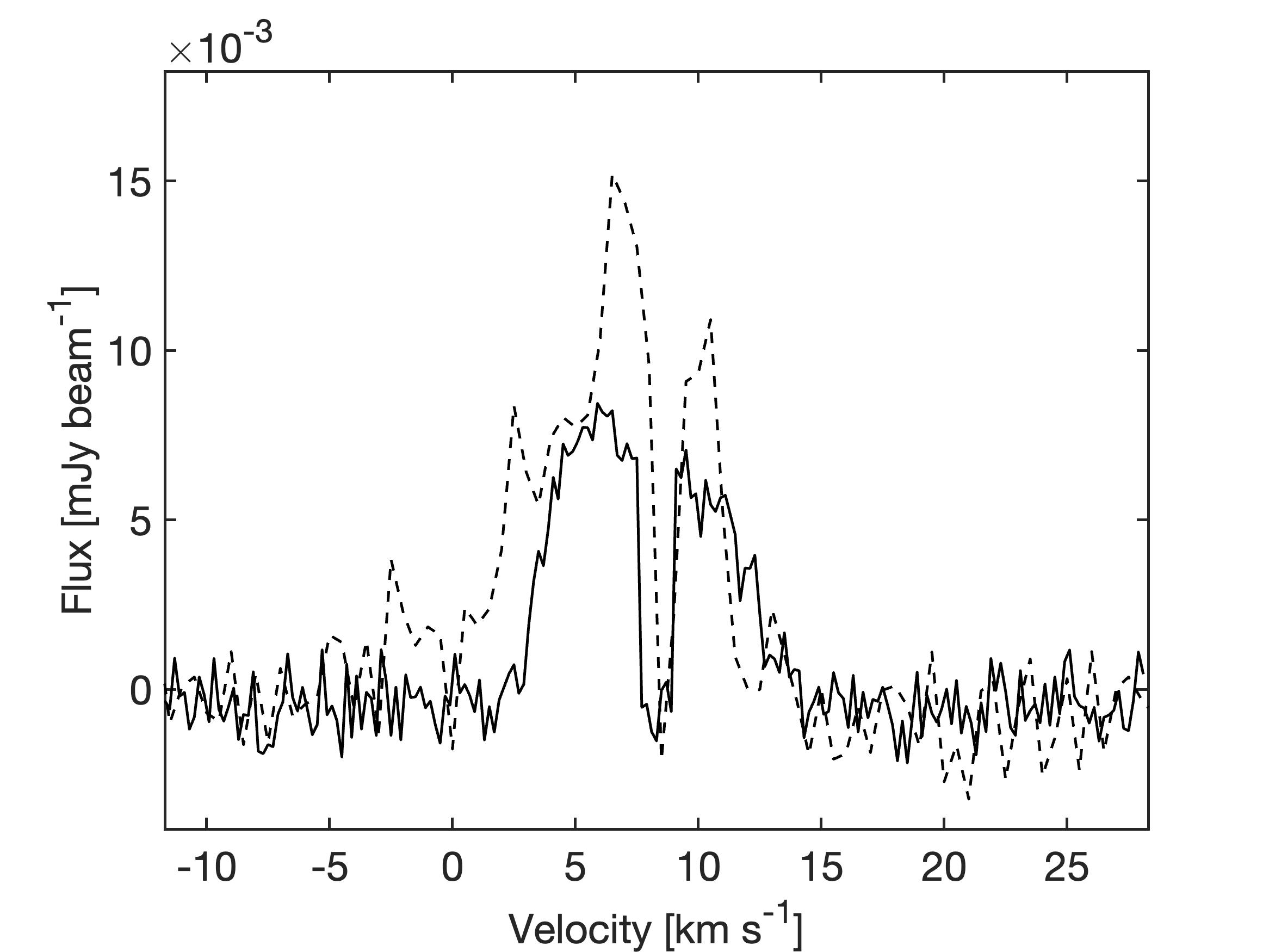}\put(-82,150){\makebox(0,0){Episodic infall, \textit{M} = \textrm{0.1}~\msun}} & \includegraphics[width=0.45\hsize]{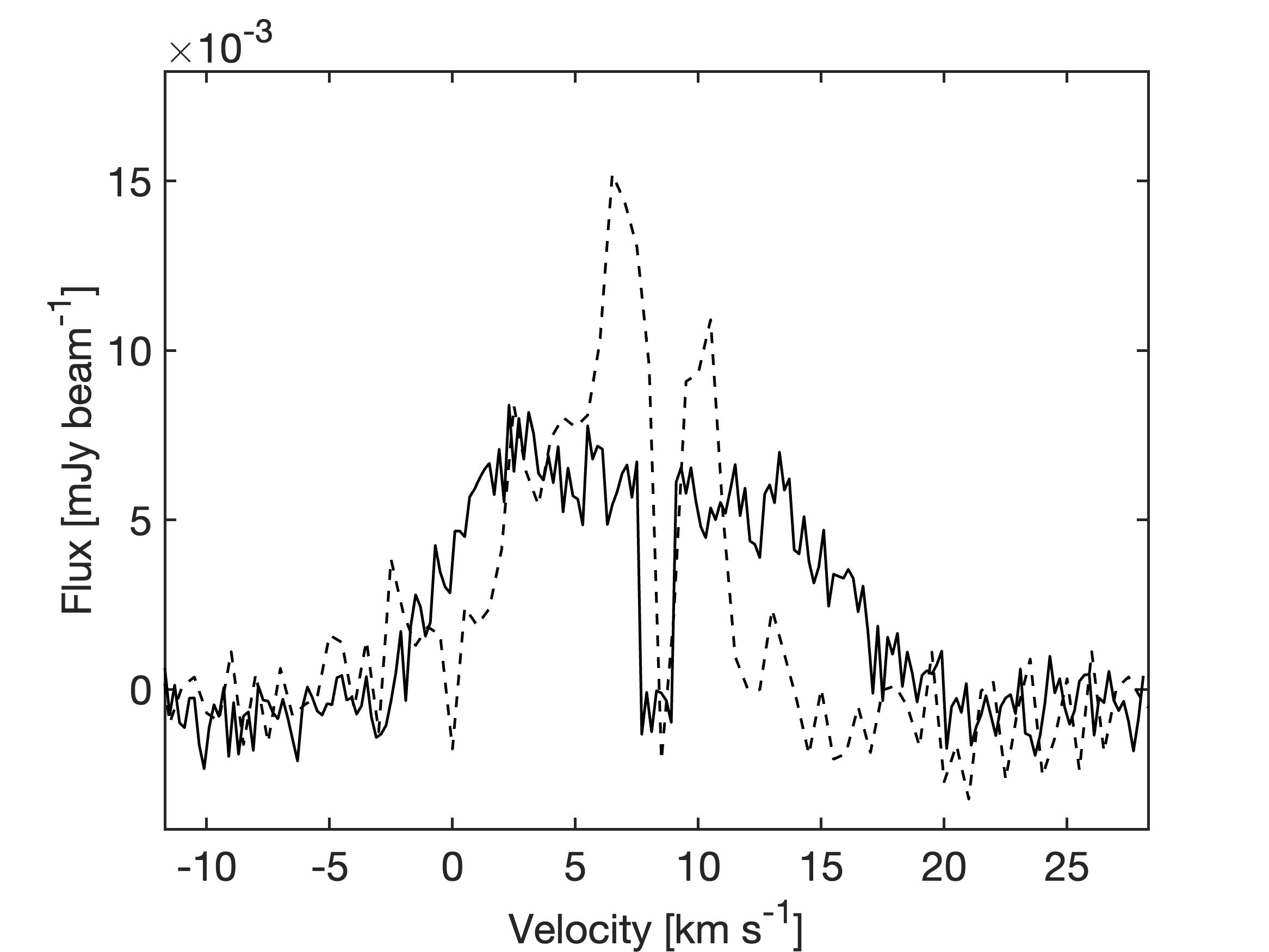}\put(-82,150){\makebox(0,0){Episodic infall, \textit{M} = \textrm{0.4}~\msun}} \\
    e & f \\
    \includegraphics[width=0.45\hsize]{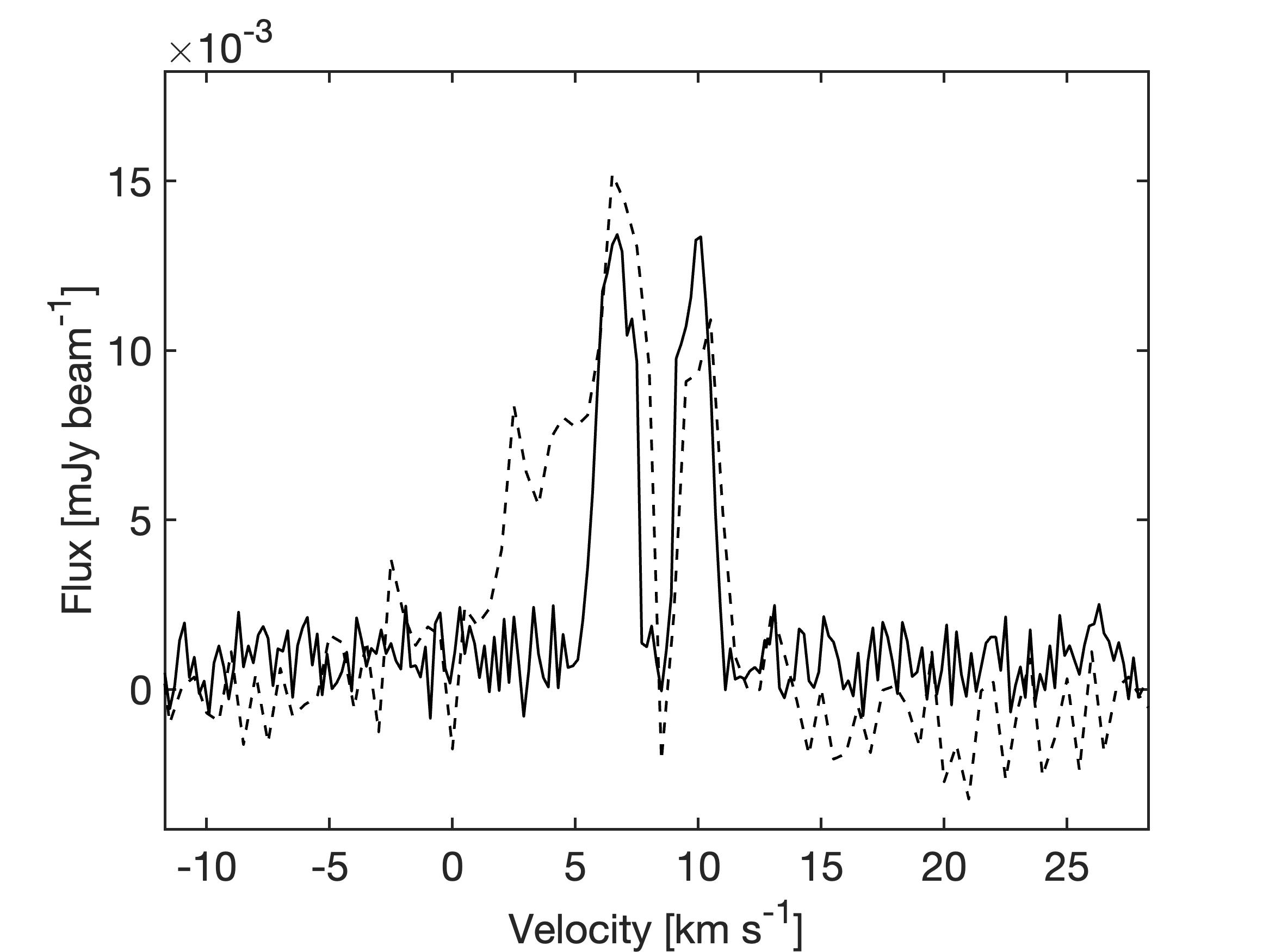}\put(-75,150){\makebox(0,0){Streamline, \textit{M} = \textrm{0.1}~\msun}} & \includegraphics[width=0.45\hsize]{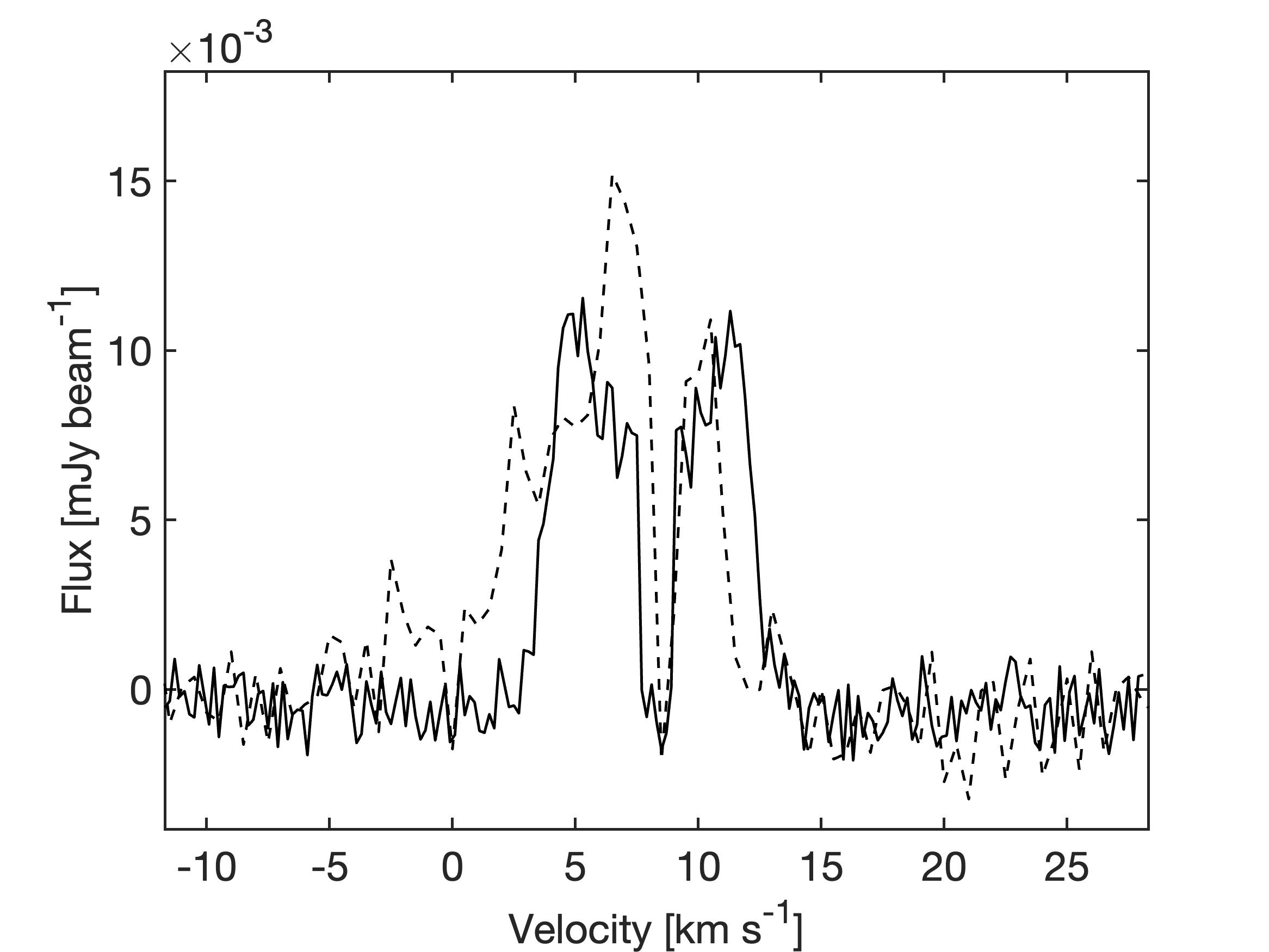}\put(-75,150){\makebox(0,0){Streamline, \textit{M} = \textrm{0.1}~\msun}} \\
   \end{tabular}
    \caption{Comparison between modelled (solid line) and observed (dashed; same in all panels) line profiles towards a position 20 au north of the protostar. a) Model free-fall line profile towards a 0.1~\msun\ protostar; b) Free-fall line profile towards a 0.4~\msun\ protostar; c) Episodic infall towards a 0.1~\msun\ protostar; d) Episodic infall towards a 0.4~\msun protostar; e) Streamline infall along cavity walls towards a 0.1~\msun protostar; f) Streamline infall along cavity walls towards a 0.4~\msun protostar.}
    \label{fig:LIMElineprofiles}
\end{figure*}

\begin{figure*}[ht]
   \centering
   \begin{tabular}{c c}
    \hspace{0cm}\includegraphics[width=0.45\hsize,trim={160 0 60 100},clip]{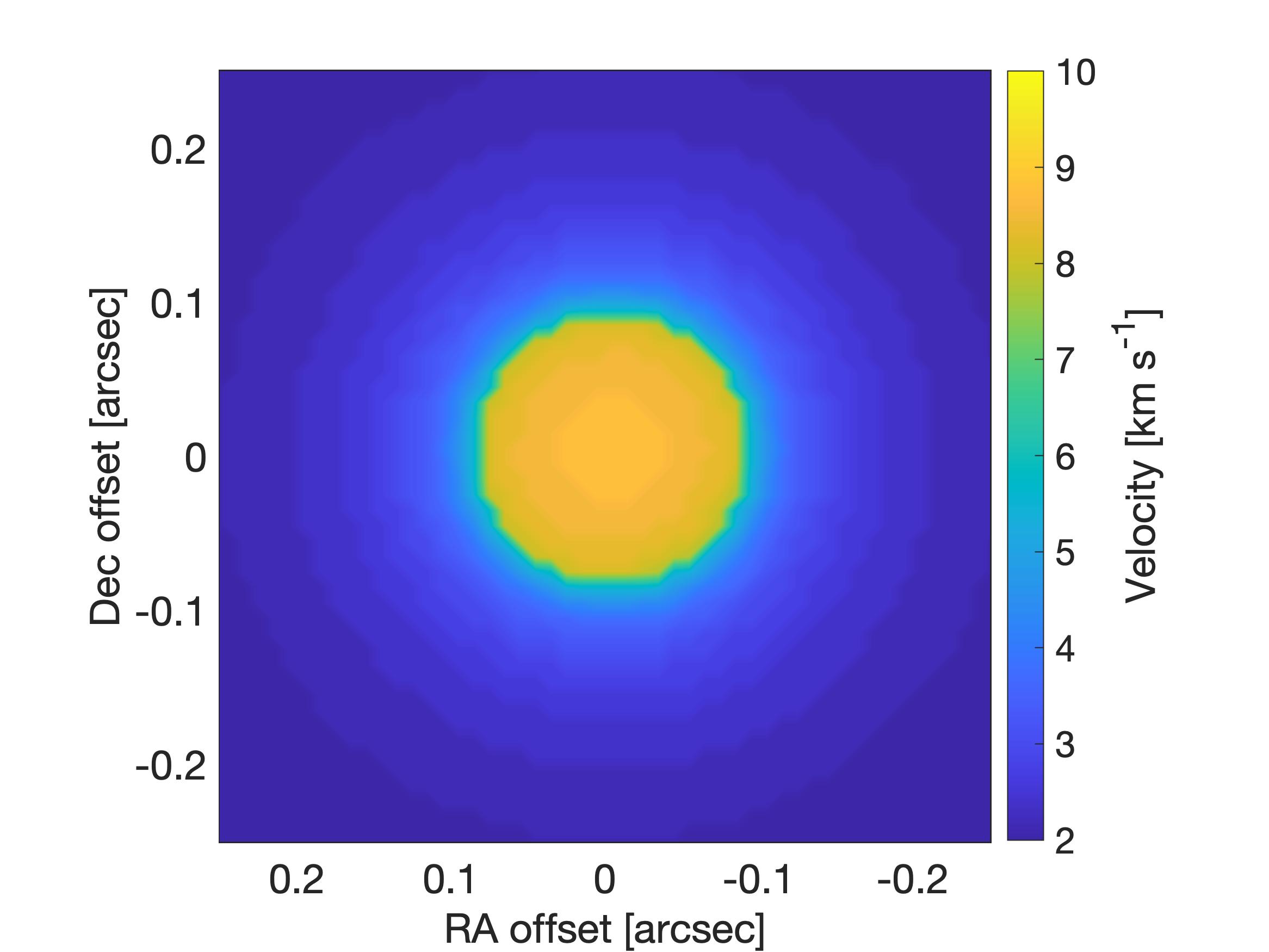}\put(-155,35){\makebox(0,0){\textcolor{white}{\sffamily Free-fall, \textit{M} = \textrm{\sffamily 0.1}~{\textit{M}$_{\odot}$}\xspace}}} & \hspace{0cm}\includegraphics[width=0.45\hsize,trim={160 0 60 100},clip]{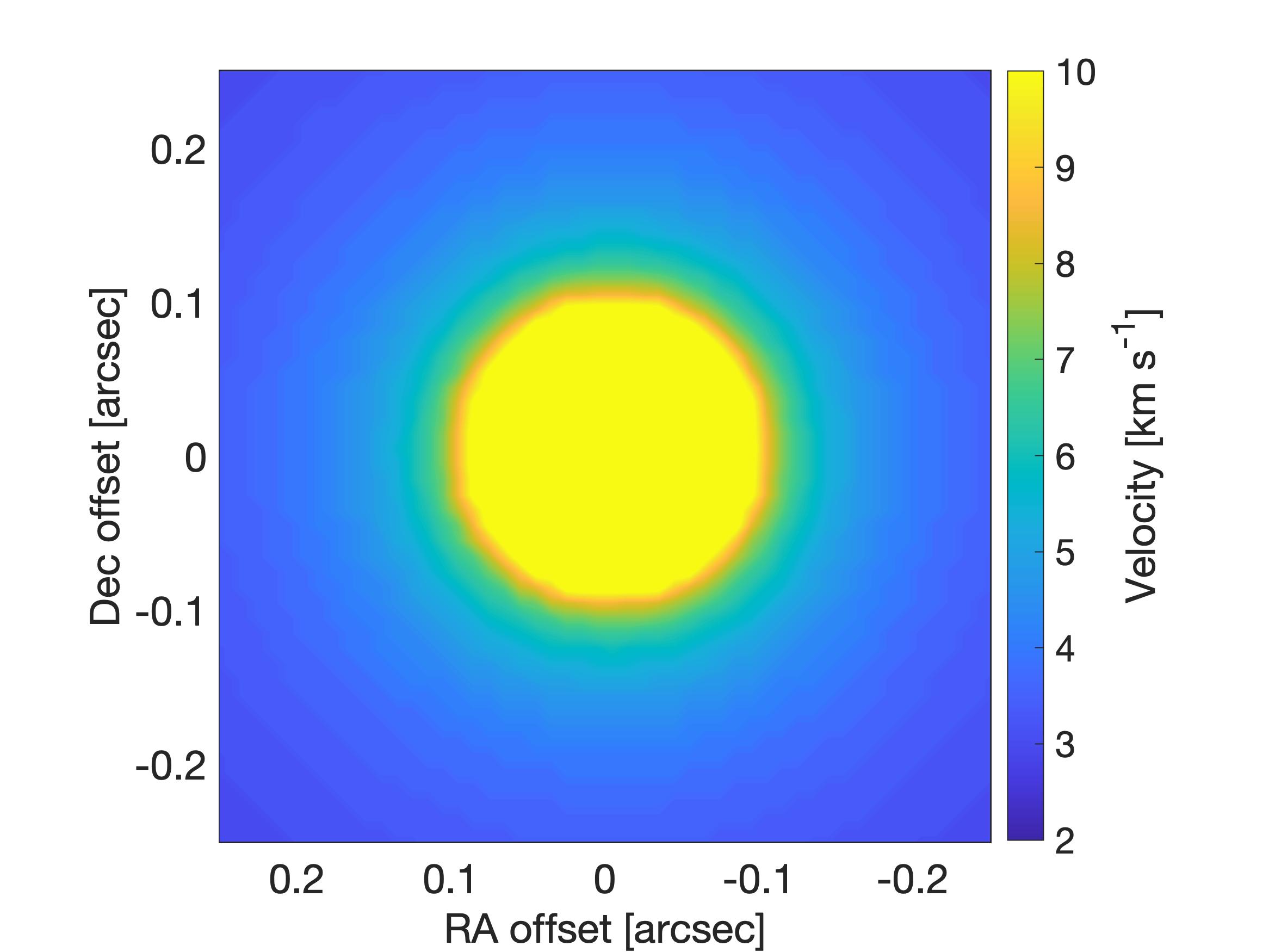}\put(-155,35){\makebox(0,0){\textcolor{white}{\sffamily Free-fall, \textit{M} = \textrm{\sffamily 0.4}~{\textit{M}$_{\odot}$}\xspace}}} \\
    \hspace{0cm}\includegraphics[width=0.45\hsize,trim={160 0 60 100},clip]{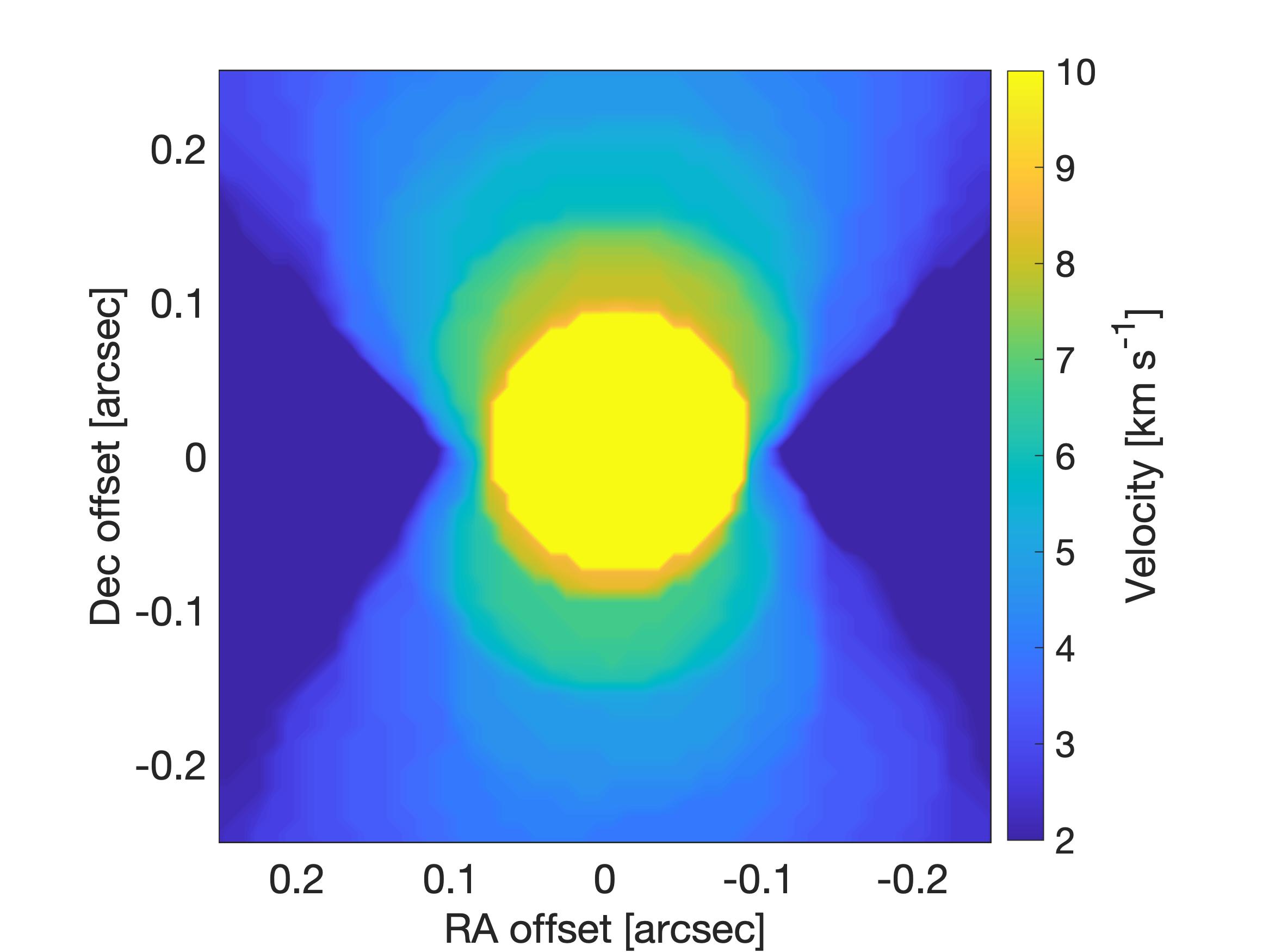}\put(-142,35){\makebox(0,0){\textcolor{white}{\sffamily Episodic infall, \textit{M} = \textrm{\sffamily 0.1}~{\textit{M}$_{\odot}$}\xspace}}} & \hspace{0cm}\includegraphics[width=0.45\hsize,trim={160 0 60 100},clip]{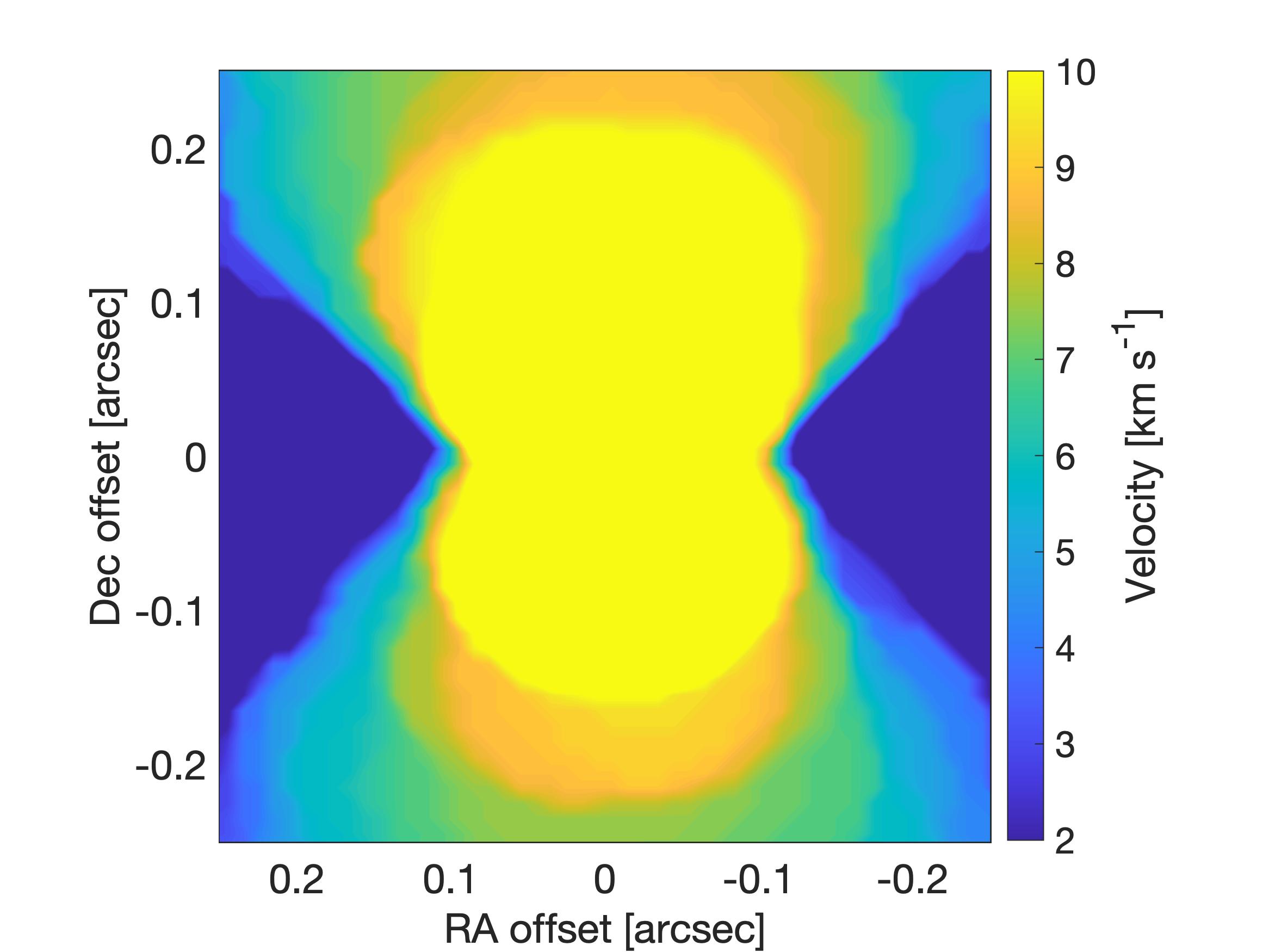}\put(-142,35){\makebox(0,0){\textcolor{black}{\sffamily Episodic infall, \textit{M} = \textrm{\sffamily 0.4}~{\textit{M}$_{\odot}$}\xspace}}} \\
    \hspace{0cm}\includegraphics[width=0.45\hsize,trim={160 0 60 100},clip]{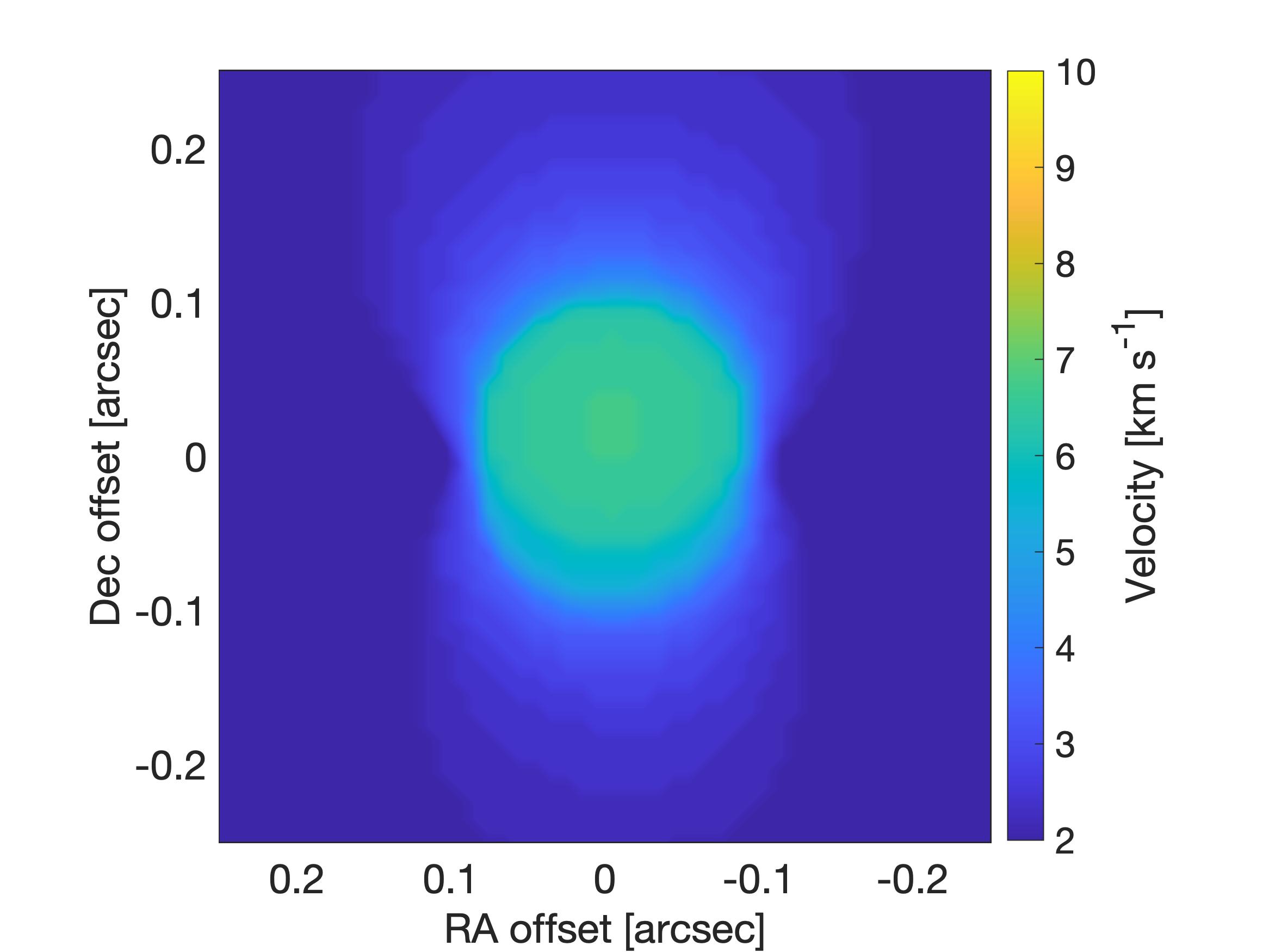}\put(-150,35){\makebox(0,0){\textcolor{white}{\sffamily Streamline, \textit{M} = \textrm{\sffamily 0.1}~{\textit{M}$_{\odot}$}\xspace}}} & \hspace{0cm}\includegraphics[width=0.45\hsize,trim={160 0 60 100},clip]{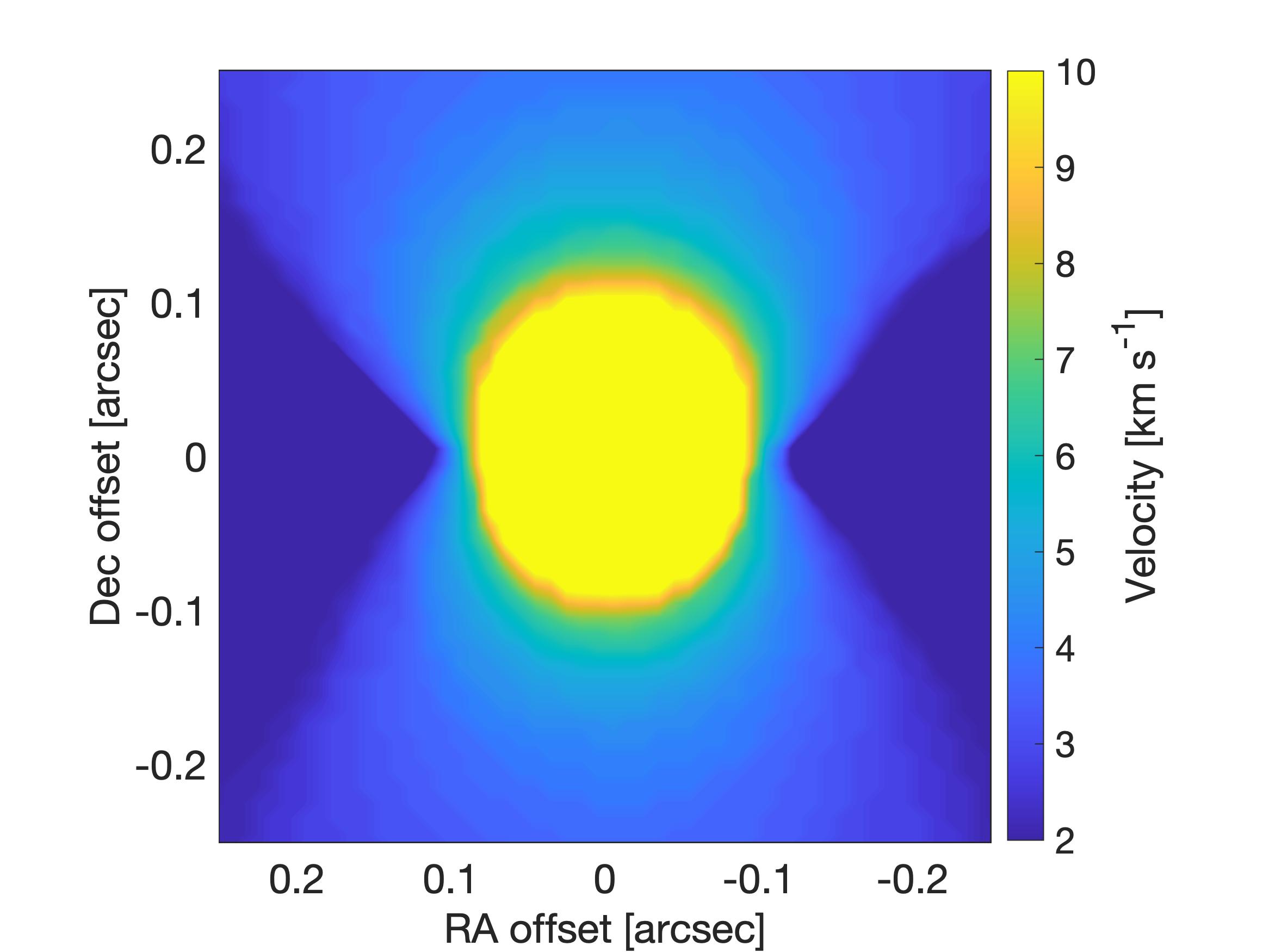}\put(-150,35){\makebox(0,0){\textcolor{white}{\sffamily Streamline, \textit{M} = \textrm{\sffamily 0.1}~{\textit{M}$_{\odot}$}\xspace}}} \\
    \end{tabular}
   \caption{\rmfamily Same as Fig.~\ref{fig:limemodels}, but \emph{before} processing the models with \texttt{simalma}.} 
   \label{fig:limemodels_simalma}
\end{figure*}

\end{appendix}

\end{document}